\documentclass[fleqn,usenatbib]{mnras}

\usepackage{graphicx}	
\usepackage{amsmath}	
\usepackage{amssymb}	
\usepackage[T1]{fontenc}
\usepackage{ae,aecompl}
\usepackage{txfonts}

\newcommand{\caii}{\hbox{Ca$\;${\sc ii}}}

\newcommand{\hal}{\hbox{H${\alpha}$}}

\def\bl{\hbox{$B_l$}}

\def\ms{\hbox{m\,s$^{-1}$}}
\def\kms{\hbox{km\,s$^{-1}$}}
\def\rad{\hbox{rad\,d$^{-1}$}}
\def\vsini{\hbox{$v \sin i$}}

\def\mearth{\hbox{${\rm M}_{\rm \oplus}$}}
\def\msun{\hbox{${\rm M}_{\odot}$}} 
\def\rsun{\hbox{${\rm R}_{\odot}$}} 
\def\lsun{\hbox{${\rm L}_{\odot}$}}

\def\mstar{\hbox{$M_{\star}$}} 
\def\rstar{\hbox{$R_{\star}$}} 
\def\lstar{\hbox{$L_{\star}$}} 

\def\pstar{\hbox{$P_{\rm rot}$}}

\def\vt{\hbox{$v_{\rm t}$}}
\def\vb{\hbox{$v_{\rm b}$}}
\def\vs{\hbox{$v_{\rm s}$}}
\def\vr{\hbox{$v_{\rm r}$}}

\title[Modeling the RV jitter using tomographic imaging]{Modeling the RV jitter of early M dwarfs using tomographic imaging}

\author[{\'E}.M. H{\'e}brard et al.]{
\'E.M. H{\'e}brard $^{1,2,3}$\thanks{E-mail: ehebrard@yorku.ca},
J.-F. Donati $^{2,3}$,
X. Delfosse $^{4,5}$,
 J. Morin $^{6}$,
  C. Moutou $^{7,8}$ and
\newauthor I. Boisse $^{8}$
\\
$^1$ Department of Physics and Astronomy, York University, Toronto, ON M3J 1P3, Canada\\
$^2$ CNRS-INSU; IRAP-UMR 5277, 14 Av.\ E.~Belin, F--31400 Toulouse, France \\
$^3$ Universit{\'e} de Toulouse; UPS-OMP; IRAP; Toulouse, France \\
$^4$ Univ. Grenoble Alpes, IPAG, F-38000 Grenoble, France \\
$^5$ CNRS, IPAG, F-38000 Grenoble, France \\
$^6$ LUPM, Universit{\'e} Montpellier II, CNRS, UMR 5299, Place E. Bataillon, 34095 Montpellier, France\\
$^7$ CNRS, Canada-France-Hawaii Telescope Corporation, 65-1238 Mamalahoa Hwy, Kamuela, HI 96743, USA \\
$^8$ Aix-Marseille Universit{\'e}, CNRS, LAM, UMR 7326, 13388 Marseille, France
}

\date{Accepted XXX. Received YYY; in original form ZZZ}

\pubyear{2015}

\begin{document}
\label{firstpage}
\pagerange{\pageref{firstpage}--\pageref{lastpage}}
\maketitle

\begin{abstract}

In this paper we show how tomographic imaging (Zeeman Doppler Imaging, ZDI) can be used to characterize stellar activity and magnetic field topologies, ultimately allowing to filter out the radial velocity (RV) activity jitter of M-dwarf moderate rotators. This work is based on spectropolarimetric observations of a sample of five weakly-active early M-dwarfs (GJ 205, GJ 358, GJ 410, GJ479, GJ 846) with HARPS-Pol and NARVAL. These stars have \vsini\ and RV jitters in the range 1-2~\kms\ and 2.7-10.0~\ms\ rms respectively. 

Using a modified version of ZDI applied to sets of phase-resolved Least-Squares- Deconvolved (LSD) profiles of unpolarized spectral lines, we are able to characterize the distribution of active regions at the stellar surfaces. We find that darks spots cover less than 2\% of the total surface of the stars of our sample. Our technique is efficient at modeling the rotationally modulated component of the activity jitter, and succeeds at decreasing the amplitude of this component by typical factors of 2-3 and up to 6 in optimal cases. From the rotationally modulated time-series of circularly polarized spectra and with ZDI, we also reconstruct the large-scale magnetic field topology. These fields suggest that bi-stability of dynamo processes observed in active M dwarfs may also be at work for moderately active M dwarfs. Comparing spot distributions with field topologies suggest that dark spots causing activity jitter concentrate at the magnetic pole and/or equator, to be confirmed with future data on a larger sample.
\end{abstract}




\section{Introduction} 
Lots of exoplanets were either detected or confirmed thanks to the radial velocity (RV) technique which allows one to detect exoplanets of various masses and sizes, from hot-Jupiters to super Earths. This is made possible thanks to the sensitivity and stability of current velocimeters. However, as an indirect method based on measuring spectral shifts, velocimetry is also sensitive to phenomena of intrinsic stellar origin capable of affecting spectra, and in particular to stellar activity. Whatever the precision of forthcoming instruments, we will remain confronted with this limitation, rendering Earth-like planets hard to detect, their spectral signatures being much smaller than the activity-induced RV jitter, even for weakly active Sun-like stars.

Signals of stellar origin can occur on different timescales; some have a short period, typically ranging from minutes to hours (\emph{e.g.,} flares, granulation), whereas some feature a longer period, ranging from days to year (\emph{e.g.,} activity cycle, spot or convection inhibition from a strong magnetic field modulated by the rotation). 
Whatever the temporal timescale, most stellar phenomena causing spectral variability are related to magnetic fields and to the associated activity demonstrations.  
The modeling of the RV jitter is essential to all extrasolar planets searches, especially when orbital periods are larger than a few days and when the host stars exhibit activity phenomena occurring on timescales commensurate with the planetary signals of interest.
The only way to improve the sensitivity of RV surveys to small planets is to characterize and model the activity jitter as well as possible. 

To diagnose the RV jitter, several complementary approaches are commonly used, mostly making use of chromospheric activity indicators like excess flux in the cores of the \hal\ and \caii\ H\&K, or measurements of spectral line asymmetries (with the bisector of the cross-correlation function).
 The idea is to check for periodic modulation of these proxies, in order to assess whether the observed RV signal is caused by activity rather than by a planet \citep[see][]{Queloz01}. 
The correlation between RVs and the slope of the bisector can in principle be used to correct for the effect of activity at a level of a few \ms\ \citep{Boisse09}. The accuracy to which the RV jitter can be corrected with this method largely depends on various parameters, e.g., the distribution of spots, the stellar inclination, the rotational broadening of line profiles. An alternative method is based on exploiting complementary information from velocimetric and high-cadence photometric simultaneous/contemporaneous observations, and make use of the predicted relationship between the photometric and RV signatures of spots \citep{Aigrain12, Haywood14}. These studies found that RV modulation caused by spots can be reliably modeled using the photometric flux $F$ and its first derivative $F'$. 
Other studies \citep{Meunier10, Borgniet15} use the Sun as a star to predict the effect of activity on conventional activity diagnostics, taking advantage of the wealth of existing data. However activity, and its correlation to RV jitter, depends strongly on spectral type, stellar mass and rotation rate, and so far, no studies are available to reliably extrapolate the solar case to all types of active stars.  

Besides, the large majority of extrasolar planets up to now was found around main-sequence stars of spectral types ranging from late-F to early-M. Despite M dwarfs are the most abundant type of stars, their intrinsic faintness in the visible domain caused them to be underrepresented in RV surveys with existing instruments. 
RV and transiting survey demonstrate than planets is very frequent around M-dwarfs, in particular Earth and Super-Earth at short period,\citep{Bonfils13, Dressing15}. 
Moreover, due to their low masses, these targets are interesting for Earth-like planet hunting; an orbiting planet of a given mass and orbital distance generates a higher reflex motion when orbiting around a M dwarf than around a solar type star. 
Therefore, observations of low-mass stars is a promising option to increase our sensitivity to Earth-like planets. Due to the low photospheric temperature of M-dwarfs, the planetary orbits that are located in the habitable zone (HZ) of the host star (\emph{i.e.}, within the proper range of orbital distances where liquid water may be stable at the planet surfaces) move closer in.
For instance, for a M-dwarf with a typical mass of 0.5~\msun\ (like those studied in our paper), the HZ lies in a range 0.2 -- 0.45 AU \citep[see][]{Kasting14}. It corresponds to orbital periods in the range 36 -- 157 d, i.e., to RV semi-amplitude of 1.5 -- 0.96 \ms\ for a planet mass of 5~\mearth\ (as opposed to ~0.4~\ms\ for a planet of the same mass orbiting in the HZ of a Sun-like star).

Despite the gain in the RV sensitivity to small planets that M dwarfs allow to achieve, modeling and filtering efficiently the activity jitter of the host stars remain essential, given that this activity jitter is still at best comparable in size to the RV signal we aim at detecting, and with a similar period as orbital periods of planets within the HZ \citep[see, \emph{e.g.,}][]{Forveille09, Robertson14}. 
So far, studies of late M dwarfs have shown that these stars exhibit significant RV jitter mostly induced by dark spots at their surfaces, implying that efficient observational strategies are mandatory to reliably disclose planets orbiting around them \citep[\emph{e.g.,}][for M5-M9]{Barnes14}. These studies rely on simulations and/or spectroscopic and/or photometric survey to diagnose the activity jitter. However, mainly due to their low luminosity in the optical to nIR domain, wether the predominant spot pattern is random, uniform or concentrated at active latitudes remains unclear \citep[][and references therein]{Barnes11, Andersen15}.
In this paper, we propose to explore a new method based on simultaneously studying the RV jitter caused by activity, and Zeeman signatures reflecting the large-scale magnetic field at the origin of activity 
to (i) investigate the level to which spot distributions causing the RV jitter relates to magnetic topologies and (ii) devise a new technique based on spectropolarimetric data to filter out activity on a sample of early M dwarfs. 

We present the results of a spectropolarimetric campaign carried out on September 2013 - September 2014. After a brief description of the stellar sample in Sec.~\ref{sec:sample} and of the data reduction procedure in Sec.~\ref{sec:outil}, we present the results obtained by analyzing circularly polarized spectra (Stokes $V$) in Sec.~\ref{sec:magn}. The stellar activity diagnostic is introduced in Sec.~\ref{sec:diag}, and is followed by the analysis of the rotational modulation of the RV jitter, and of its modeling in Sec.~\ref{sec:rv}. 
The magnetic field and brightness reconstruction procedure using Zeeman-Doppler imaging (ZDI) are presented in Sec.~\ref{sec:zdi} and \ref{sec:di}. We summarize the main outcome of this analysis and discuss its implications in Sec.~\ref{sec:dis}.

\section{Spectropolarimetric observations}
\subsection{Stellar sample}
\label{sec:sample}
Our stellar sample includes five weakly-active, early-M dwarfs with different rotation periods (spanning 11-33~d) and stellar masses (0.35-0.61~$\msun$). 
The selected targets are among the most observed and best characterized ones in the ESO/HARPS RV survey of M dwarfs \citep{Bonfils13}, guaranteeing that their activity jitters are known (with rms in the range 2.7-10.0~\ms) and detectable. So far no planets are detected for the stars of the sample.
The five targets are known to show RV variations mostly caused by activity  \citep{Bonfils13, Donati08c}. 
The main properties of this stellar sample, both inferred from this work or extracted from previous publications, are listed in Table~\ref{tab:starparam}. The sample is complementary to those studied in spectropolarimetry by \cite{Donati08c, Morin08b, Morin10}

Stellar masses are derived from the empirical mass-luminosity relationship of \citet[][]{Delfosse00} together with parallaxes and K-band photometry (both taken from $Hipparcos$ catalogue, \citealt{Koen10}).
The luminosity is deduced from the infrared K band photometry and J-K colors are converted into luminosities with the bolometric correction of \citet{Leggett01}.
The stellar radius \rstar\ is estimated from the mass-radius relation given in \citet{Baraffe15}.

The line-of-sight-projected equatorial rotation velocity value (\vsini) is either taken from the literature \citep{Bonfils12, Forveille09, Donati08c, Bonfils07}, or constrained thanks to the ZDI code (see Sec.~\ref{sec:zdi}). For the whole sample, the \vsini\ is lower than 2~\kms\ (see Table~\ref{tab:starparam}) and the precision on \vsini\ estimate does not exceed 0.5~\kms. The \vsini\ values are compatible with the amount of rotational broadening observed in the spectra of our sample of stars.
The measurement of the stellar rotation period \pstar\ is presented in detail in Sec.~\ref{sec:period}. We found rotation periods ranging from 13.83 to 33.63~d.
Finally, the inclination of the rotation axis with respect to the line of sight, $i$, is estimated from the tomographic technique, with a precision of typically $\pm$~10\degr\ \citep[][for more details]{Morin10}.

\begin{table*}
\begin{center}
\begin{tabular}{ccccccccccccc}
\hline
Nom & ST & J &K& distance &\mstar    & \lstar     &   \rstar   & \vsini  & $i$ & \pstar  & rms$_0$ & $\sigma_0$\\
            &         && & pc & \msun & \lsun  & \rsun  & \kms & \degr & d & \ms& \ms\\
\hline
GJ 205  & M1   &5.0 	& 3.90 	&5.66$\pm$0.04	& 0.63$\pm$0.06 & 0.061$\pm$0.006  & 0.55$\pm$0.08 & 1 &  60 & 33.63 $\pm$ 0.37$^a$ & 2.71 & 1.45\\
GJ 358  & M2   &6.90  & 6.06  	&9.47$\pm$0.15 	&0.42$\pm$0.04 	& 0.023$\pm$0.002   & 0.41$\pm$0.06 & 1 &  60& 25.37 $\pm$ 0.32$^a$ &  5.10 & 2.08 \\
GJ 410  & M0   & 6.52 & 5.68  	&11.77$\pm$0.15	& 0.58$\pm$ 0.06 & 0.055$\pm$0.005   & 0.53$\pm$0.08 & 2 &  60& 13.83 $\pm$ 0.10$^c$ & 10.0 & 3.28\\
GJ 479  & M2   & 6.86 & 6.02	&9.69$\pm$0.22 	&0.43$\pm$0.04 	& 0.025$\pm$0.003   & 0.42$\pm$0.06 & 1 &  60 & 24.04 $\pm$ 0.75$^b$ &   5.45 & 2.02\\
GJ 846  & M0   &6.20 & 5.56	& 10.24$\pm$0.16	& 0.60$\pm$0.06	&  0.059$\pm$0.006 & 0.54$\pm$0.08 & 2 &  60 &  10.73 $\pm$ 0.10$^b$ &  3.30 & 2.45\\
\hline
\end{tabular}
\caption[]{Stellar parameters of the M dwarfs sample. 
Columns 1-8 list the star name, its spectral type (ST), its J \& K band magnitude and its distance (coming from the Hipparcos catalogue \citealt{Koen10}), the stellar mass, luminosity and theoretical radius \citep{Baraffe15}.
The columns 9-12 respectively list the measured \vsini\ (with a estimated error of $\pm$0.5~\kms) , the assessment of the stellar inclination angle $i$ (with a estimated error of $\pm$10\degr), the rotation period of the star \pstar, the rms of RV measurements and the average noise $\sigma_0$ on the RV measurements.
These four last parameters come from this study.\\
$^a$ compatible with \cite{Kiraga07}, 
$^b$ compatible with \cite{Bonfils12}, 
$^c$ compatible with \cite{Donati08c}.
}
\label{tab:starparam}
\end{center}
\end{table*}


\subsection{Instrumental set-up and data reduction}
\label{sec:outil}
Observations presented here were collected during two observing campaigns with the HARPS\footnote{High Accuracy Radial velocity Planet Searcher and spectropolarimeter at the ESO/3.6m telescope in La Silla (Chile)} velocimeter \citep{Mayor03, Snik11} used in spectropolarimetric mode and in a smaller extent with the NARVAL\footnote{the ESPaDOnS twin at the 2m Telescope Bernard Lyot (TBL) atop Pic-du-Midi (France)} spectropolarimeter \citep{Donati03, Donati09}.

We observed from October 2013 to September 2014 with HARPS-Pol. In this instrument two optical fibers convey the stellar light, split into two orthogonal polarization states, from the Cassegrain focus of the telescope to the spectrograph. The instrument covers the 368-691~nm wavelength domain in a single exposure, at a resolving power of 100~000.
An additional campaign was carried out from September 2013 to April 2014 with NARVAL, providing full coverage of the optical domain from 350  to 1050 nm in a single exposure, at a resolving power of 65 000, and into two orthogonal polarization states. The main characteristic of the instruments are listed in Table~\ref{tab:instru}.
\begin{table}
\begin{center}
\begin{tabular}{ccccccc}
\hline
Instrument & Tel.  & Spectral    &   $R$ &      $\eta$          \\
                 & (m)         &        domain (nm)                &                                 &        (\%)          \\
\hline
NARVAL & 2 & 350 - 1050 & 68 000 & 10-15                \\
HARPS-Pol & 3.6 & 368 - 691 & 100 000 &  2-3        \\
\hline
\end{tabular}
\caption{Main characteristics of NARVAL \citep{Donati03} and HARPS-Pol \citep{Snik11}: Column 1 gives the instrument name, column 2 the diameter of the telescope primary mirror, column 3 the spectral domain (covered in a single exposure), column 4 the resolving power $R$ and column 5 the estimated peak instrument throughput $\eta$ (at $\sim$ 550nm).}
\label{tab:instru}
\end{center}
\end{table}

A spectropolarimetric observation consists of four sub-exposures taken at different azimuths of the quarter-wave plate (for HARPS-Pol) / half-wave rhombs (for NARVAL) relative to the optical axis of the beam splitter. The corresponding frames are combined together to produce a set of Stokes $I$ (unpolarized intensity) and Stokes $V$ (circularly polarized) spectra. Although it is possible to extract polarization spectra from two sub-exposures only, using four allows us to eliminate all systematic errors or spurious polarization signatures at first order \citep{Donati97}.

The peak signal-to-noise ratios (S/N) per CCD pixel range from 70 to 200 at 600~nm for HARPS-Pol spectra (for which the CCD pixel size is 0.85~\kms), and from 230 to 480 at 700~nm for NARVAL spectra (for which the CCD pixel size is 2.6~\kms). It mostly depends on the star magnitude and weather/seeing conditions.  An overview of the observations is presented in Table~\ref{tab:obs}, and the detail journal of observations of each star is given in Appendix~\ref{ann:journal}.
\begin{table}
\begin{center}
\begin{tabular}{ccccc}
\hline
Name & BJD$_{0}$   & $n_{\rm obs} $    &   S/N  &  Cycle \\
 &(+ 2456000) &  &  & \\
\hline
GJ 205  & 569.88    &  22  & 170 - 228 &  0.000 - 3.523\\
GJ 358  & 675.70    &  23  & 70 - 133 &  0.000 - 2.880\\
GJ 410  & 673.88    &  29  &  79 - 125 &  1.000 - 6.199 \\
GJ 479  & 778.00    &  23   & 63 - 146   & 0.024 - 2.684\\
GJ 846 (\#2)  & 829.87   &  11  & 189 - 318  & 25.764 - 31.848  \\
\hline
\hline
GJ 205  & 569.88    &   4   & 308 - 454 &  1.623 - 2.186\\
GJ 410  & 673.88    &  13 &  169 - 303 &  0.558 - 7.542 \\
GJ 846  (\#1) & 546.46   &  15  & 91 - 158  & 0.000 - 8.709  \\
\hline
\end{tabular}
\caption{Synthetic journal of HARPS-Pol (top panel) and NARVAL (bottom panel) observations. The first days of observation is given in columns 2. Column 3 indicates the number of collected spectra. Columns 4 lists the peak S/N (resp., per 0.85 and 2.6 \kms\ velocity bin for HARPS-Pol (at 650~nm) and NARVAL (at 750~nm))  - we precise the minimum and maximum obtained values. Column 5 indicates the rotational cycle bounds (computed with the rotation period mentioned in Table~\ref{tab:starparam} according to ephemeris given by Eq~\ref{eq:eph}).
}
\label{tab:obs}
\end{center}
\end{table}

Rotational cycles of each target are computed from Barycentric Julian Dates (BJDs) according to the ephemeris:
\begin{equation}
\rm BJD~(d) = BJD_{0} + \pstar.E,
\label{eq:eph}
\end{equation}
in which E is the rotational cycle, $\rm BJD_{0}$ is the initial date chosen arbitrarily and \pstar\ is the stellar rotation period derived from the magnetic analysis (see Sec.~\ref{sec:period}).

The data extraction is carried out with \textsc{\small{Libre-Esprit}}, a fully automated dedicated pipeline that performs optical extraction of the spectra. The initial procedure is described in \cite{Donati97}, and was adapted to HARPS-Pol data to make it compliant with precision velocimetry.    

We apply Least-Squares Deconvolution (LSD, \citealt{Donati97}) to all the observations in order to gather all the available polarimetric information into a single synthetic profile.
LSD is similar to cross-correlation in the sense that it extracts information from a large number of spectral lines through a deconvolution procedure (see \citealt{Donati97} for more details). To extract Stokes $V$ LSD profiles from circular polarization spectra, we use a mask of atomic lines computed with an \textsc{\small{Atlas}} local thermodynamic equilibrium (LTE) model of the stellar atmosphere matching the properties of our whole sample \citep{Kurucz93}. The final mask contains about 4000 moderate to strong atomic lines, with a known Land\'e factor, from 350~nm to 1082~nm. The use of atomic lines only for the LSD masks relies on former studies of early and mid M dwarfs \citep{Donati08c, Morin08b}. 
Zeeman signatures are clearly detected in Stokes V LSD profiles for all stars of our sample with a maximum peak-to-peak amplitude varying from 0.1\% to 0.5\% of the unpolarized continuum level. We observe temporal variations of the intensity and of the shape of the Stokes $V$ LSD profile due to rotational modulation for the whole stars of the sample (see Sec~\ref{sec:magn}).

For the unpolarized spectra, we use a denser line mask to increase our sensitivity to profile distortions and to RV variations of these five slow rotators. The mask is derived from M-dwarf spectra previously collected with HARPS \citep{Bonfils13}, and contains 9000 lines from 440 to 686~nm. With this procedure, Stokes $I$ LSD profiles distortions are detected with a maximum amplitude varying from 0.001\% to 0.01\% of the unpolarized continuum level (see Sec.~\ref{sec:rv}).

 For the stars observed with both HARPS-Pol and NARVAL (i.e., GJ 205, GJ 410 and GJ 846), we can use the collected spectra to compare the instrument efficiency. 
NARVAL being on a 2m telescope and HARPS-Pol on a 3.6m telescope, the ratio of the collected flux is about 0.31 at the telescope. The NARVAL peak throughput at 550~nm is thus, \emph{in fine}, about 5.0 times higher than that of HARPS-Pol, once the pixel size is taken into account (see Table~\ref{tab:eff}, second column). LSD allows to add up information from the whole observed spectral domain. Including the gain associated to the larger spectral domain, NARVAL is in average 8.4 times more efficient than HARPS-Pol (see Table~\ref{tab:eff}, third column). This explains why longitudinal field measurements secured with NARVAL are significantly more accurate than those derived from HARPS-Pol spectra despite the large ratio in telescope photon collecting power in favour of HARPS-Pol. For RV measurements, only HARPS-Pol spectra are used, NARVAL being limited to typical RV precisions of ~20 m/s \citep{Moutou07}. 
 \begin{table}
 \begin{center}
  \begin{tabular}{|ccccccccc}
  \hline
& $\rho$ in spectrum  & $\rho$ in LSD profile \\           
\hline
 GJ 205& 3.57 &5.76  \\
GJ 410 & 6.32 & 10.61 \\
GJ846 & 5.04 & 8.70 \\
\hline
\end{tabular}
\caption{
Resulting peak flux ratio $\rho$ between NARVAL and Harps-Pol.
Column 1 indicates the star name. Column 2 gives $\rho$ computed from spectra, after having taken into account the pixel size differences and the telescope photon collecting power. Column 3 lists $\rho$ obtained from LSD profiles and then this value also takes into account the size of the spectral domain.
}
\label{tab:eff}
\end{center}
\end{table}

\section{Magnetic analysis}
\label{sec:magn}
The goal of this section is to characterize the large-scale magnetic fields of the observed stars from the collected Stokes $V$ LSD profiles. To get an overview of the magnetic data, we start by simply computing the longitudinal field \bl, \emph{i.e.}, the line-of-sight-projected magnetic vector averaged over the visible stellar hemisphere (Sec~\ref{sec:bl}). 
From temporal variations of \bl\ and its rotational modulation, one can derive a reliable estimate of the stellar rotation period \pstar\ (Sec~\ref{sec:period}, and \emph{e.g.}, \citealt{Morin08b}).  
In a second step, we apply ZDI to our time series of Stokes $V$ LSD profiles, in order to recover the topology of the large-scale field that generates the observed Zeeman signatures and their rotational modulation (see Sec~\ref{sec:zdi}).

\subsection{Longitudinal magnetic field}
\label{sec:bl}

From each pair of Stokes $I$ and $V$ LSD profiles, we compute \bl\ (in Gauss) as follow \citep{Donati97}: 
\begin{equation}
B_l = \frac{-2.14 \times 10^{11}}{\lambda_0 g_{\rm eff}c} \frac{\int v V(v)dv}{\int \left[I_c - I(v)\right]dv},
\label{equ:bl}
\end{equation}
with $I$ and $V$ denoting the unpolarized and circularly polarized LSD profiles, $I_c$ the continuum level, $v$ the radial velocity in \kms, $c$, the speed of light in \kms, $\lambda_0$ the central wavelength in nm and $g_{\rm eff}$ the effective Land\'e factor.
\bl\ is a simple magnetic field proxy one can easily extract, but which conveys little information on the likely complexity of the magnetic field geometry.

\subsection{Period determination}
\label{sec:period}

To estimate the stellar rotation period we first fit \bl\ with a multiple sine fit (fundamental period + the first harmonic).
The explored period range spans 0.5$\times$ to 2$\times$ the value found in the literature. We choose \pstar\ that minimizes $\chi^2_r$, defined as the reduced $\chi^2$ of the multiple sine fit to the \bl\ data.
We compare this value to the period found computing the Lomb-Scargle periodogram of the \bl\ data \citep{Lomb76, Scargle82, Zechmeister09}. This periodogram estimates the power associated to each period in the explored \pstar\ interval. 
To assess the chance that the strongest peak of the derived periodogram is caused by noise in the observations rather than by a true signal, we compute the 10\% and 1\% false alarm probabilities (FAPs) as defined in \cite{Zechmeister09}. \\

\begin{figure}
\begin{center}
\includegraphics[scale=0.22,angle=0]{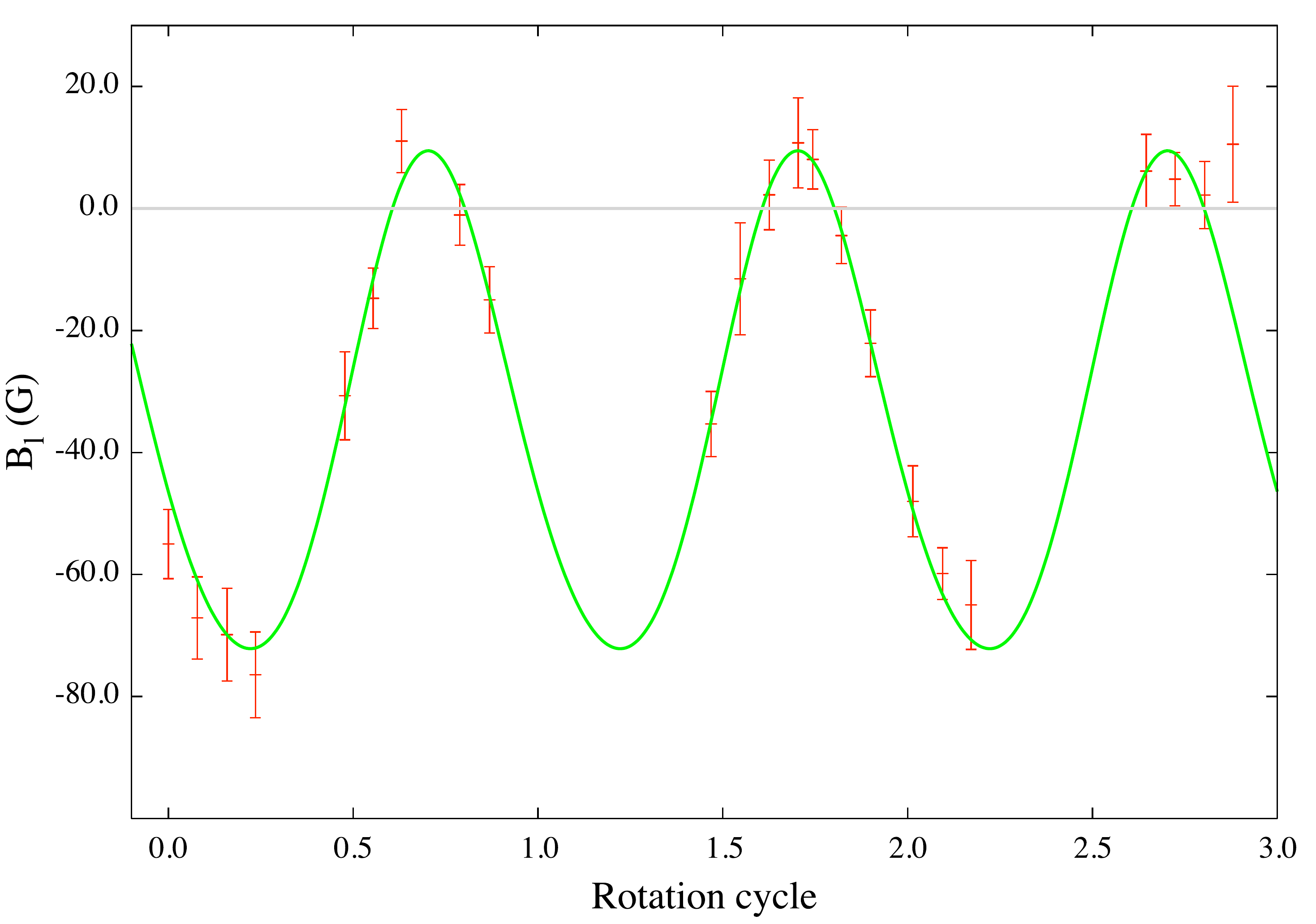} 
\includegraphics[scale=0.24,angle=0]{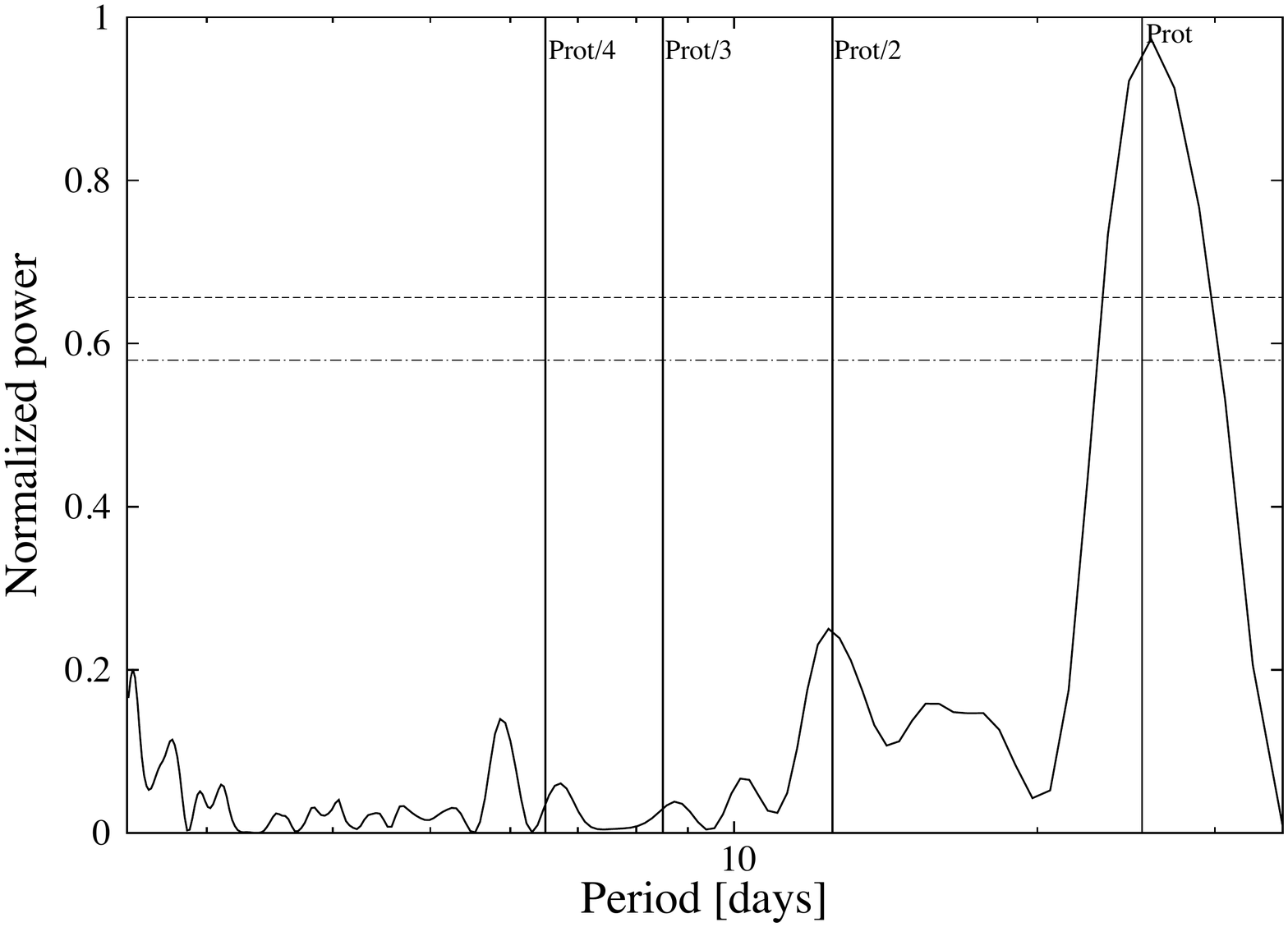} 
\caption{
\textit{Top:} 
\bl\ measurements of GJ 358 from HARPS-Pol spectra are shown as red dots (with $\pm$1$\sigma$ error bars).  The green line depicts a multiple sine fit (fundamental + 1rst harmonic) to the \bl\ measurements.  
The horizontal grey line represent the 0~G level.
\textit{Bottom:} Lomb-Scargle periodogram of \bl\ and the FAP at 1\% (dashed line) and 10\% (dotted-dashed line). The vertical lines depict \pstar\ and its three first harmonics \pstar/2, \pstar/3 and \pstar/4.}
\label{fig:bl358}
\end{center}
\end{figure}

$\bullet$ GJ 358 : 
The resulting curves for GJ~358 are presented Fig.~\ref{fig:bl358}. We note that \bl\ remains mainly negative (averaged value of --32.0~$\pm$~1.5~G), with a peak-to-peak amplitude of 70~G. 
The variations are periodic and well-fitted ($\chi^2_r$~=~1.0) with a multiple sine fit at \pstar~=~25.37~$\pm$~0.32~d (1$\sigma$ error bar). This period is in agreement with the period of $\sim$25.26~d given by \cite{Kiraga07} from a photometric survey, as well as with the period found computing the Lomb-Scargle periodogram and associated with a FAP much lower than 1\% (Fig.~\ref{fig:bl358}, bottom panel). \\

$\bullet$ GJ 479 : 
We observe a similar behaviour for \bl\ of GJ~479, with a rotation period of 24.04~$\pm$~0.75~d (see Fig.~\ref{fig:bl479}), in good agreement with the period estimated in a range 23-24~d by \cite{Bonfils12}. As indicated in the periodogram of \bl\ data, and contrary to the previous case, the first harmonic is essential to fit the data down to $\chi^2_r$~=~1.0. \\

\begin{figure}
\begin{center}
\includegraphics[scale=0.22,angle=0]{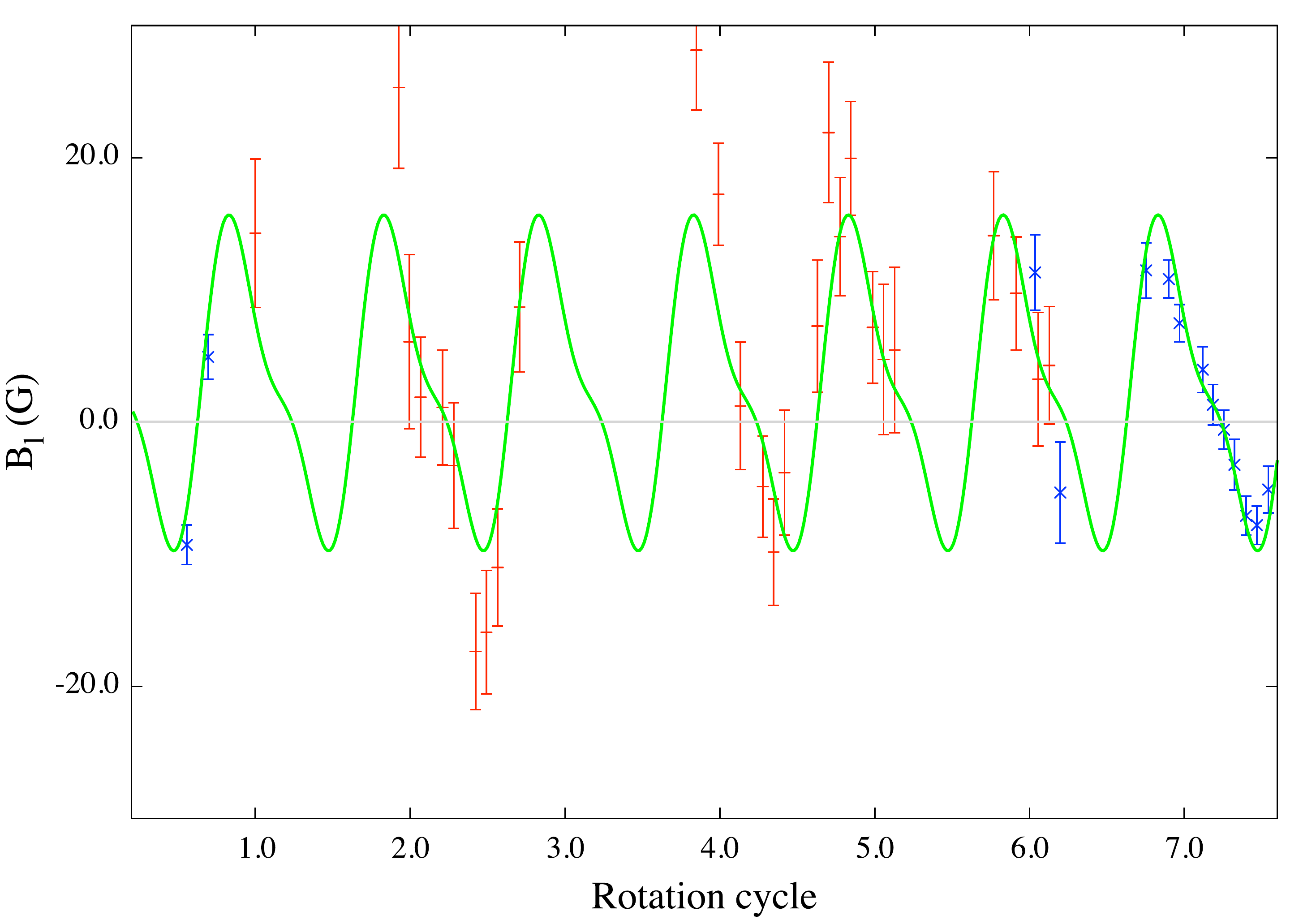} 
\includegraphics[scale=0.24,angle=0]{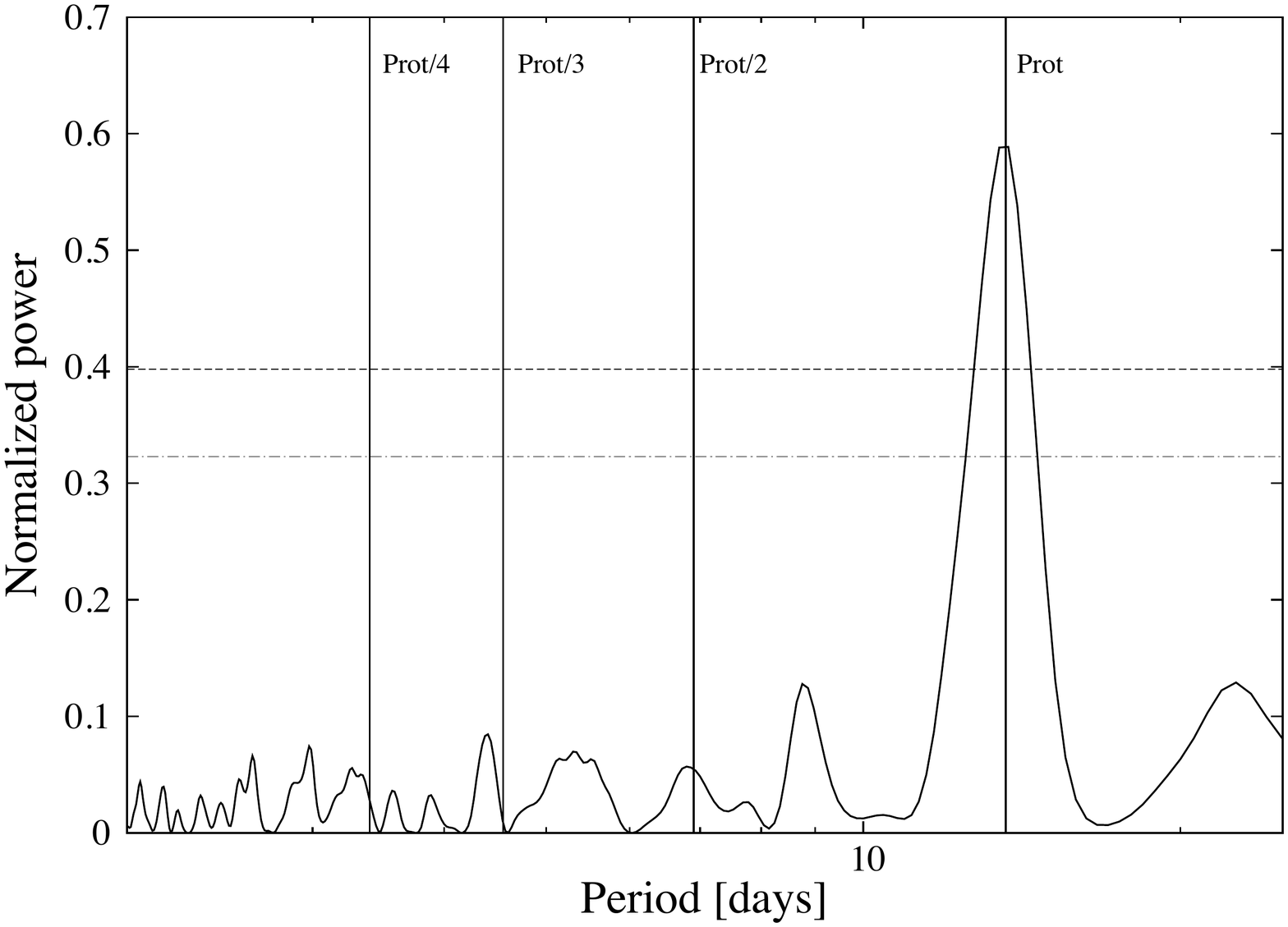} 
\caption{
Same as Fig.~\ref{fig:bl358} for GJ~410. Measurements from HARPS-Pol spectra are shown in red, while those from NARVAL spectra are shown in blue. Note the much smaller error bars on NARVAL \bl\ measurements, despite the 3.2$\times$ smaller photon collecting power of TBL}
\label{fig:bl410}
\end{center}
\end{figure}

$\bullet$ GJ 410 : 
For GJ~410 (Fig.~\ref{fig:bl410}), \bl\ varies periodically and exhibits regular sign switches; the averaged value is 3.0~$\pm$~0.5~G. The best period we derive from fitting \bl\ measurements is equal to 13.83~$\pm$~0.10~d, in agreement with the former study of \cite{Donati08c} (13.51~$\pm$~0.12~d) and the Lomb-Scargle periodogram (see Fig.~\ref{fig:bl410}, FAP < 1\%). This is one of the most active stars of our sample.\\

$\bullet$ GJ 205 : 
For GJ~205, we derive \pstar~=~33.63~$\pm$~0.37~d. To fit the data down to $\chi^2_r$~=~1.0, the fundamental period and its first harmonic \pstar/2 are needed. This is confirmed with the Lomb-Scargle periodogram whose strongest peak is at 16.8~d, i.e., \pstar/2 (FAP < 1\%, see Fig.~\ref{fig:bl205}), and with the former photometric study of \cite{Kiraga07} ($\sim$~33.61~d). \\

$\bullet$ GJ 846 : 
For GJ~846, we only secured 11 measurements in July-September 2014 with HARPS-Pol  and 15 measurements in September-December 2013 with NARVAL, spread over 6 and 9 rotation cycles respectively. The amplitude of the \bl\ variations changes between the two observation epochs: we first observe variations with a peak-to-peak amplitude of 10~$\pm$~2~G (averaged value 1.4~$\pm$~0.5~G), then variations with a peak-to-peak amplitude of 20~$\pm$~4~G (averaged value = 6.0~$\pm$~1.5~G). \bl\ keeps the same (positive) sign during the two runs. We derive a period equal to 10.73~$\pm$~0.10~d. This period is in good agreement with the periodicity of 10.7~d found in \cite{Bonfils12}.
\\

Our observations thus demonstrate clearly that the spectropolarimetric data provides us with an accurate measurement of \pstar.  
In Sec.~\ref{sec:diag} we will demonstrate that spectropolrimetry is more efficient that usual proxies (\hal\ or the full width at half-maximum FWHM) to determine the rotational period, and that \pstar\ is a key parameter to track the origin of the activity signal (\emph{i.e..}, the magnetic field)

\subsection{Magnetic imaging}
\label{sec:zdi}

To recover the parent large-scale magnetic field from time series of rotationally-modulated Zeeman signatures, we use the ZDI tomographic imaging technique.
ZDI has been largely tested and improved since the initial release of \citet{Brown91} \citep[see, \emph{e.g.},][]{Donati06b, Morin08b}. ZDI assumes that profile variations are mainly due to rotational modulation (plus some amount of differential rotation, if needed), and can turn series of circular polarization Stokes $V$ LSD profiles into maps of the parent magnetic topology. 
The longitudinal and latitudinal resolution depends mainly on \vsini, $i$, and the phase coverage of the observations.

In our imaging procedure, we use spherical harmonics (SH) to describe the large-scale field, allowing in particular to reliably recover both simple and complex topologies \citep[see, \emph{e.g.},][]{Donati01, Donati06b}.  
$\textbf{B}$ can be written as the sum of a poloidal and a toroidal component; their expressions can be found in \citet{Donati06a}. The parameters $\alpha$$_{l,m}$, $\beta$$_{l,m}$ and $\gamma$$_{l,m}$, noting the SH coefficients (with $l$ denoting the degree of the mode, and $m$~$\in$~$\lbrack$~0~;~$l$~$\rbrack$ its order), describe, respectively, the radial poloidal, non-radial poloidal and toroidal components of the magnetic field. The energy associated with the axisymmetric ($m$~<~$l$/2) and non-axisymmetric modes of the poloidal field component, as well as that of the poloidal and toroidal components, can be estimated directly from the coefficients of the SH expansion.  For the slow rotators considered here, most spatial information we can recover about the field concentrates in modes with orders equal to or lower than 5 \citep[see, \emph{e.g.},][]{Morin08b}.

To compute synthetic circular polarization profiles, the surface of the star is divided into 5000 cells of similar projected areas (at maximum visibility), whose contribution to the integrated Stokes $I$ and $V$ LSD profiles depends on the RV of each cell, on the local field strength and orientation, on the location of the cell and its projected area, on the rotation cycle, and on the local surface brightness of the photosphere (assumed to be uniform at this stage).
To model the local unpolarized Stokes $I$ and the local circular polarized Stokes $V$ profiles (resp. $I_{l,j}$ and $V_{l,j}$) at each cell $j$ in presence of magnetic fields, we use Unno-Rachkovsky's (UR) equations (e.g. Landi degl'Innocenti \& Landolfi 2004). We set the central wavelength, the Doppler width and the Land\'e factor of the equivalent line to 650 nm, 1.6~\kms\ and 1.25, respectively, and we adjust the average line-equivalent width to the observed value. Summing the spectral contributions of all grid cells yield the synthetic Stokes V profiles at a given rotation phase. 

ZDI proceeds by iteratively comparing the synthetic profiles to the observed ones, until they match within the error bars. Since the inversion problem is ill-posed, ZDI uses the principles of maximum-entropy image reconstruction to retrieve the simplest image compatible with the data. The form we use for the regularization function is $S = \sum_{l,m}l(\alpha_{l,m}^2 + \beta_{l,m}^2 + \gamma_{l,m}^2)$ (more details in \citealt{Donati01}).

ZDI depending on the assumed rotation period, it can be used to confirm and often to improve the accuracy of the estimate derived from \bl\ curves, as Stokes $V$ profiles intrinsically contain more information than \bl\ curves.  In some cases, surface differential rotation (DR) is required in order to fit Stokes $V$ data down to the noise level. To achieve this, we assume that the rotation rate at the surface of the star depends on latitude and can be expressed as $\Omega(\theta) = \Omega_{\rm eq}-$d$\Omega \sin^2(\theta)$, with $\theta$ denoting the latitude, $\Omega_{\rm eq}$, the rotation rate at the equator and d$\Omega$, the difference in rotation rate between the equator and the pole. This law is used to compute the phase shift of each ring of the grid at any observation epoch with respect to its position at a reference epoch.
We carry out reconstructions for a range of $\Omega_{\rm eq}$ and $d\Omega$ values; the optimum DR parameters are those minimizing the information content. They are obtained by fitting the surface of the $\chi^2_r$ map with a paraboloid around the minimum value \citep{Donati03b}.

\subsection{Results}
\label{sec:magn_result}
The M dwarfs of our sample exhibit magnetic fields with Zeeman signatures that do not exceed 0.5\% of the unpolarized continuum. We distinguish two kinds of magnetic topologies (see Table~\ref{tab:p_magn}) : two stars harbour a large-scale magnetic field dominated by an axial dipole (GJ~358, GJ~205), whereas three stars exhibit a more complex field featuring a significant - in most case dominant - toroidal axisymmetric component (GJ~410, GJ~479 \& GJ~846).
\begin{table}
\begin{center}
\begin{tabular}{ccccccccccc}
\hline
Nom &  \mstar     & \pstar & $B_V$  & Pol. & Dip. & Axi \\
        &    (\msun) &  (d)    & (G)  & (\%)& (\%) & (\%)\\
\hline
GJ 205  & 0.61 &   33.63 $\pm$ 0.37 &    20  &   99    &  90  &    73       \\
GJ 358  & 0.41 &   25.37 $\pm$ 0.32 &   130 &   97    &  98  &   85   \\
GJ 410  & 0.58 &  13.83 $\pm$ 0.10 &    65   &   25    &   88 & 11        \\
GJ 479  & 0.43 &   24.04 $\pm$ 0.75 &   65   &   37     &  74 &   29       \\
GJ 846 (\#1)& 0.59 &  10.73 $\pm$ 0.10 & 45 & 27 & 69 & 68  \\
GJ 846 (\#2) &  0.59 &  10.73 $\pm$ 0.10    &    30  &     63  &   52  &  86 \\
\hline
\end{tabular}
\caption{
Properties of the large-scale magnetic field topologies of the moderately active M dwarfs sample. In columns 1-3 we report the name of the star (with runs \#1 and \#2 for GJ~846 corresponding to the first and second observing epochs, see Table~\ref{tab:obs846}), the mass and the rotation period, initially presented in Table~\ref{tab:starparam}. Column 4 mentions the assessment of the average magnetic flux reconstructed from the Zeeman signatures. Column 5 lists the magnetic energy lying in poloidal component. Columns 6-7 present the magnetic energy reconstructed as a poloidal dipole and the percentage of poloidal energy in axisymmetric modes (defined as $m$ < $l$/2).
}
\label{tab:p_magn}
\end{center}
\end{table}
\\

\begin{figure*}
\includegraphics[scale=0.5,angle=0]{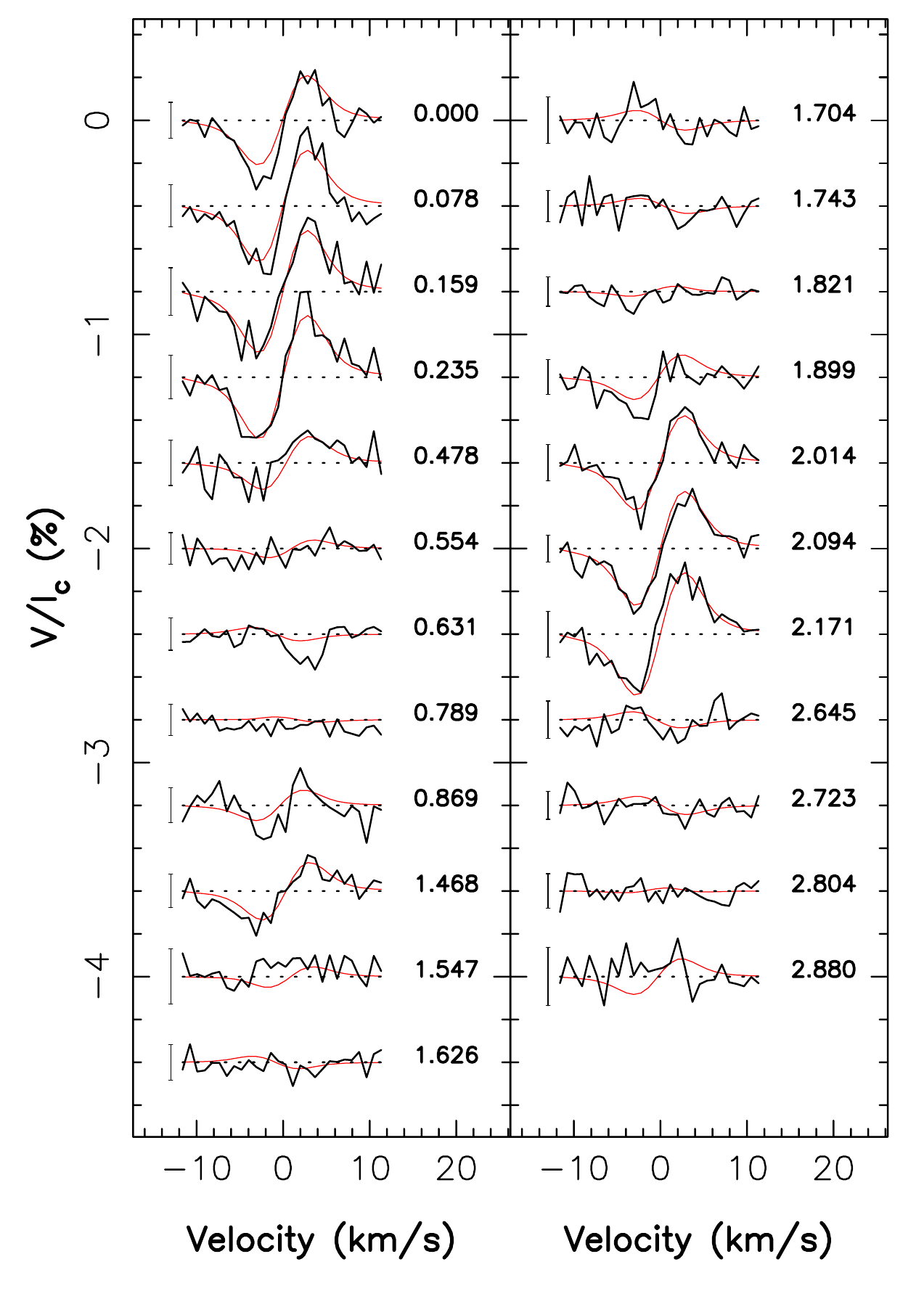}
\includegraphics[scale=0.45,angle=0]{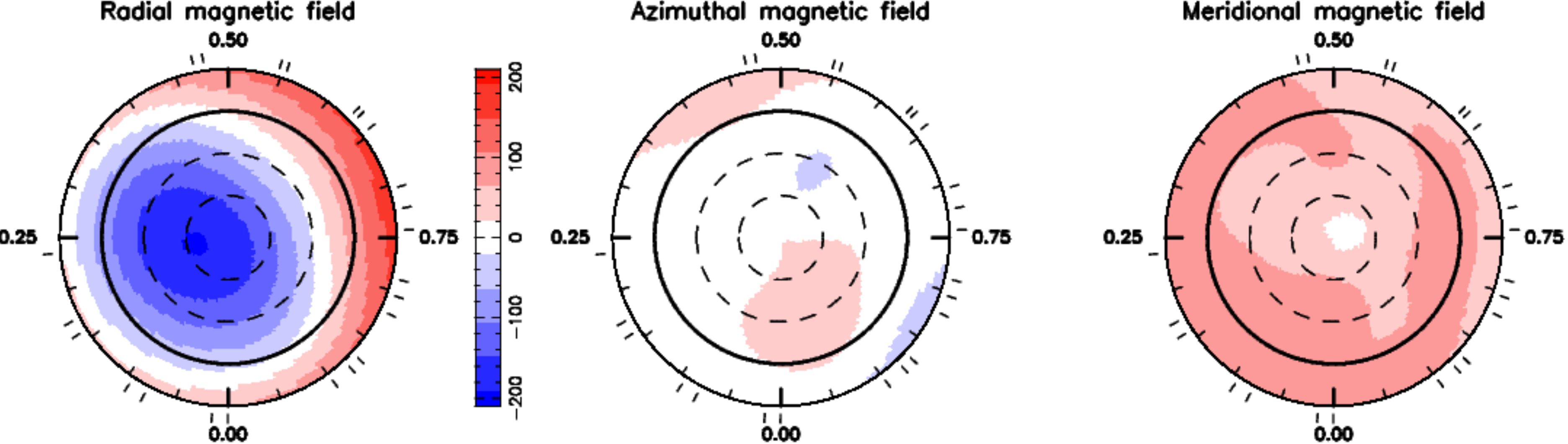} 
\caption{\textit{Top:} maximum-entropy fit (thin red line) to the observed (thick black line) Stokes $V$ LSD photospheric profiles of GJ~358. Rotational cycles and 3$\sigma$ error bars are also shown next to each profile.
\textit{Bottom:} map of the large-scale magnetic field at the surface of GJ~358.
The radial (left corner), azimuthal (center) and meridional (right corner) components of the magnetic field $B$ are shown. Magnetic fluxes are labelled in G. The star is shown in a flattened polar projection down to latitude -30\degr, with the equator depicted as a bold circle and parallels as dashed circles. Radial ticks around each plot indicate phases of observations.
This figure is best viewed in color.}
\label{fig:map358}
\end{figure*}
$\bullet$ GJ 358 \& GJ 205 : 
For the stars whose large-scale field is mostly poloidal, the Stokes $V$ LSD signatures are mainly anti-symmetric with respect to the line center. 
Their shape remain mostly constant (see, e.g., GJ~358 in Fig.~\ref{fig:map358}), but their intensity vary significantly as the star rotates (see, \emph {e.g.}, the rotation cycles 0.235 and 0.789 in Fig.~\ref{fig:map358} top panel).
For these stars, the reconstructed magnetic topologies are simple with more than 90\% of the magnetic energy concentrated in a dipolar poloidal component (\emph{i.e.}, SH mode with $l$~=~1, see column 6 of Table~\ref{tab:p_magn}). 
The Stokes $V$ LSD signatures are fitted down to a $\chi^2_r \sim$ 1, from an initial value (corresponding to a null field map) of $\sim$~3.5-3.1 (depending on the S/N ratio and on the number of collected data).
In the most magnetic regions, the flux reaches 230~G at the surface of GJ~358, but only 30~G at the surface of GJ~205 (see Fig.~\ref{fig:map205}).
\\

\begin{figure*}
\includegraphics[scale=0.5,angle=0]{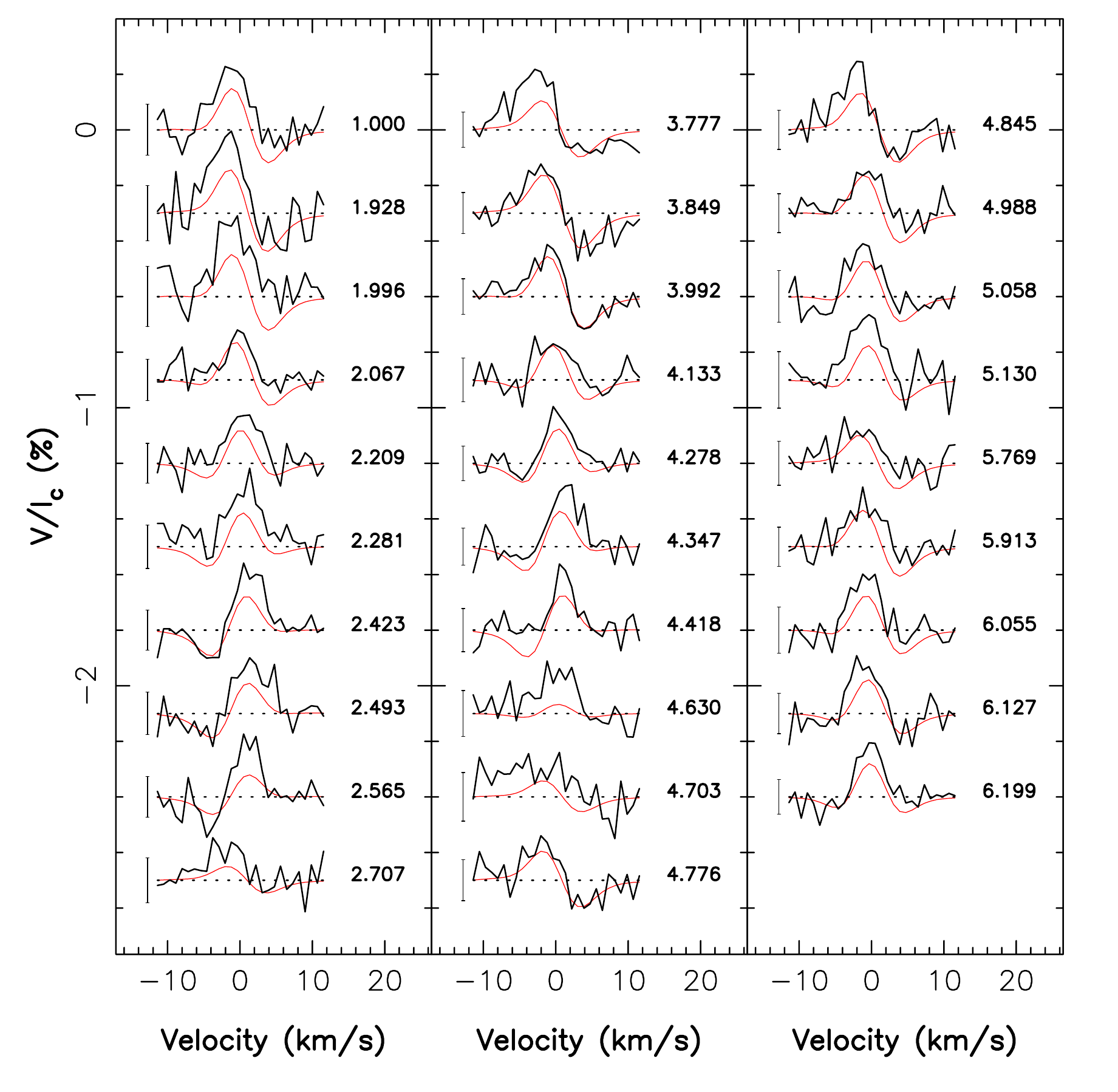} 
\includegraphics[scale=0.5,angle=0]{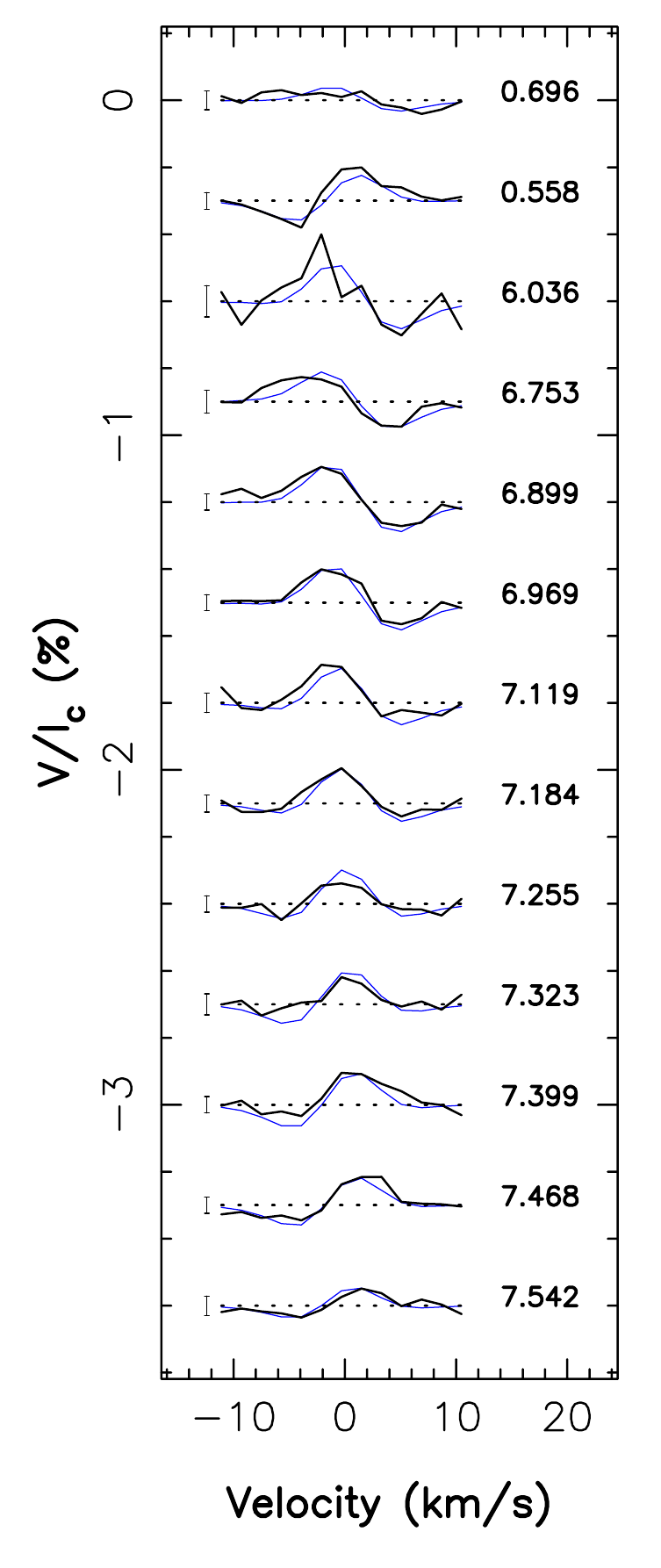} 
\includegraphics[scale=0.59,angle=0]{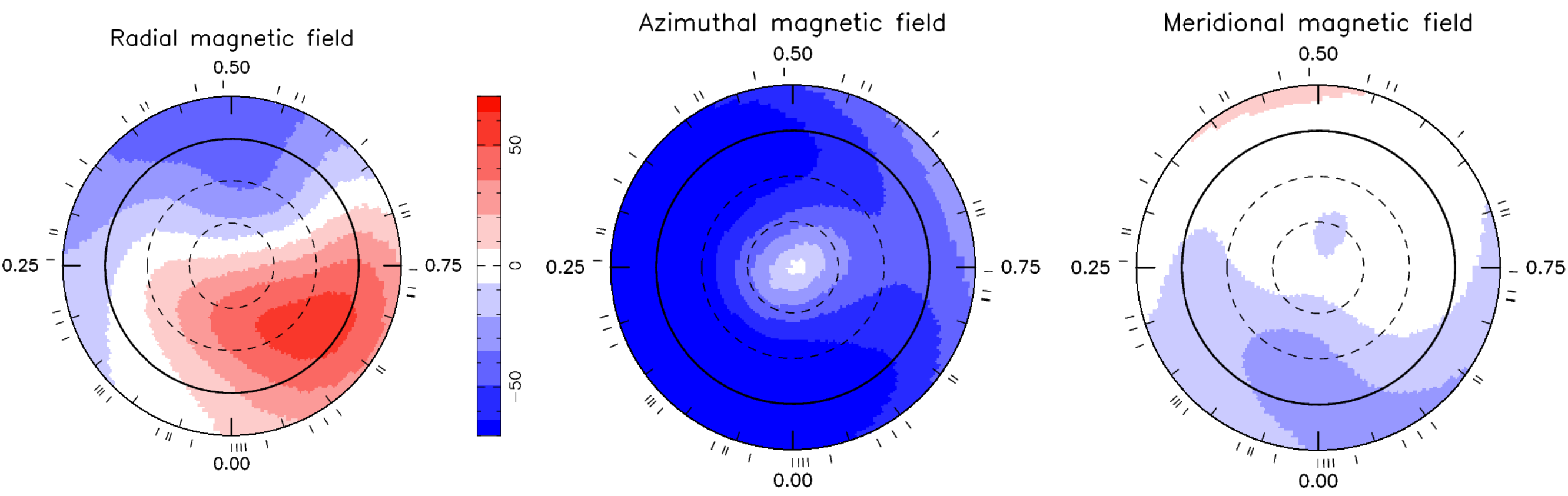} 
\caption{\textit{}
Same as Fig.~\ref{fig:map358} for GJ~410. LSD Stokes $V$ profiles in the top left and top right panels correspond to HARPS-Pol and NARVAL observations respectively.}
\label{fig:map410}
\end{figure*}

$\bullet$ GJ 479 \& GJ 410 : 
For the stars with a significant toroidal component, the Stokes $V$ LSD profiles have a sign that varies during the stellar rotation (see, \emph {e.g.}, the rotation cycles 2.067 and 2.493 of GJ~410 in Fig.~\ref{fig:map410} top left panel). The large-scale magnetic field reconstruction indicates that the axisymmetric poloidal component includes less than 40\% of the magnetic energy (see column 7 of Table~\ref{tab:p_magn}), and features a mostly dipolar structure ; the toroidal component includes more than 60\% of the reconstructed magnetic energy, and is mostly axisymmetric, showing up as an azimuthal field ring of $\sim$~80~G encircling the star at equatorial or intermediate latitudes (see Table~\ref{tab:p_magn}, two last rows, and Fig.~\ref{fig:map410} and Fig.~\ref{fig:map479} bottom panel, for, resp., GJ~410 and GJ~479).
The magnetic field flux is moderate, reaching $\sim$~70~G in the in the strongest field regions.

\begin{figure}
\includegraphics[scale=0.33,angle=0]{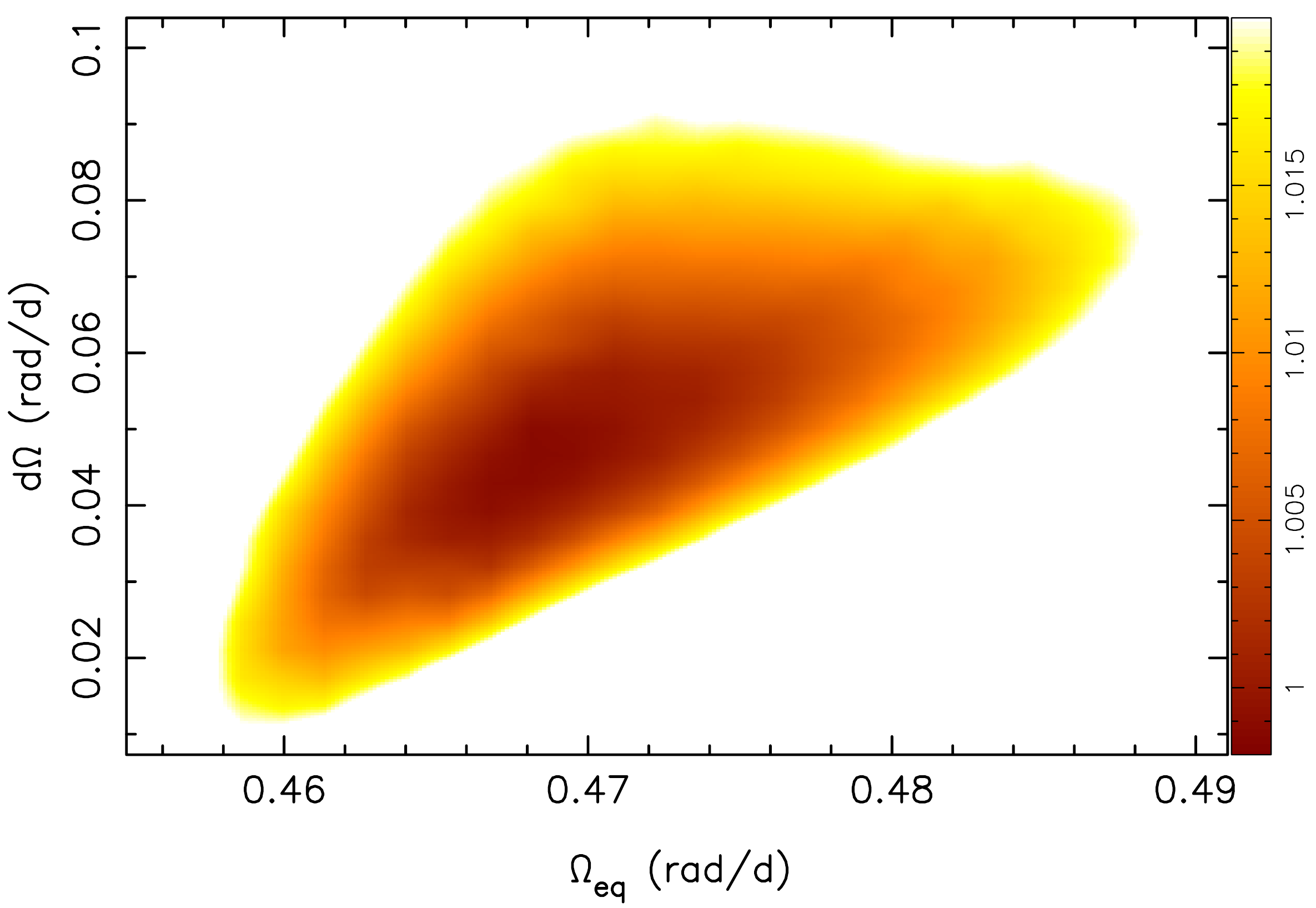} 
\caption{\textit{}
Variations of $\chi^2_r$ as a function of $\Omega_{eq}$ and d$\Omega$, derived from the modelling of GJ~410 Stokes $V$ LSD profiles at constant information content. The outer colour contour traces the 1.75\% increase in $\chi^2_r$ that corresponds to a 3$\sigma$ ellipse for both parameters as a pair.}
\label{fig:dr410}
\end{figure}

Moreover, thanks to the dense spectropolarimetric data set of GJ~410 (42 measurements spread over 7.5 rotation cycles), we can easily estimate the amount of latitudinal DR shearing the magnetic maps. Indeed, the Stokes $V$ LSD data set of all stars of our sample can be fitted down to $\chi^2_r$~=~1 when assuming solid body rotation, except for GJ~410 ($\chi^2_r$~=~1.6). Assuming DR, we are able to fit the data of this early-M dwarf down to $\chi^2_r$~=~1.0, with $\Omega_{\rm eq}$~=~0.47~$\pm$~0.03~\rad\ and $d\Omega$~=~0.05~$\pm$~0.03~\rad\ (see Fig.~\ref{fig:dr410}), corresponding to rotation periods at the equator and pole of 13.37~$\pm$~0.86 and 14.96~$\pm$~1.25~d, respectively. This result is in good agreement with \pstar\ previously found (13.83~$\pm$~0.10~d, Sec.~\ref{sec:period}), and with the former DR estimate of GJ~410 \citep[see][]{Donati08c}.

Finally, we note that the large-scale field of GJ~410 significantly evolved between 2007-2008 \citep{Donati08c} and 2014 (our data), both in strength (decreasing from 100 to 60~G) and in topology (the energy in the dipolar component increased from 50 to 88\%).
\\

$\bullet$ GJ 846 : 
For GJ~846, we notice a variability of the large-scale magnetic topology between 2013 and 2014, as expected from the \bl\ analysis (Sec.~\ref{sec:period}).
More specifically, the energy in the poloidal component increases from 27\% in 2013 to 63\% in 2014 (see Fig.~\ref{fig:map846}, middle and bottom panels), 

Early-M dwarfs like GJ~410 and GJ~846 were already reported to be prone to increased variability, probably as a result of their stronger surface differential rotation \citep[see][]{Morin08a, Morin08b, Donati08c}.  

\section{Characterization of the RV jitter}
\label{sec:diag}
Our sample stars are known to exhibit RV variations caused by stellar activity. To characterize the origin of the RV modulation, we compute the bisector, the full width at half-maximum FWHM, and the \hal\ index as described in section~\ref{sec:proxies}. 
We then analyse how these quantities vary with time through their Lomb-Scargle periodograms, and compare with temporal variations of the RV itself (see Sec.~\ref{sec:diagac}). 
As the model we propose aims at modeling the component of the RV signal that is rotationally modulated (see Sec.~\ref{sec:rv}), the first step is to assess quantitatively the amount to which the RVs of our sample stars are periodic (see Sec.~\ref{sec:detec}).

\subsection{Computing RVs and activity proxies}
\label{sec:proxies}
RVs are computed by fitting a Gaussian to the Stokes $I$ LSD profiles (equivalent to the CCF), the Gaussian centroid giving the RV estimate \vr. 
The FWHM measurements is directly computed from the Gaussian fit to the Stokes $I$ LSD profiles.

To derive the bisector, we first interpolate the CCF profile using a cubic spline interpolation; we then compute the set of midpoints of horizontal line segments extending across the profile \citep{Gray82}. To assess temporal evolution of the line profile, we calculate the velocity span \citep[as introduced, e.g., by][]{Gray82, Queloz01} \vs, given by $\vt-\vb$, where \vt\ and \vb\ are respectively the average velocity at the top and bottom parts of the bisector\footnote{The top and bottom parts include all points within 10-40\% and 60-90\% of the full line depth, respectively.}. 
For RV variations caused by stellar activity, we commonly observe an anti-correlation between \vs\ and \vr\ \citep[see, e.g.,][]{Queloz01}. However, as expected for slow rotators whose rotation profile is not resolved by the velocimeter (typically \vsini~<~2~\kms, see \emph{e.g.}, \citealt{Desort07}), this \vs\ vs. \vr\ anti-correlation is not observed in our sample, \vs\ exhibiting no variations (for example, for GJ~358, \vs\ has a peak-to-peak amplitude of $\sim$~10~\ms\ and a rms of 3.6~\ms, see Fig.~\ref{fig:gj358}). 
For this reason, this proxy is not discussed in the following sections, even though \vs\ is computed and shown on Fig.~\ref{fig:gj358} and similar following graphs.  

The \hal\ index is also often used to characterize RV variations caused by activity.  This index is defined as the ratio between the flux in the \hal\ absorption line and that in the surrounding continuum, as described in \cite{Boisse11}. We use a 0.16~nm window centered at 656.2808~nm for the central line, and two windows of  1.075 and 0.875~nm around 655.087 and 658.031~nm respectively for the continuum as presented in \cite{DaSilva11}. 

\subsection{Activity jitter in the M dwarfs sample}
\label{sec:activity}
\subsubsection{Diagnostic of the activity}
\label{sec:diagac}
Only 11 RV measurements spanning 5.4 rotation cycles were collected for GJ~846 (run \#2) - too sparse a set for a reliable periodogram analysis.  As a result, the following sections concentrate only on the 4 other stars of the sample, namely GJ~358, GJ~479, GJ~410 and GJ~205.\\

\begin{figure*}
\includegraphics[scale=0.5,angle=0]{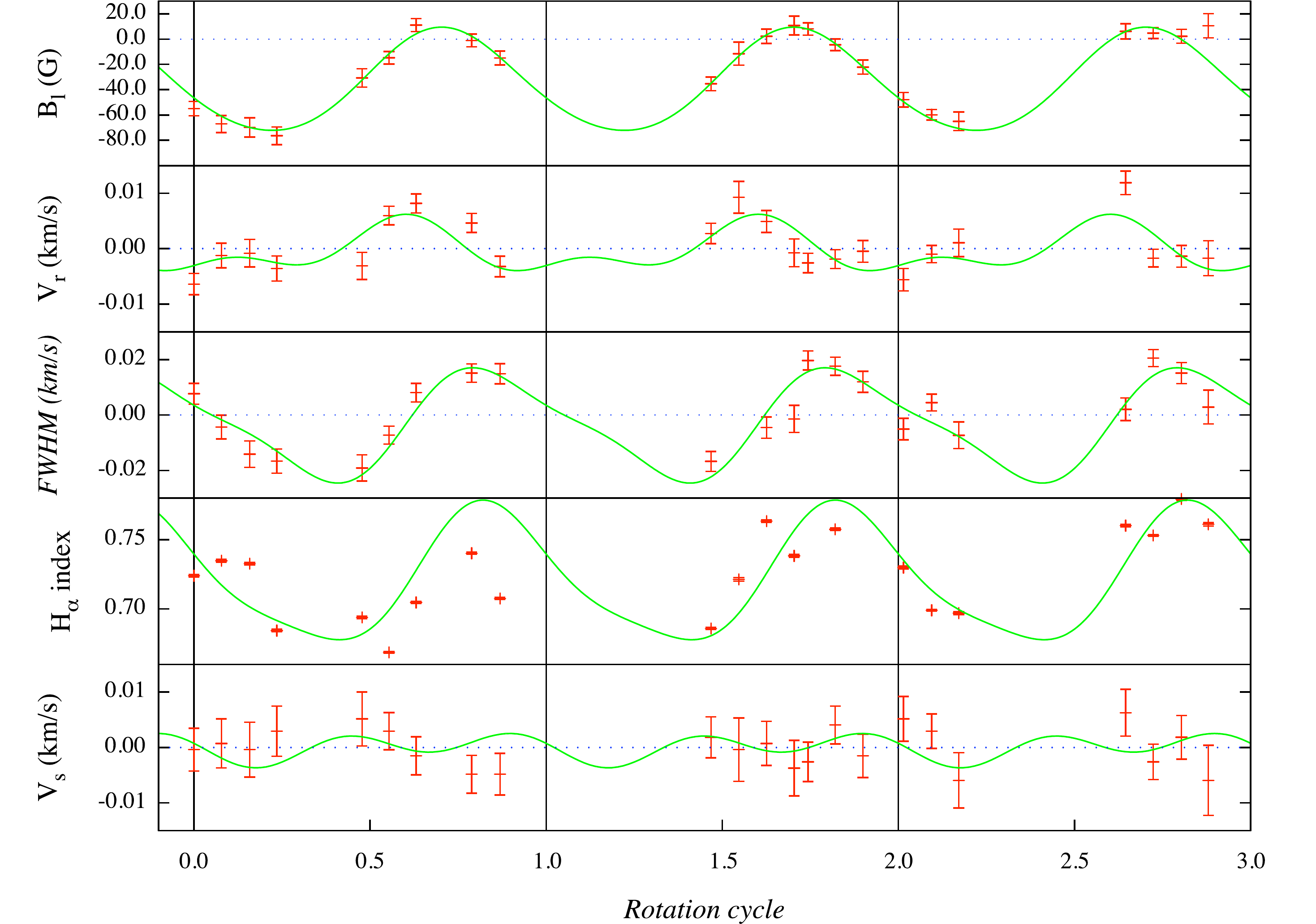} 
\includegraphics[scale=0.5,angle=0]{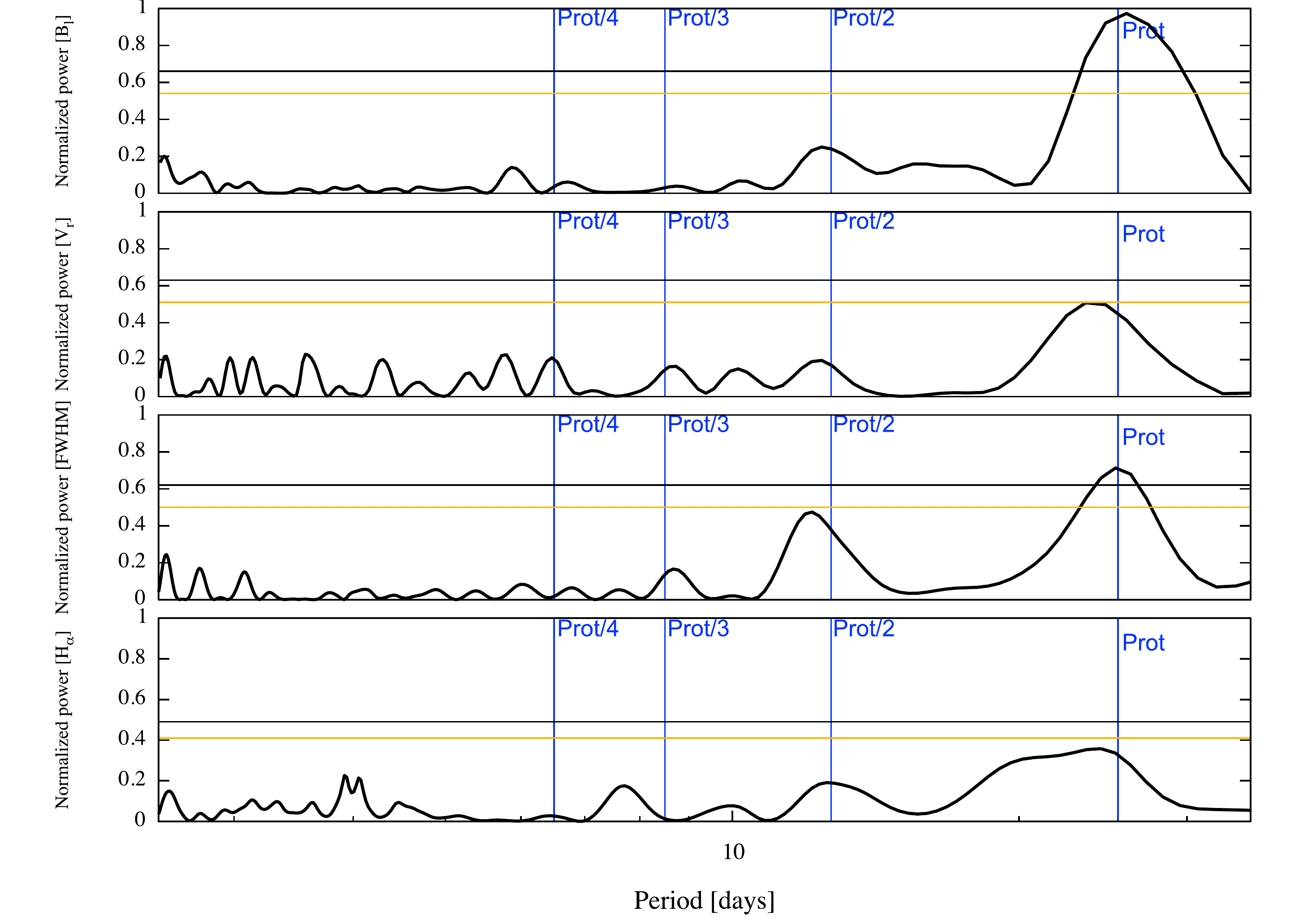} 
\caption{
\textit{Top: }Temporal variations of \bl, \vr, FWHM, \hal\ and \vs\ for GJ~358. Data and their error bars are in red. For all plots, the zero level is depicted by a dotted line. The green lines depict a multiple sine fit (including the fundamental at \pstar\ and first harmonic at \pstar/2) to the data points. The vertical black lines outline the beginning of each rotation cycle.
\textit{Bottom: }Lomb-Scargle periodograms of \bl, \vr, FWHM and \hal\ for GJ~358. The blue vertical lines outline the rotation period \pstar\ and its first 3 harmonics at \pstar/2, \pstar/3 and \pstar/4. The yellow and black horizontal lines respectively mark FAP levels of 10\% and 1\%.
}
\label{fig:gj358}
\end{figure*}

$\bullet$ GJ 358 : 
The \bl, RV, FWHM, \hal\ and \vs\ curves as well as the periodograms of \bl, RV, FWHM and \hal\ are presented in Fig.~\ref{fig:gj358}.
The periodograms of both \hal\ and FWHM show that the period \pstar\ previously identified with \bl\ has significantly more power than its harmonics (FAP < 1\% for FWHM, < 15\% for \hal). It is a further confirmation that the observed RV modulation is mainly caused by activity.
The periodogram of \vr\ indicates a period of $P$~=~24.47~$\pm$~0.60~d, in agreement with \pstar, but with a FAP of only 10\%.
Moreover, we notice that \vr\ and \bl\ vary in quadrature : when \bl\ reaches its maximum value (of about +10~G), \vr\ is at mid-distance between its maximum and minimum (see, \emph{e.g.}, phases 0.70-0.75).\\

$\bullet$ GJ 479 : 
\hal\ and FWHM show variations with a period in the range 23-25~d, in agreement with the \pstar\ that we previously derived from our \bl\ data. RVs allows to measure a period of 23.2~$\pm$~1.9~d, again fully compatible with \pstar\ (see Fig.~\ref{fig:gj479}).
Moreover the \hal\ periodogram exhibits a similar harmonics spectrum (from \pstar\ to \pstar/4) as those of \bl\ and \vr.
Furthermore the shape of \vr\ and \bl\ curve are very similar, in particular \vr\ crosses its median value when \bl\ is close to zero (see, \emph{e.g.}, phase 0.45).\\

\begin{figure*}
\includegraphics[scale=0.5,angle=0]{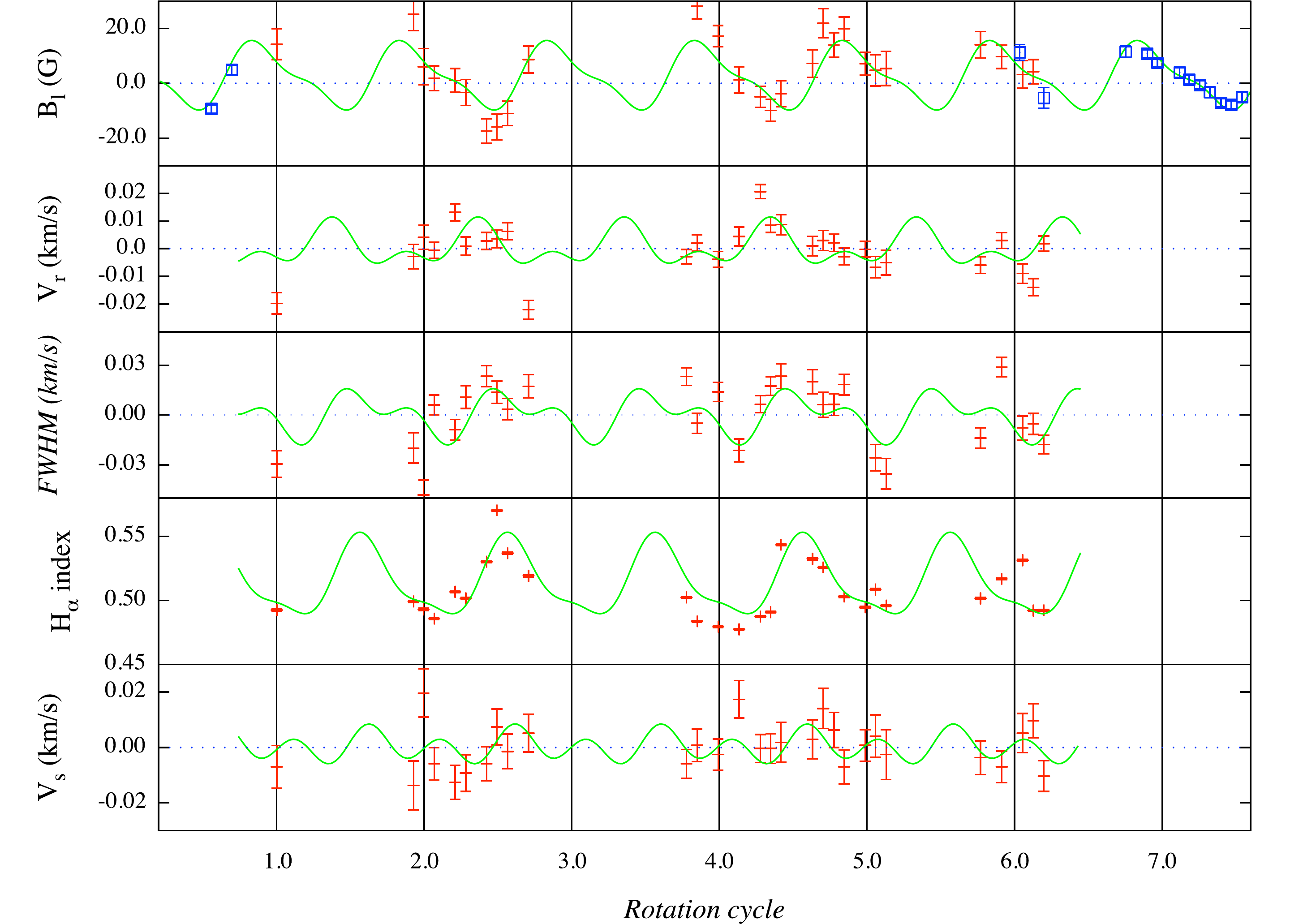} 
\includegraphics[scale=0.5,angle=0]{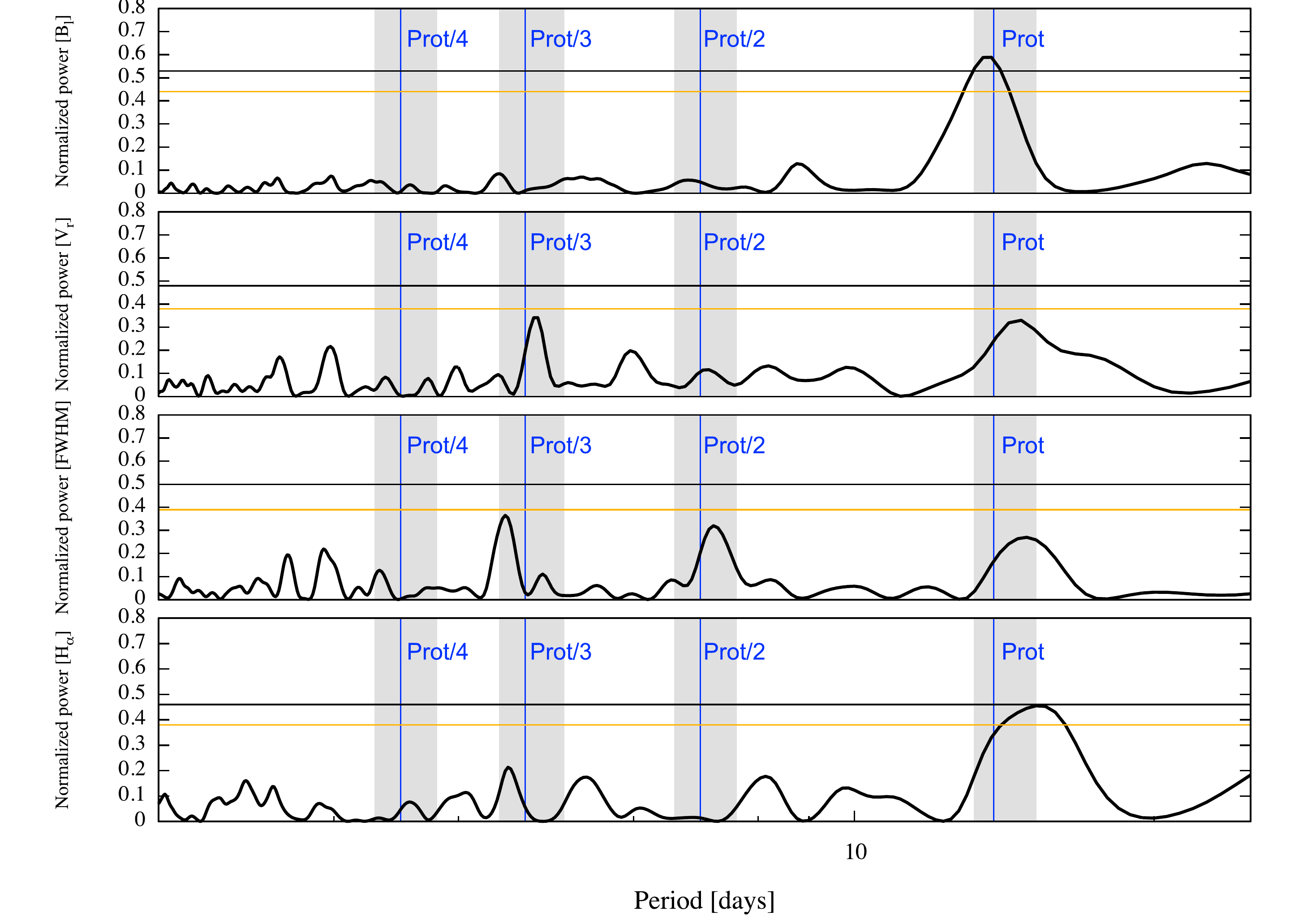} 
\caption{As Figure~\ref{fig:gj479} for GJ 410. The blue data are the \bl\ values computed from NARVAL LSD profiles. The grey bands depict the range of periods at the surface of the star as a result of DR.}
\label{fig:gj410}
\end{figure*}

$\bullet$ GJ 410 : 
Being the most active star of our sample, GJ~410 exhibits the largest temporal variations for all proxies (typically $\times$1.5, see Fig.~\ref{fig:gj410}).
The periodogram of \vr\ indicates a period $P$~=~14.20~$\pm$~0.20~d, within the range of surface periods that differential rotation triggers (13.4-15.3~d, see Sec.~\ref{sec:magn_result}). The period measured with \vr\ being higher than \pstar\ measured with \bl, this suggests that the surface spots generating the observed RV variations are located at mid to high latitudes.\\

$\bullet$ GJ 205 : 
The data and their periodogams are presented in Fig.~\ref{fig:gj205}. 
The \hal\ periodogram shows a main peak at 33.46~d with a FAP~$\leqslant$~15\%. This period is consistent with the \pstar\ derived from \bl\ and mentioned in \cite{Kiraga07} (33.63~$\pm$~0.37~d and 33.61~d resp.), and confirms that the observed signals are due to stellar activity.
FWHM measurements do not allow to firmly identify the rotation period of the star, and no signal with the \pstar\ derived from \bl\ is detected in \vr\ with a FAP of 98\%. The strongest peak in \vr\ periodogram is at 39.70~$\pm$~0.85~d and not compatible with \pstar. 
However, GJ~205 being an early-M dwarf, one can assume that it features a similar amount of DR to that observed on GJ~410, \emph{i.e.}, $d\Omega$~=~0.05~\rad. This level of differential rotation would correspond to a difference of 10~d between the polar and the equatorial rotation periods, and would thus allow to reconcile the observed peaks in the different periodograms. If confirmed, this would suggest that dark spots are located at high latitudes. Unfortunately, the present data set does not allow to measure the DR, and to further confirm this assumption.\\

Comparing the period derived from \bl\  and those derived from RV, FWHM and \hal\ (see Table~\ref{tab:all_period}) demonstrates that the rotation period of the star is most of the time more efficiently determined through \bl\ data than through RVs or other usual activity proxies (high FAP). 
In our small sample, we find that the different period values are in agreement for GJ~358 and GJ~479. We thus can suspect a solid rotation and/or that the strongest magnetic area and the spots are at similar latitudes. For GJ~410, the periods differ but are in agreement with the differential rotation we measured from the magnetic data. Finally, for GJ~205, we suspect a differential rotation effect, but we cannot measure with the current dataset.

This preliminary analysis demonstrates that most of our sample stars show rotationally modulated RVs, whose variations are obviously linked with those of the longitudinal field and other activity proxies. 
By comparing these different values of period, we can thus further investigate the origin of the observed RV jitter.

\subsubsection{RV signal detection}
\label{sec:detec}

The rms of the data (rms$_0$) is 2-3$\times$ higher than the average noise $\sigma_0$. The multiple sine fit (including the two first harmonics) to the RV data allows to improve $\chi^2_r$ with respect to a fit with a constant RV, however we never reach $\chi^2_r$~=~1.0 (see Table~\ref{tab:fitvr2}, 7 first columns).
This suggests that the RV jitter, J$_{\rm tot}$, includes both a rotationally modulated component J$_{\rm m}$ (due, \emph{e.g.}, to long-lived spots at the stellar surface), as well as a randomly varying one J$_{\rm r}$ (of yet unclear origin, \emph{e.g.}, spots with lifetimes shorter than the rotation cycle). Their respective strengths vary from one star to the other. For example, the poorest fit to the data are that of GJ~205, whose period in the RV data significantly differs from \pstar\ (determined from the magnetic data, see Sec.~\ref{sec:period}) and for which periodicity in the \vr\ signal is not really detected.

Multiple sine fits and Doppler-imaging (see Sec.~\ref{sec:rv}) can only succeed at modeling signals varying periodically with \pstar;  our first task is thus to quantify the extent to which the observed RV signals are indeed mostly modulated by rotation.
We thus compute the probability that a multiple sine fit provides a significantly better match to the observed RV variations than does a constant RV. We use the incomplete Gamma function to assess this probability $p$, given both the number of degrees of freedom and the improvement in $\chi^2_r$ that a multiple sine fit (including 2 harmonics) provides with respect to a constant RV. The closer $p$ gets to 1.0 and the false alarm probability (FAP = 1-$p$) to 0, the more reliably the rotational modulation of the RV signal is detected and dominates the RV variations.
As we test the ability of the model to fit the rotationally modulated component J$_m$, we use a scaled $\Delta\chi^2$ given by:
\begin{equation}
\Delta\chi^2 = \frac{\chi^2_{r,0} - \chi^2_{r,1}}{\chi^2_{r,1}}.N,
\label{eq:deltachi2}
\end{equation}
where N denotes the number of measurements. The resulting FAP are gathered in Table~\ref{tab:fitvr2}.
We note that for GJ~358, GJ~479 and GJ~410, J$_{\rm m}$ is dominant with a FAP level of $<$~1\%, whereas for GJ~205 J$_{\rm r}$ largely dominates the signal (with a FAP level of $\sim$~73\%\footnote{However if we assume that DR is present at the surface of GJ~205 (at a level similar to that reported for GJ~410) and is responsible for modulating the RV data with a period of 39.70~$\pm$~0.35~d (rather than that of 33.63~$\pm$~0.37~d derived from the magnetic data), the FAP level drops down to 2\%.  
}), so that for this star no coherent signal is detected at the rotation period measured from \bl.

To quantify the strength of J$_{\rm m}$ and J$_{\rm r}$, we compute their rms, once having quadratically subtracted the noise (see Table~\ref{tab:jmjr}). 
Whereas J$_{\rm m}$ is the major component for GJ~358 and, in a smaller extend for GJ~479, the trend is reversed for GJ~205 and GJ~410, where $J_{\rm r}$ becomes dominant. The Doppler imaging being able to model the rotational modulation only, we aim at reduce the activity jitter by a factor A$_1$.
\begin{table*}
  \begin{tabular}{cccc||ccc||ccc|ccc}
\cline{2-13}
   & \multicolumn{3}{c||}{ Raw RV} &  \multicolumn{3}{c||}{Multiple sine } & \multicolumn{5}{c}{  DI modeling} \\
   & \multicolumn{3}{c||}{ data} &  \multicolumn{3}{c||}{ fit} & \multicolumn{5}{c}{} \\
   & \multicolumn{3}{c||}{ } &  \multicolumn{3}{c||}{ (fond. + 2 harm.) } & \multicolumn{3}{c|}{RV curve} & \multicolumn{2}{c}{LSD profiles}\\
\cline{2-13}
   & $\sigma_0$ & rms$_0$ &$\chi^2_{r,0}$  & rms$_1$    & $\chi^2_{r,1}$ & FAP  &  rms$_2$    & $\chi^2_{r,2}$ & FAP & $\chi^2_{r,i}$ & $\chi^2_{r,f}$ & FAP \\             
   &  (\ms)  & (\ms)  && (\ms)  & &(\%) & (\ms)    &&(\%)  & & & (\%)\\             
\hline
\multicolumn{1}{c|}{GJ~205} &  1.45 & 3.52 & 7.81 & 3.05 & 6.27 & 73 & 3.17&6.31& 98 & 4.3 & 3.8 & 75 \\
\multicolumn{1}{c|}{GJ~410} &  3.28 & 8.84 &  7.85 & 6.55 & 3.96 & 0.04 & 6.78 & 4.14 & 2.8& 2.9 & 2.0 & $\leqslant$ 0.01\\
\multicolumn{1}{c|}{GJ~479} &  2.02 & 5.29 &7.71 & 3.65  & 3.58 & 0.09 &3.93&4.05 & 7.4 &5.0 & 2.9 &$\leqslant$ 0.01\\
\multicolumn{1}{c|}{GJ~358} &  2.08 & 4.79 & 5.59 & 2.47 & 1.69 & $\leqslant$ 0.01 &2.88& 2.05 & $\leqslant$ 0.01 & 3.8 & 2.1 & $\leqslant$ 0.01\\
\hline
\end{tabular}
\caption{
Table of the parameter that characterize the detection and the multiple sine fit of the RV activity jitter. The first column gives the name of the star, columns 2-4 give the observed average RV noise $\sigma_0$, the rms of the RV data rms$_0$, and the associated $\chi^2_r$, $\chi^2_{r,0}$.
Columns 5-6 mention the rms of the RV residual obtained after a multiple sine-fit (fundamental + 2 harmonics), the $\chi^2_r$ associated to the fit $\chi^2_{r,1}$.
Column 7 lists the estimate of the likelihood of the fit (FAP, see text). Columns 8-10 give the rms of the RV residuals after the DI modeling, the associated $\chi^2_r$, and the FAP. Columns 11-13 list the initial and final $\chi^2_r$ linked to the $RI$ reconstruction, and the associated FAP. 
}
\label{tab:fitvr2}
\end{table*}

 \begin{table} 
\begin{center}
  \begin{tabular}{c|ccccc}
  \hline
   & J$_{\rm tot}$ & $J_{\rm m}$ & $J_{\rm r}$ &A$_1$\\             
   &  (\ms)  & (\ms)  & (\ms)  && \\           
\hline
GJ~205 &  3.21 & 1.75 &  2.68 & 1.2   \\
GJ~410 &  8.21 & 5.67 &  5.93 & 1.4\\
GJ~479 &  4.89 & 3.83 & 3.04 &1.7\\
GJ~358 &  4.31 & 4.10 &  1.33  & 3.2\\
\hline
\end{tabular}
\caption{
The first column indicates the name of the star. Column 2 gives rms$_{J, tot}$, the rms of the RV data, once the noise has been quadratically subtracted ($J_{\rm tot}$~=~$\sqrt{ \hbox{rms}^{2}_0 - \sigma_0^{2}}$, with rms$_0$, the rms of the data, see Table~\ref{tab:fitvr2})). Column 3 mentions $J_{\rm m}$, the rms of the RV jitter due to rotational modulation ($J_{\rm m}$~=~$\sqrt{ \hbox{rms}^{2}_0 - \hbox{rms}^{2}_1}$). Column 4 indicates the rms of the random component  of the RV jitter, $J_{\rm r}$ ($J_{\rm r}$~=~$\sqrt{ \hbox{rms}^{2}_1 - \sigma_0^{2}}$). With a model of the rotational modulation only, we can reduce the activity jitter by a factor A$_1$~=~$J_{\rm tot}$/ $J_{\rm r}$ (column 5). 
}
\label{tab:jmjr}
\end{center}
\end{table}

\section{Modeling of the RV jitter }
\label{sec:rv}
The goal of this section is to consistently model the rotationally modulated component of the activity jitter (called J$_{\rm m}$ in Sec.~\ref{sec:detec}) and translate it into a distribution of surface features, whose relation to the parent magnetic topology (described in Sec.~\ref{sec:magn_result}) can be studied - at least on a statistical point of view.

For this first approach, we assume that the distorsions observed in Stokes $I$ LSD profiles are only due to rotational modulation induced by spots.
On the stellar surface, two kinds of features can be found: hot/bright plages and cool/dark spots. These features induce a RV activity jitter and variations of the bisector span.
When the contribution of spots is dominant, the amplitude of the RV variations is higher than those of the bisector span, and when the contribution of plages is dominant, the amplitude of the RV variations is similar to or smaller than those of the bisector span (\citealt{Dumusque14}, Eq.10).  
In our sample, \vs\ data do not exhibit a peak-to-peak amplitude higher than 15~\ms\ and we do not observe any clear \vs\ variations (see Sec.\ref{sec:proxies}), whereas the peak-to-peak amplitude of RV variations is always higher than 15~\ms.
Moreover, thanks to 3D-simulations of the near-surface convection of M dwarfs that take into account the small-scale magnetic field, \cite{Beeck15} show that dark spots are much more abondant than plages.
Thus, at first order, we consider dark spots only as the main origin of the observed rotationally modulated RV variations. 

In the imaging procedure, we characterize a spot with its relative brightness $b$, and its local profile $I_{\rm s}$. This two parameters being fixed, we adopt a simple two-temperature model (warm photosphere, cool spots) for the stellar surface and we choose the spot covering fraction as image parameter.

\subsection{Method}
\label{sec:di}
As previously presented for the magnetic field reconstruction, the stellar surface is divided into 5000 cells, and the Stokes $I$ profile at a given rotation phase is computed as the sum of all local Stokes $I$ profiles from the different cells.
With the spot description we chose, the parameter we reconstruct during the ZDI process is $1 - C_j$, with $C_j$ denoting the proportion of photosphere inside each cell ($C_j$~=~0 and $C_j$~=~1, respectively, corresponding to a spotted cell, and to an unspotted cell), and therefore, the local profile $I_{j}$ of the cell $j$ is given by : 
\begin{equation}
I_{j} = C_j I_{\rm p} + b(1 - C_j )I_{\rm s}
\end{equation}
where $I_{\rm  p}$ is the local unpolarized profile within the photosphere, $I_{\rm s}$ that within the spot, and $b$ the relative spot to photosphere brightness contrast. To compute $I_{\rm p}$, we use the profile given by UR's analytical solution of the polarized radiative transfer equation in a Milne Eddington's atmosphere (see \citealt{Hebrard14} fo the values of the different parametrers) and we adjust the average line-equivalent width to the observed value only. 
Following \cite{Dumusque14}, the local profile within the spot $I_{\rm s}$ is simply a broadened version (by a Gaussian of FWHM $w$~=~2-3~\kms, depending on the stars) of that in the photosphere $I_{\rm p}$. We also have the option of red shifting $I_{\rm s}$ with respect to $I_{\rm p}$ (to simulate the inhibition of the convective blue shift within the spot). However, we did not use this option for the present study given that convective blue shifts of M dwarfs are expected to be quite small. 
\\

As a result of their low \vsini, our sample stars feature spectral lines that mostly reflect their intrinsic profiles rather than their Doppler broadening (as opposed with most stars studied to date with conventional Doppler imaging, \emph{e.g.}, \citealt{Cameron92, Morin08a}). The consequence is that a direct modeling of the observed profiles would critically depend on our ability to achieve a detailed description of the local profile.  

To overcome this limitation, we propose a novel technique, based on interpreting the residuals with respect to the average profile, rather than the profiles themselves. Practically speaking, we start the process by computing the average profile over the whole data set <$I$>. We then subtract <$I$> from each individual Stokes $I$ profile of the time series to derive the profile residuals $RI$ that directly reflect the profile distorsions and include most information about the spot distribution to be reconstructed.  
In parallel, we model <$I$> by adjusting the parameters of the local profile $I_{\rm p}$ until we obtain a good fit (including the Doppler broadening);  we call this model average profile <$I'$>.  We then sum the $RI$ residuals to <$I'$> and obtain a new data set $I'$.  Since <$I'$> is now perfectly known, the imaging code can concentrate its efforts on reproducing the $RI$ residuals, i.e. the core material of our data set.

\subsection{Simulations}
\label{sec:simu}

We performed a set of simulations to test the performances of our novel reconstruction method. From an initial brightness map, we compute the associated Stokes $I$ and $RI$ data set for a given \vsini, stellar inclinaison $i$, and spectral resolution. The objective is to retrieve both the brightness map and the quantities derived from the reconstructed profiles : the RV curve \vr, FWHM and \vs. 

We present below the simulation results obtained in the case of slow rotators (\vsini~$\leqslant$~4~\kms), and derived assuming a spectral resolving power of 10$^5$ (\emph{i.e.,} the resolution of HARPS-Pol). We further assume that the S/N of the LSD profile residuals $RI$ is equal to 4,000 (value close to the observed S/N). Two different cases are studied : (i) a dense and regular sampling to test more specifically the use of the residuals (simulation A), and (ii) irregular sampling based on the observation of GJ~479, to mainly estimate the impact of a realistic phase coverage on the determination of the average profile (simulation B).

Dark spots are assumed to be circular with a relative size $f$ \footnote{defined as the fractional area of the star covered by a spot, $\frac{1-\rm cos \alpha}{2}$, see \cite{Hebrard14}} with respect to the overall stellar surface. The total equivalent spot area, $\epsilon$, is thus defined as 
$\epsilon = f \times (1 - C) \times b$. For our simulation, we set b~=~0.5.
We consider two dark spots : spot \#1 has a relative area of $f_1$~=~3\% with $C_1$~=~0.4 and thus $\epsilon_1$~=~0.9\%, and is located at 20\degr\ of latitude, spot \#2 is characterized by $f_2$~=~1.5\% with $C_2$~=~0.2, and thus $\epsilon_2$~=~0.6\%, and is at 50\degr\ of latitude. The full equivalent spot area $\epsilon$ is equal to $\epsilon_1 + \epsilon_2$~=~1.5\% . The \vsini\ of the star ranges from 1 to 4~\kms, and the stellar inclination is $i$~=~60\degr. The local profile within a spot is 15\% larger than in the quiet photosphere.
\\
\begin{table}
\begin{center}
\begin{tabular}{ccccccccc}
\hline
simu &\vsini\ (\kms) & initial $\chi^2_r$ & final $\chi^2_r$ & spotted area (\%)\\
\hline
\hline
\multicolumn{5}{c}{Imaging from $I$ = conventional method} \\
\hline
A & 1 &             6.5          &          1.0             &        
  1.35      \\
   & 2 &              12.2         &             1.0           &            
   1.45                  \\
   & 4 &              24.9           &           1.0             &             
   1.50                 \\
   \hline
B & 1 &         4.7                &              1.0                 &       
1.30                 \\
   & 2 &         8.1                 &             1.0            &              1.45            \\
   & 4 &         17.4             &                1.0        &        
   1.50      \\
\hline
\hline
\multicolumn{5}{c}{Imaging from $I'$ = residual method} \\
\hline
A   & 1   & \textcolor{blue}{4.7}  & \textcolor{blue}{1.13}  & \textcolor{blue}{
1.35}  \\
   &  2  & \textcolor{blue}{11.5}  & \textcolor{blue}{1.11}  & \textcolor{blue}{
   1.40}  \\
   &   4 & \textcolor{blue}{24.7}  & \textcolor{blue}{1.0}  & \textcolor{blue}{
   1.50}  \\
\hline
B   &  1  & \textcolor{blue}{3.3} & \textcolor{blue}{1.15}  & \textcolor{blue}{
1.25}  \\
   &   2 & \textcolor{blue}{8.2} & \textcolor{blue}{1.1}  & \textcolor{blue}{
   1.40}  \\
   &   4 & \textcolor{blue}{18.0} & \textcolor{blue}{1.0}  & \textcolor{blue}{
   1.45}  \\
\hline
\end{tabular}
\caption{Parameters of the reconstructed map for a star with \vsini~=~1~\kms\ and $i$~=~60\degr\ with 2 spots covering 1.5\% of the stellar surface. Column 1 gives the considered simulations, and column 2 the \vsini\ of the stars. Columns 3-4 indicates the initial and final $\chi^2_r$ associated with the reconstruction. Column 5 give the spotted reconstructed area. The results obtained from imaging using directly $I$ are given in black, the results obtained from $RI$ are given in blue.
Simulation A : reconstruction from a dense and regular sampling. 
Simulation B : reconstruction from a random sampling. 
}
\label{tab:simu3}
\end{center}
\end{table}
\begin{figure*}
\begin{center}
\includegraphics[scale=0.4,angle=0]{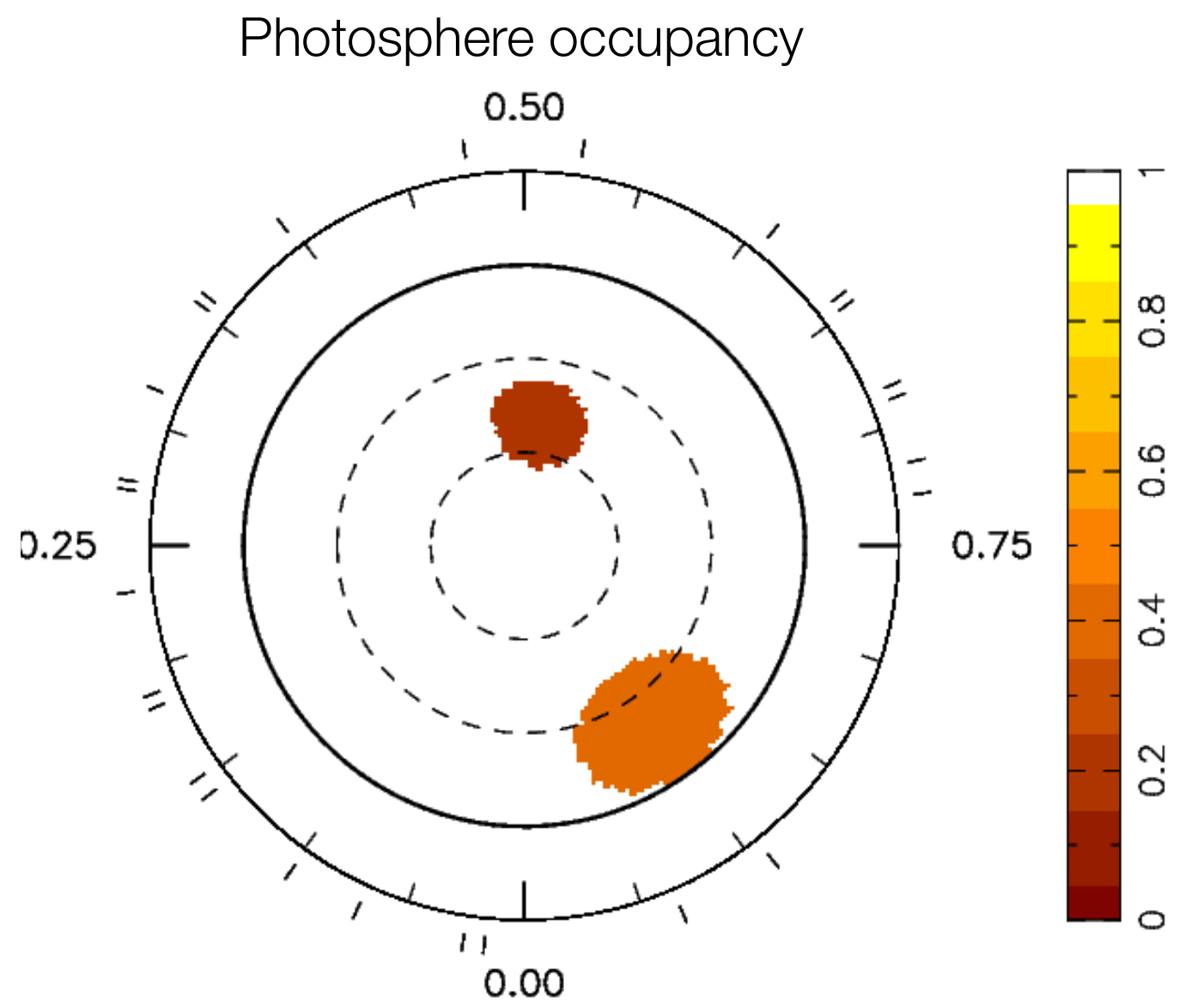}\\
\includegraphics[scale=0.4,angle=0]{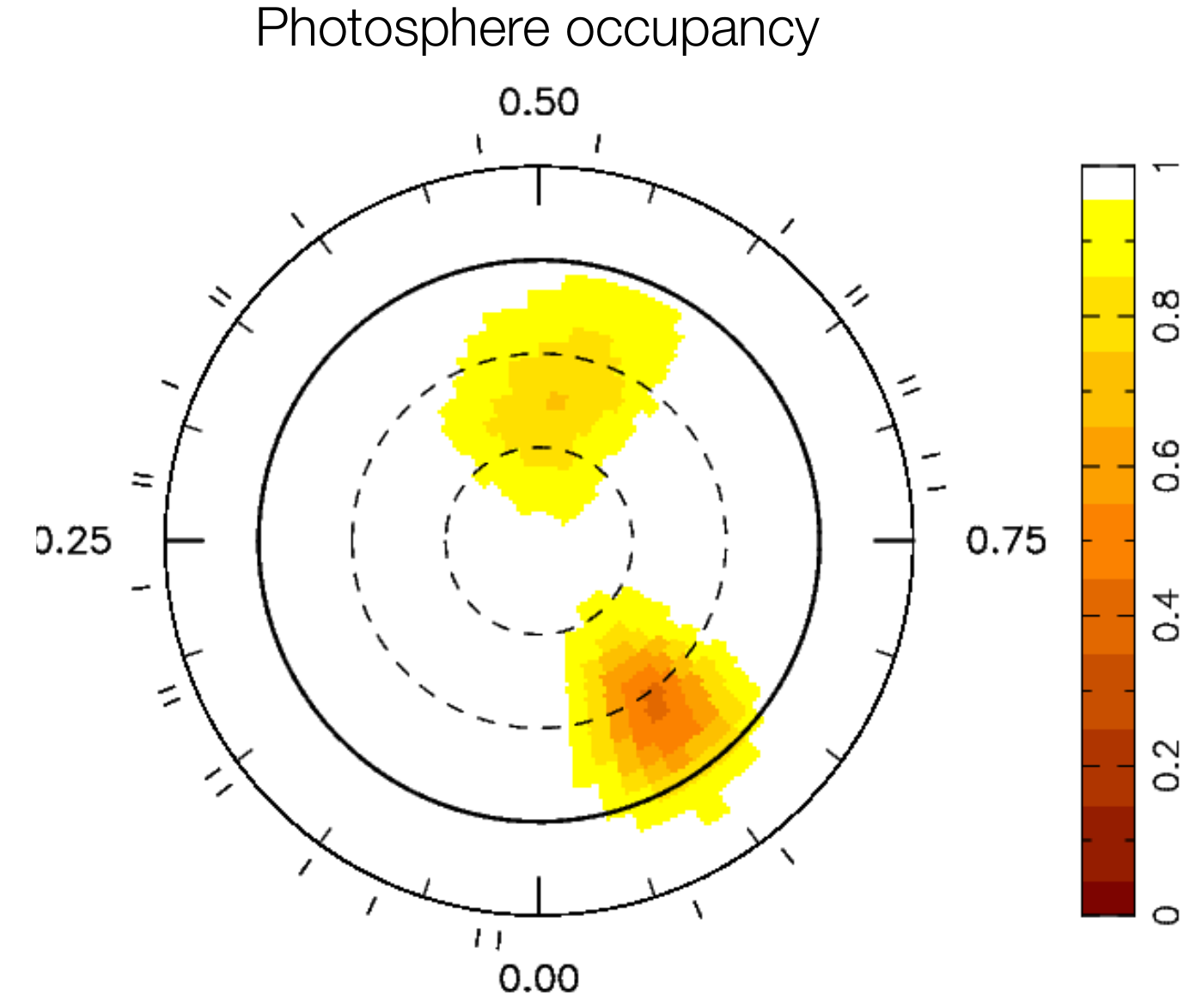}
\includegraphics[scale=0.2,angle=0]{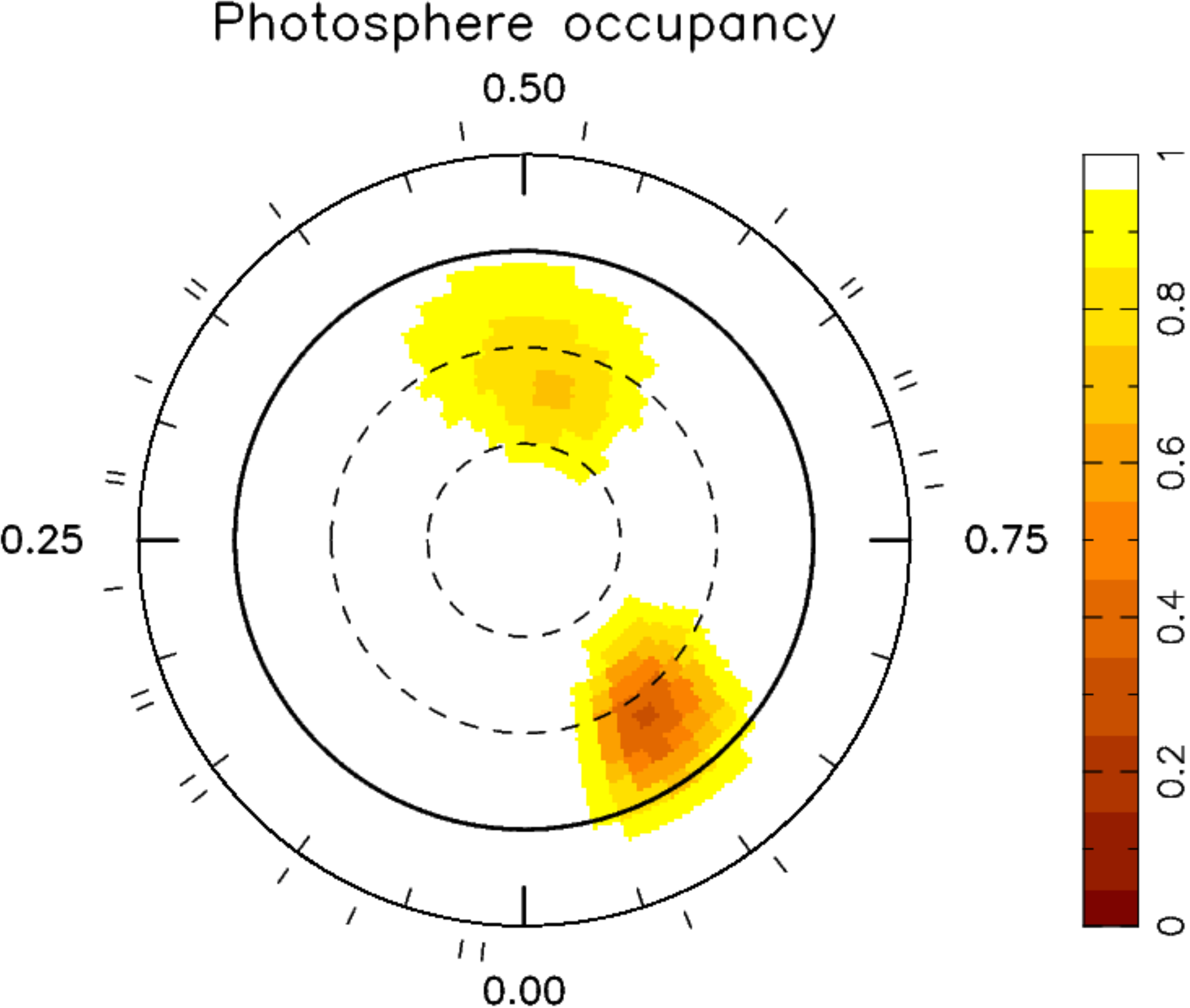}
\includegraphics[scale=0.24]{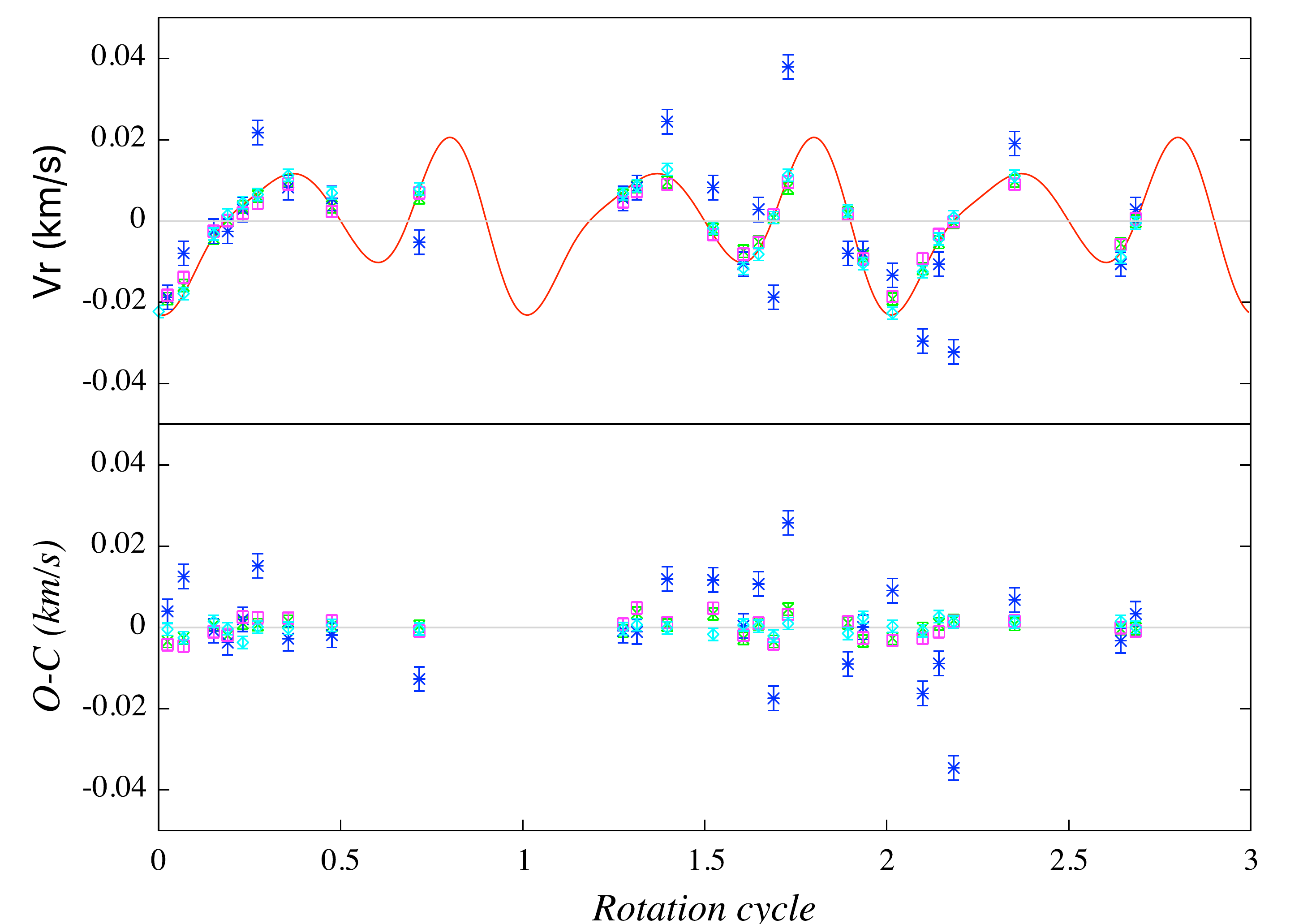}
\includegraphics[scale=0.24]{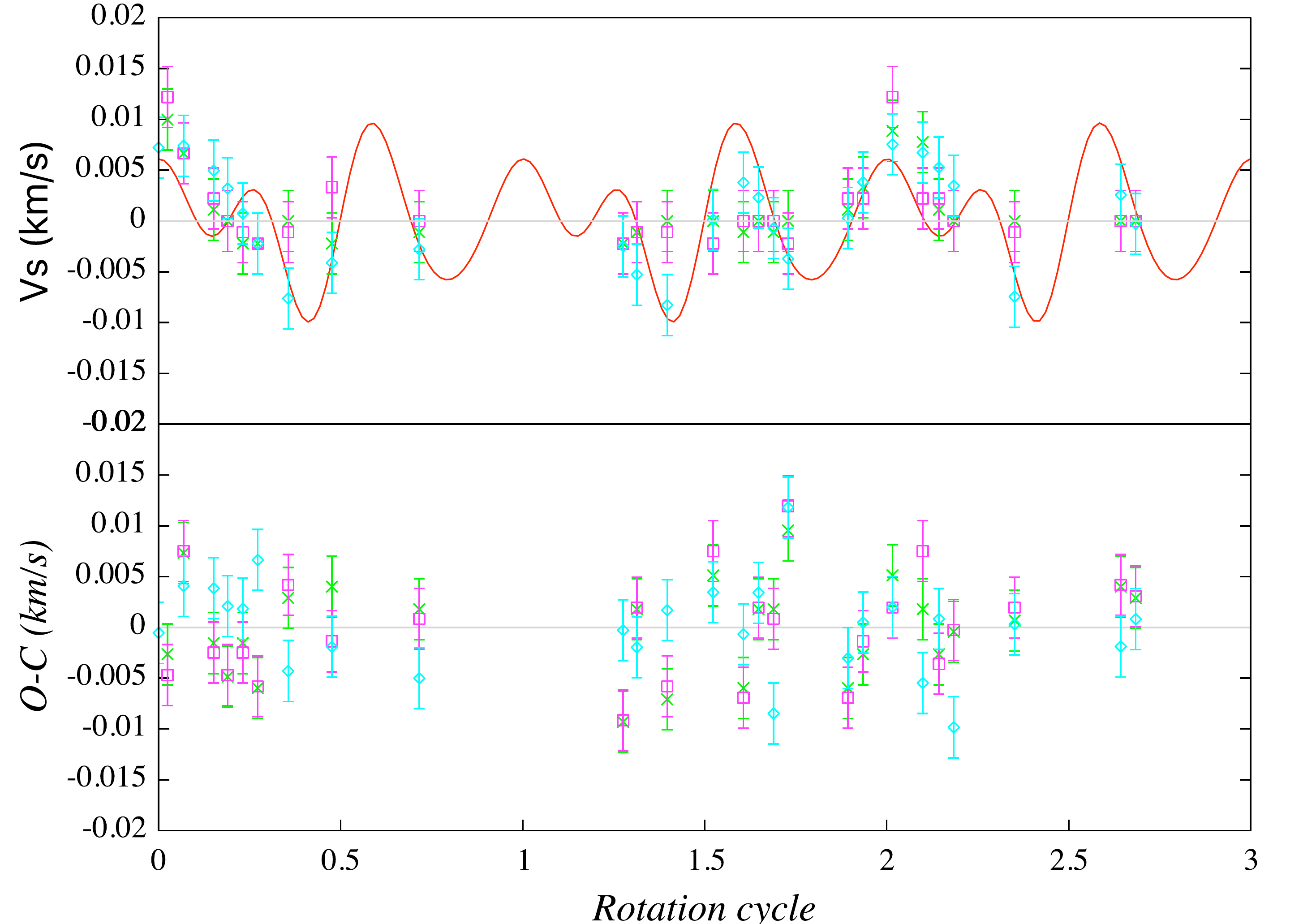}
\includegraphics[scale=0.24]{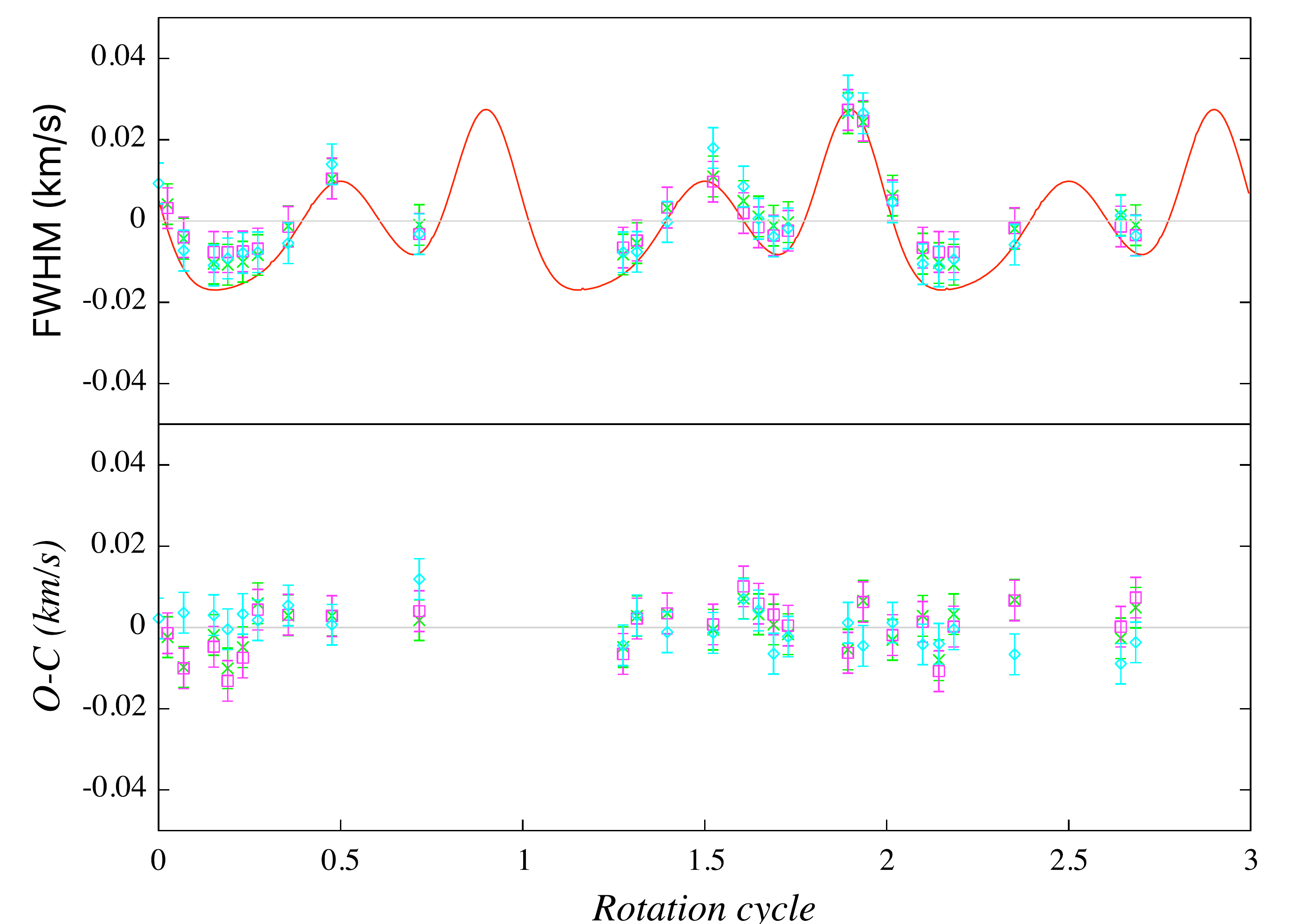}
\includegraphics[scale=0.24]{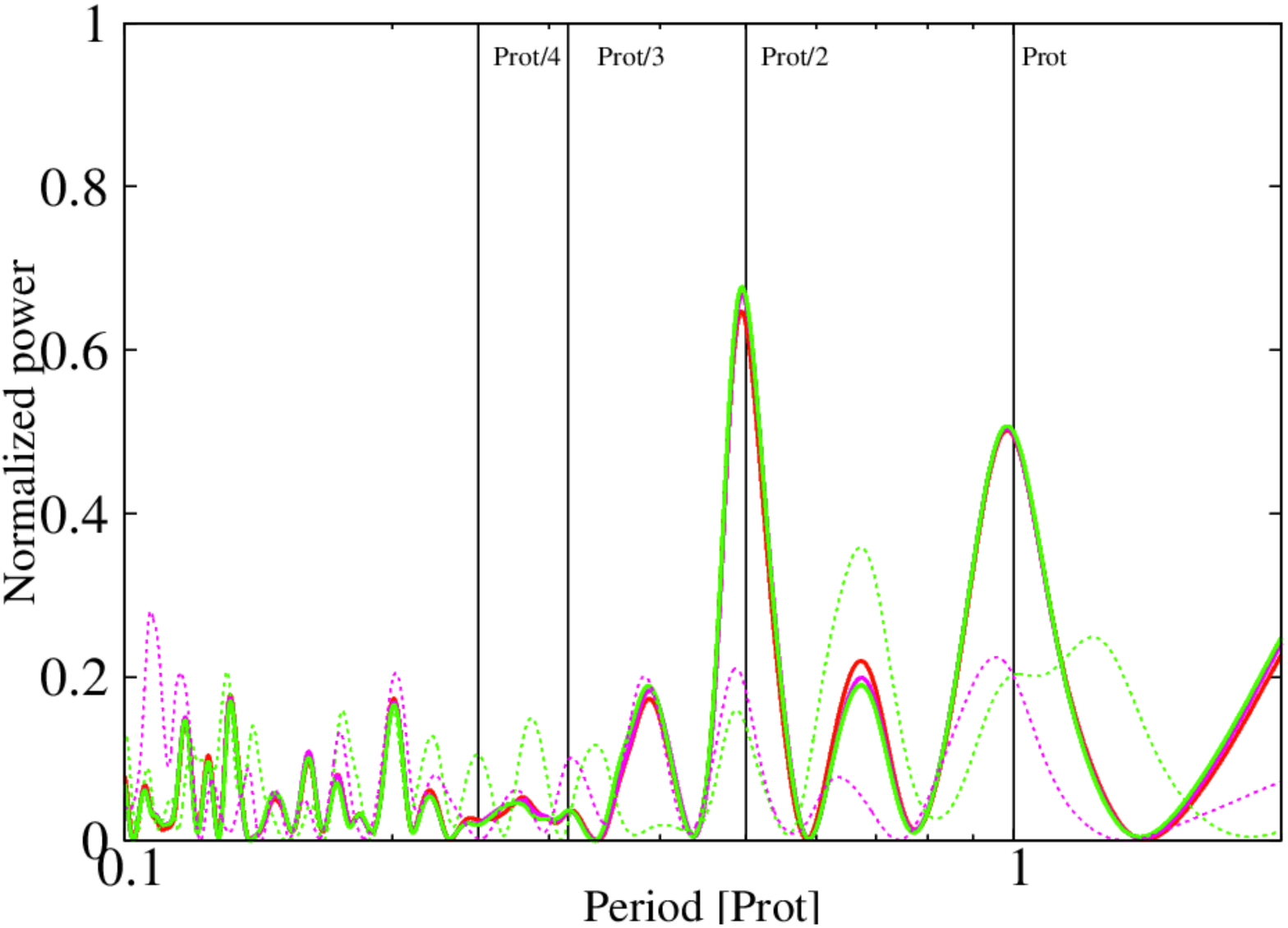}
\caption{
Reconstructed map obtained for a simulated star with \vsini~=~1~\kms and $i$~=~60\degr\ featuring an equivalent spot area $\epsilon$ of 1.5\%. 
\textit{Top}: Spot distribution to reconstruct.
\textit{2nd row, left}: Reconstructed map from $I$ with the realistic sampling B. 
\textit{2nd row, right}: Same as \textit{left} but reconstructed from RI residuals.
The colour-scale depicts the photosphere filling factor of each cell $C_j$ (white corresponding to an unspotted cell).
\textit{3rd row, left:} original \vr's (red solid line) compared with those reconstructed using either the conventional (pink open squares) or residual (green crosses) imaging method respectively, and with those derived from the \vr vs \vs anticorrelation (blue asterisks). The multiple sine fit to the data (cyan open diamonds) is also shown. The corresponding O-C residuals are presented on the bottom curve. The gray line depicts the 0~\ms\ level. 
\textit{3rd row, right:} Same as \textit{3rd row, left} but for \vs.
\textit{Bottom left:} Same as \textit{3rd row, left} but for FWHM.
\textit{Bottom right:} Periodograms of \vr\ (solid red line), of the \vr\ computed from data set obtained with the conventional method (solid pink line), and with the residual method (solid green line). The periodograms of the filtered RVs (using either the conventional or the residual method) are respectively shown with the red and green dashed lines.  
The vertical lines outline the rotation period (in unit of \pstar) and its 3 first harmonics (\pstar/4, \pstar/3, \pstar/2).
This figure is best viewed in colour.
}
\label{fig:vrvs_lent}
\end{center}
\end{figure*}
\subsubsection{Reconstructed map}
Figure~\ref{fig:vrvs_lent} (top part) and Table~\ref{tab:simu3} show, respectively, the maps and their associated reconstructed characteristics. To test the impact of using the profile residuals $RI$ instead of Stokes $I$, we compare the maps obtained using Stokes $I$ profiles directly (called hereafter 'the conventional method') with those obtained from the $I'$ replacement data set described above (called below 'the residual method').

We note that the global spot distribution is recovered, whatever the technique we used.
With the conventional method, $\chi^2_r$~=~1.0 is reached whatever the \vsini\ and the phase coverage. However the spotted area is roughly under-estimated with decreasing \vsini.
A similar but amplified behavior is observed with the residual method. The use of the average profile <$I$> to compute the $I'$ dataset mainly affects the reconstructed spotted equivalent spot coverage $\epsilon$, which ends up being underestimated (1.25-1.45\% instead of 1.5\% depending on the \vsini). 
This loss of accuracy when \vsini\ decreases mainly reflects that information gets increasingly blurred in longitude as stellar rotation slows down, thus weakening profile distortions and making them harder to reconstruct for the code.  \\

\subsubsection{Model of the RV jitter}
The main parameter we aim at recovering is the RV curve shown in Figure~\ref{fig:vrvs_lent} (3rd row, on the left) in the case of \vsini~=~1~\kms.
\begin{table}
\begin{center}
\begin{tabular}{cccccccccc}
\hline
Simu& \vsini & (a) &  (b) &  (c) &  (d) &  (e) &  (f) & (g) \\
& (\kms)&  &  & &   &  &  & \\
\cline{3-9}
\multicolumn{9}{c}{~~~~~~~~~~~~~~~~~~~~~~~~~~(\ms)} \\
\hline
\hline
A & 1  & 1.5 & \textcolor{blue}{2.0} & \textcolor{blue}{2.4}  & \textcolor{blue}{4.9}  & \textcolor{blue}{1.6}  &  \textcolor{blue}{1.5} & \textcolor{blue}{12.9}\\
  &  2 & 1.7 & \textcolor{blue}{2.4} & \textcolor{blue}{2.4}  & \textcolor{blue}{10.4}  & \textcolor{blue}{2.1}  &  \textcolor{blue}{1.5} & \textcolor{blue}{11.9}\\
  &  4 & 2.9 & \textcolor{blue}{3.2} & \textcolor{blue}{3.2}  & \textcolor{blue}{25.3}  & \textcolor{blue}{6.3}  &  \textcolor{blue}{2.8} & \textcolor{blue}{27.3}\\
 \hline
B  & 1& 1.6 & \textcolor{blue}{1.8} & \textcolor{blue}{2.1}  & \textcolor{blue}{3.6}  & \textcolor{blue}{1.7}  &  \textcolor{blue}{1.4} & \textcolor{blue}{17.7}\\
  & 2  & 1.8 & \textcolor{blue}{1.8} & \textcolor{blue}{2.1}  & \textcolor{blue}{8.3}  & \textcolor{blue}{2.0}  &  \textcolor{blue}{1.2} & \textcolor{blue}{13.4}\\
  &  4 & 3.2 & \textcolor{blue}{3.2} & \textcolor{blue}{3.3}  & \textcolor{blue}{18.5}  & \textcolor{blue}{4.6}  &  \textcolor{blue}{2.6} & \textcolor{blue}{31.3}\\  
  \hline
\end{tabular}
\caption
{ Column 1 gives the studied simulation, column 2 the \vsini\ of the star. Columns 3 indicates the rms RV noise, whose increase with \vsini\ reflects the decrease in RV precision resulting from the shallower and broader line profiles of faster rotators. Columns 4-9 give the rms of the RV residual, after a filtering from the direct method (b), from the indirect method (c), from a multiple sine-fit with, respectively, 1, 2 and 3 harmonics (d)-(f) , and from the anti-correlation \vr-\vs\ (g) .
}
\label{tab:simu4}
\end{center}
\end{table}

First, we note that RV variations are fitted down to the noise level, with both methods. 
For this spot configuration, the periodogram exhibits conspicuous peaks at \pstar\ and \pstar/2.  
We find that both imaging methods provide similar results in the sense that they are quite successful at filtering the rotationally modulated activity jitter; we do not observe any strong peaks in the periodogram of the RV residuals O-C (= observed - computed, see periodograms Figure~\ref{fig:vrvs_lent}, right bottom panel).

To quantify the model efficiency, we compare the rms of O-C data (see Table~\ref{tab:simu4}) using the two different imaging methods (based on $I$ and $I'$) with that derived from the multiple sine fits of \vr\ (with fundamental + 1 to 3 harmonics), and from the usual anti-correlation observed between \vs\ and \vr\ \citep[\emph{e.g.}, see][for more details]{Melo07}.
From these results, we clearly see that the quality of the filtering based on Doppler imaging is similar to the one obtained from the multiple sine-fit (fundamental + two first harmonics), and much better than that based on the anti-correlation between \vs\ and \vr. More specifically and with respect to the latter case, we decrease the rms dispersion by a factor of 5 to 8 (depending on the \vsini).

The use of the profile residuals $RI$ can lead to a small underestimate of the equivalent spot area epsilon (for smaller values of \vsini\ in particular), however the RV filtering is not affected reaching down to almost the noise level with both methods.  

Finally, the density of the sampling does not affect much the quality of the reconstructed RVs, as long as it is dense and even enough (typically a few tens of observing points covering a few rotation cycles). The lowest the \vsini, the stronger the importance of the sampling. 

\subsection{Application to M dwarfs}
\label{sec:mapcq}

We apply the residual imaging method presented and tested in Sec.~\ref{sec:di} \& \ref{sec:simu} to recover the parent spot distribution generating the observed RV activity jitter for the various stars of our sample.  

To assess the likelihood of the RV fit we obtain from the map, we compute the FAP as presented in Sec.~\ref{sec:detec}. We take the multiple sine fit of \vr\ (fundamental + 2 first harmonics) as a reference to compute $\Delta\chi^2$; we then obtain (with a formula resembling Eq.\ref{eq:deltachi2}) :
\begin{figure}
\begin{center}
\includegraphics[scale = 0.5]{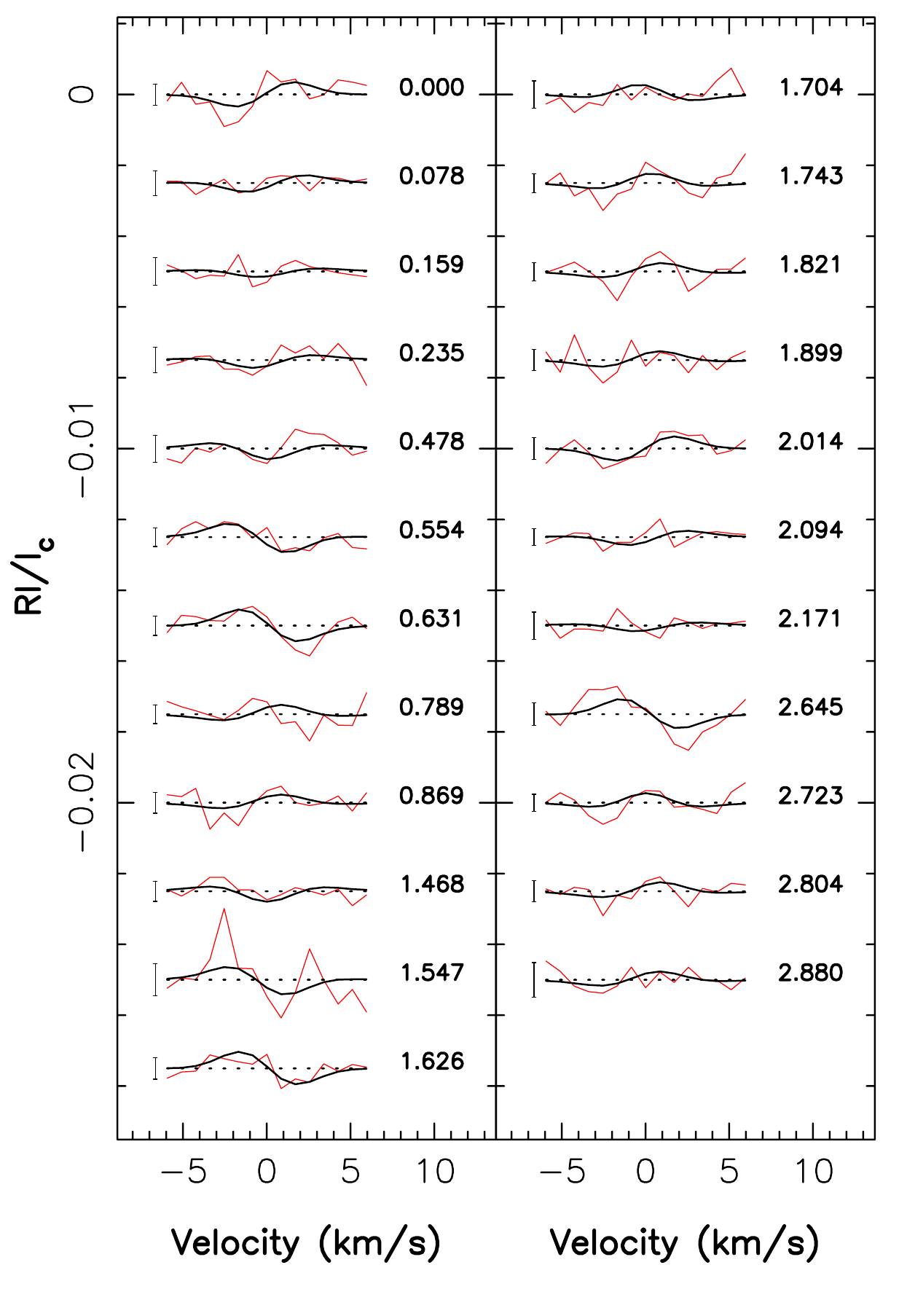}
\caption{
Temporal series of $RI$ of GJ~358. Data are in red, the modeled $RI$ are in black. On the right of each spectrum, we indicate the observation phase, on the left the 1-$\sigma$ error bars.
}
\label{fig:res358}
\end{center}
\end{figure}
\begin{figure*}
\includegraphics[scale = 0.4]{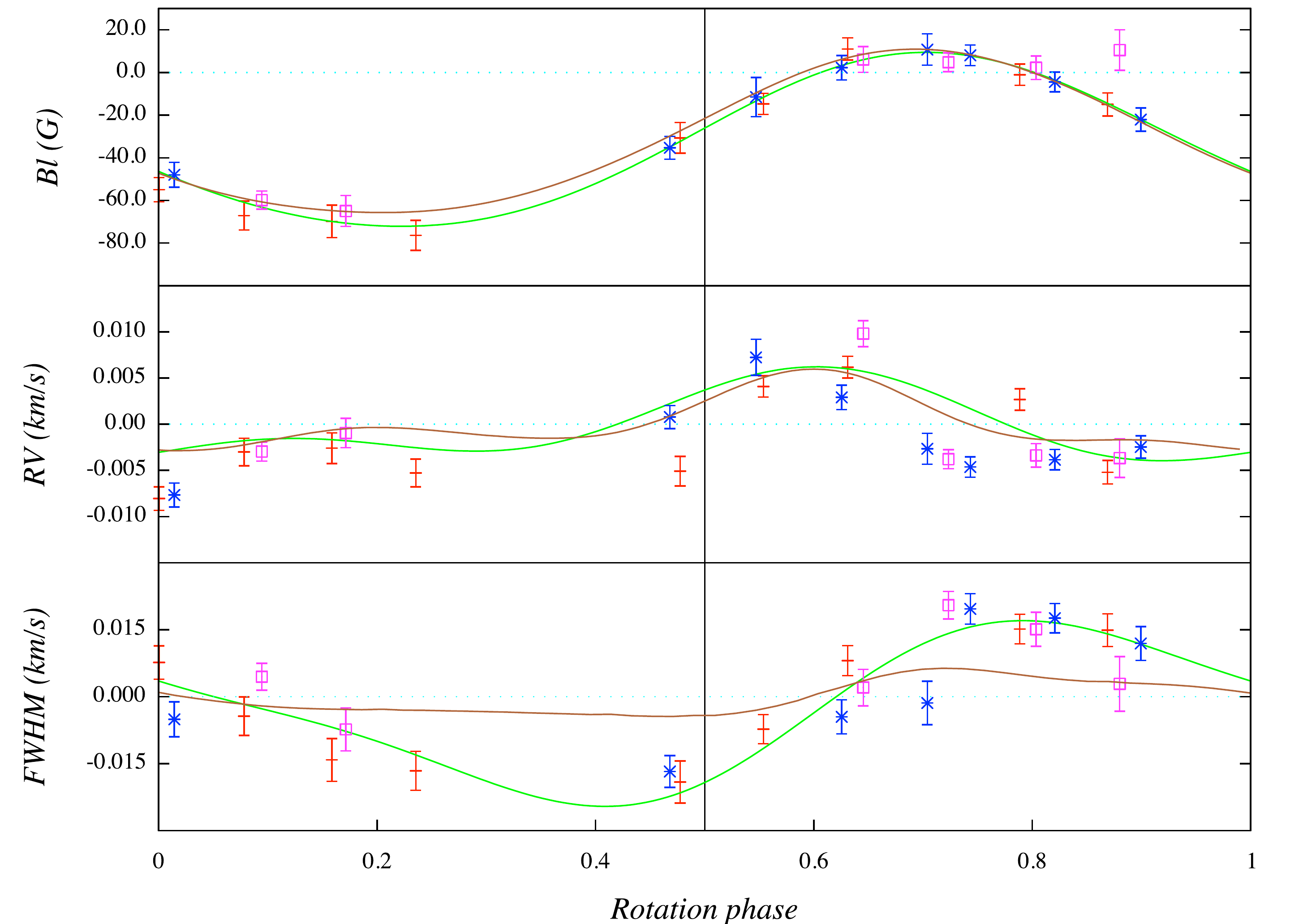}\\
\includegraphics[scale = 0.47]{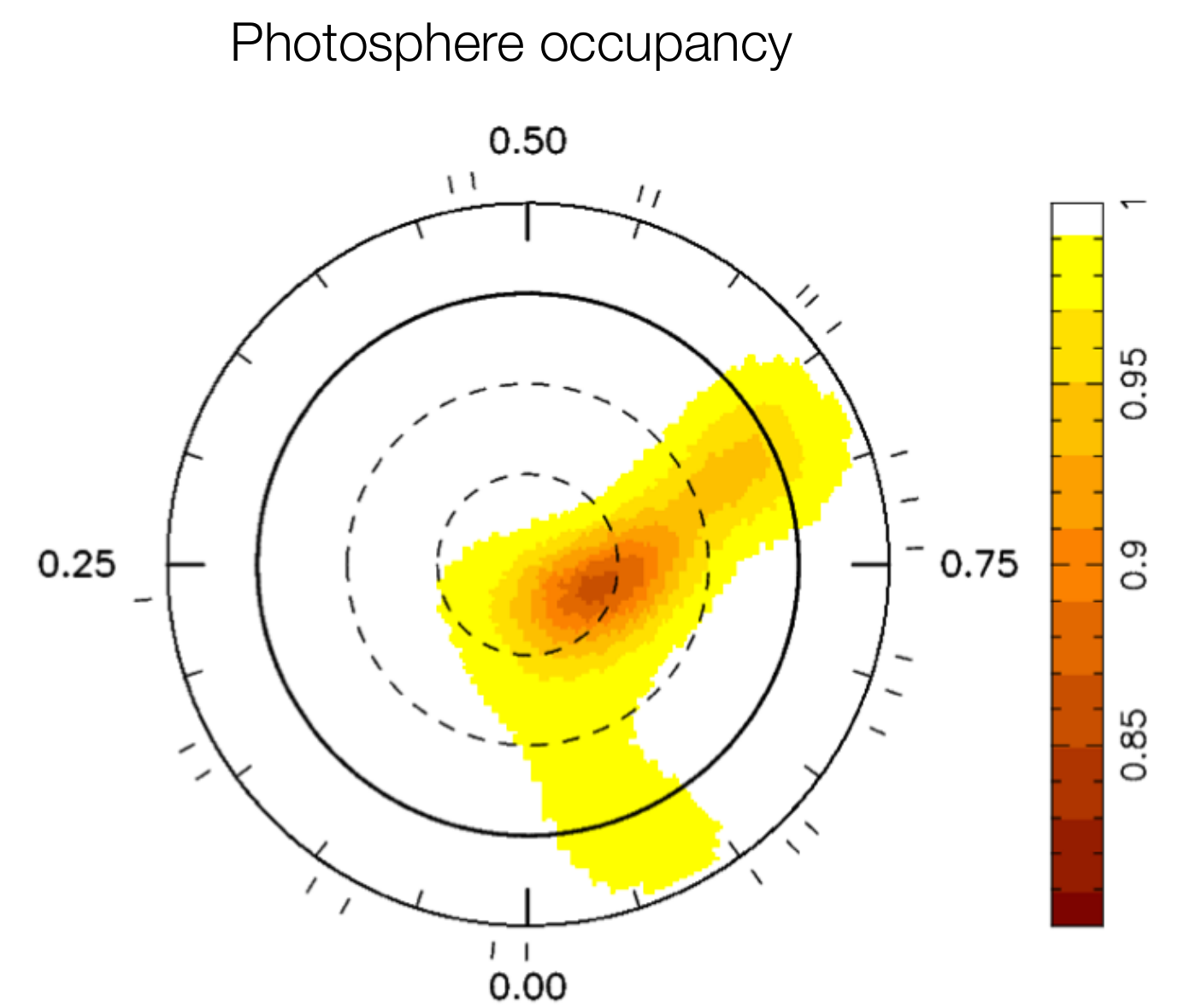}
\includegraphics[scale = 0.25]{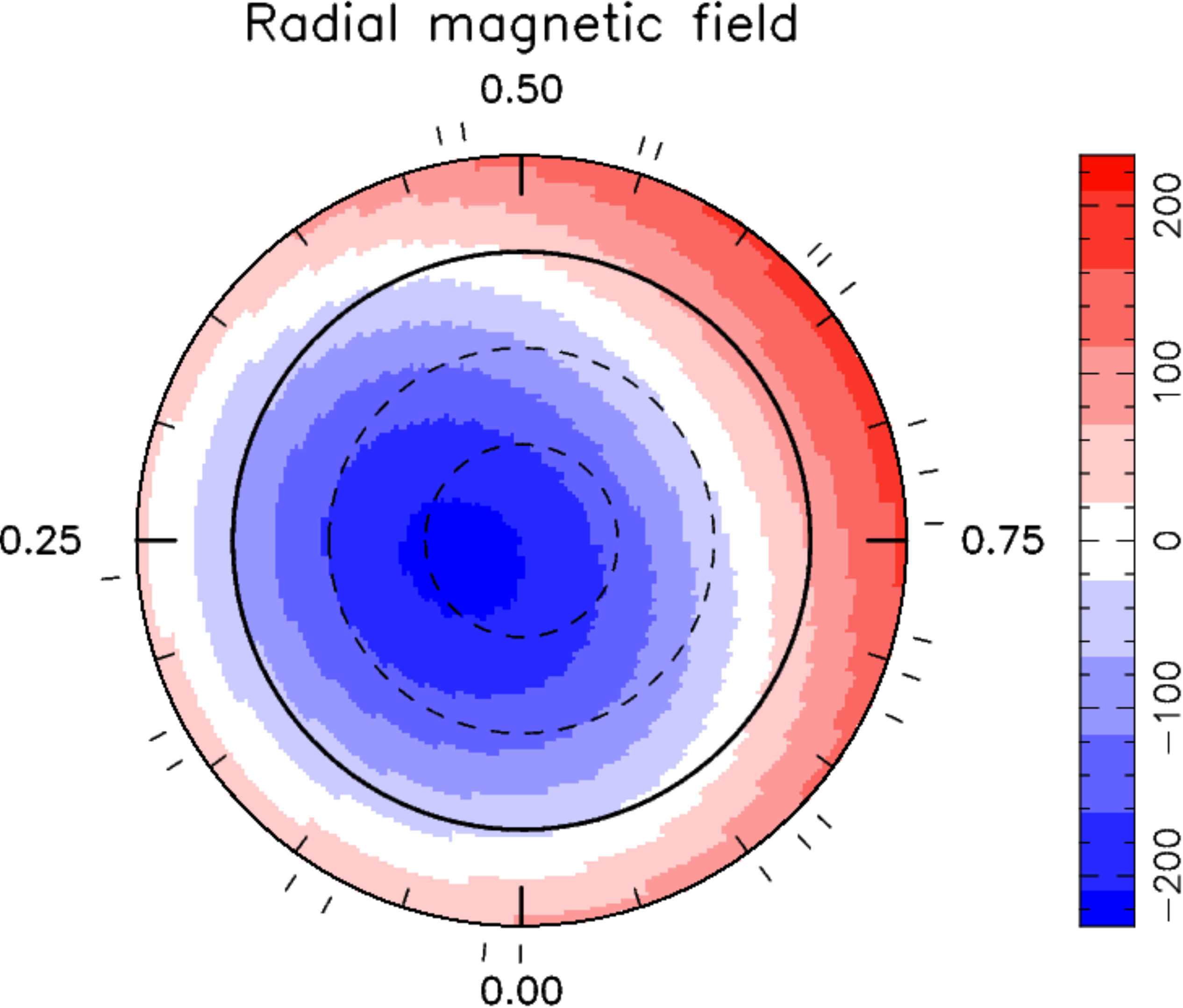}
\includegraphics[height = 6.6cm, width=16cm]{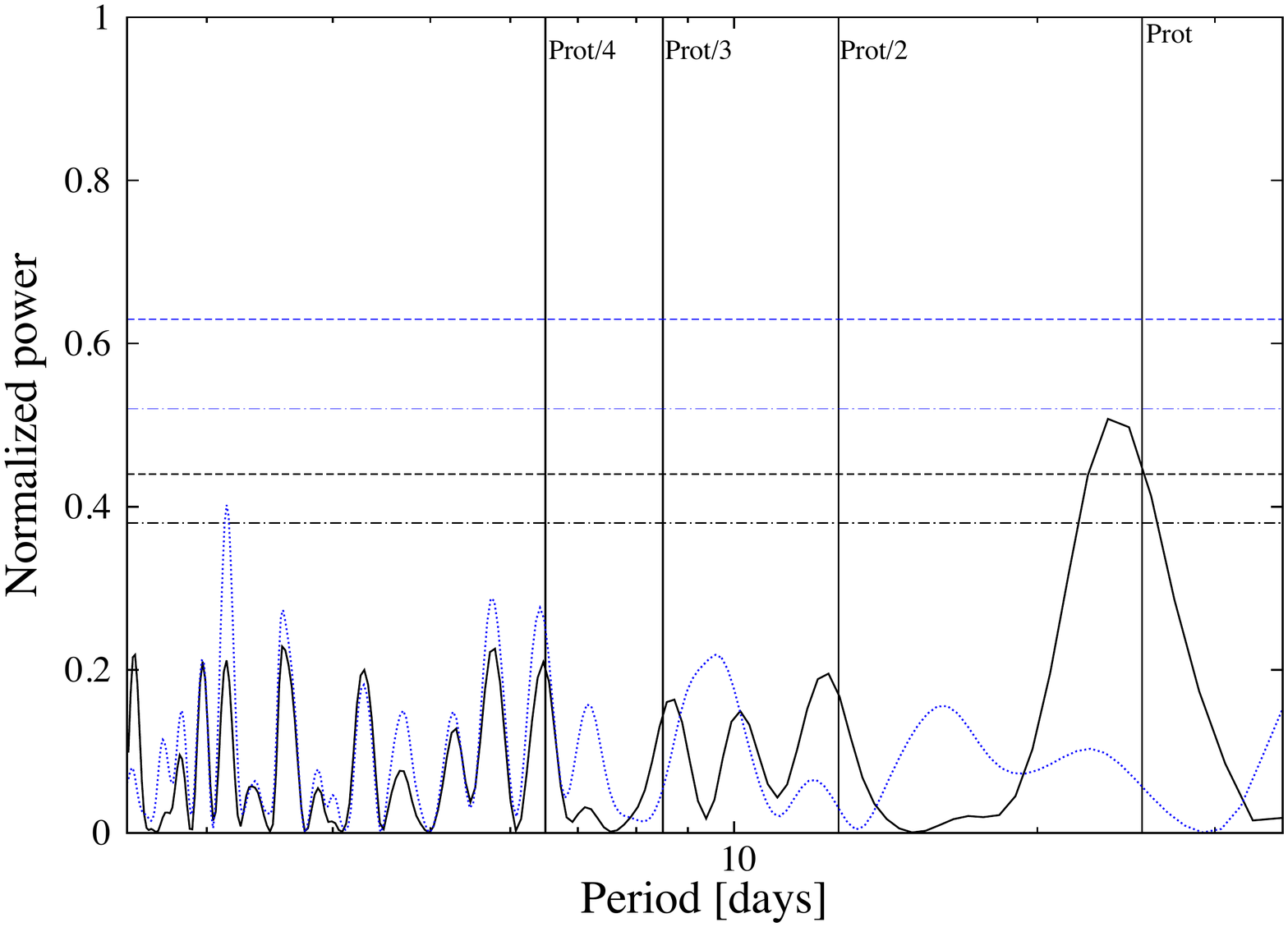}
\caption{
\textit{Top:} Temporal evolution of \bl, \vr\ and FWHM (with respect to the average value) of GJ~358.
Data and their error bars are represented in red, blue and pink according the rotation cycle (cycle 1 in red, cycle 2 in blue and cycle 3 in pink). The green curves corresponds to the sine fit, and the brown curves represents the RVs computed from the DI map.
\textit{Middle left:} Maps of the filling factor of the photosphere (white means that there is only quiet photosphere, brown means there is only spot in the cell), and \textit{Middle right:} Map of the radial large-scale magnetic field.
\textit{Bottom:} Periodograms of observed RVs (black), and of the RVs after the RV filtering from DI (blue). The FAP at 1\% and 10\% are represented in dotted lines and dot-dashed lines.
}
\label{fig:mapcq358}
\end{figure*}
\begin{equation}
\Delta\chi^2 = \frac{\chi^2_{r,0} - \chi^2_{r,2}}{\chi^2_{r,1}}.N
\end{equation}
 with $\chi^2_{r,1}$ and $\chi^2_{r,2}$, respectively, corresponding to the $\chi^2_{r}$ of the multiple sine fit and of the fit obtained with the Doppler imaging based on residual reconstruction (hereafter DI fit).
The number of degrees of freedom associated with the imaging process is estimated from the number of parameters associated with the non-axisymmetric SH modes needed to describe the observed variations, \emph{i.e.}, $\sim$20 for $l$~$\leqslant$~4.
Results are presented in Table~\ref{tab:fitvr2}.\\

\subsubsection{GJ 358} 
From the $RI$ profiles (see Fig.~\ref{fig:res358}), we reconstruct the map shown in Fig.~\ref{fig:mapcq358}, featuring an equivalent spot coverage $\epsilon$ of $\sim$1\% (with $b$~=~0.5). The initial $\chi^2_r$ is 3.8 and corresponds to the fit to the $RI$ spectra with an unspotted star. Adding spots on the stellar surface allows the code to reduce $\chi^2_r$ down to 2.1, with a main spot at high latitude ($\sim$~60\degr), and extending towards the equator.

Synthetic RV curve derived from this brightness map exhibits a full amplitude of 8.5~\ms, and matches the data down to $\chi^2_{r,2}$~=~2.05. The rms of the RV residuals is 2.88~\ms\ (whereas $\sigma_0$~=~2.08~\ms). The low FAP (<~0.01\%) demonstrates that the imaging process provides a very significant improvement in the quality of the fit to the data. 
Moreover, in the \vr\ periodograms we clearly see that (i) the signals at \pstar, \pstar/2 and \pstar/3 have been removed, and (ii) no major periodic signal remains.\\

\subsubsection{GJ~479} The reconstructed spots have a equivalent surface of $\sim$1.4\%, and are located at mid-latitude ($\sim$~40\degr, see Fig.~\ref{fig:mapcq479}). It corresponds to a final $\chi^2_r$ of 2.9 (starting from $\chi^2_r$~=~5.0).

The J$_m$ component of the RV jitter deduced from this map has a peak-to-peak amplitude of 11~\ms, and the rms of RV residuals is 3.93~\ms. 
Once J$_m$ is subtracted from RV data, the periodogram does not exhibit any strong peak anymore: the filtering allows to clean up the signals whose periods are \pstar, \pstar/2, \pstar/3.\\

\subsubsection{GJ 410} 
\begin{figure}
\begin{center}
\includegraphics[scale = 0.41]{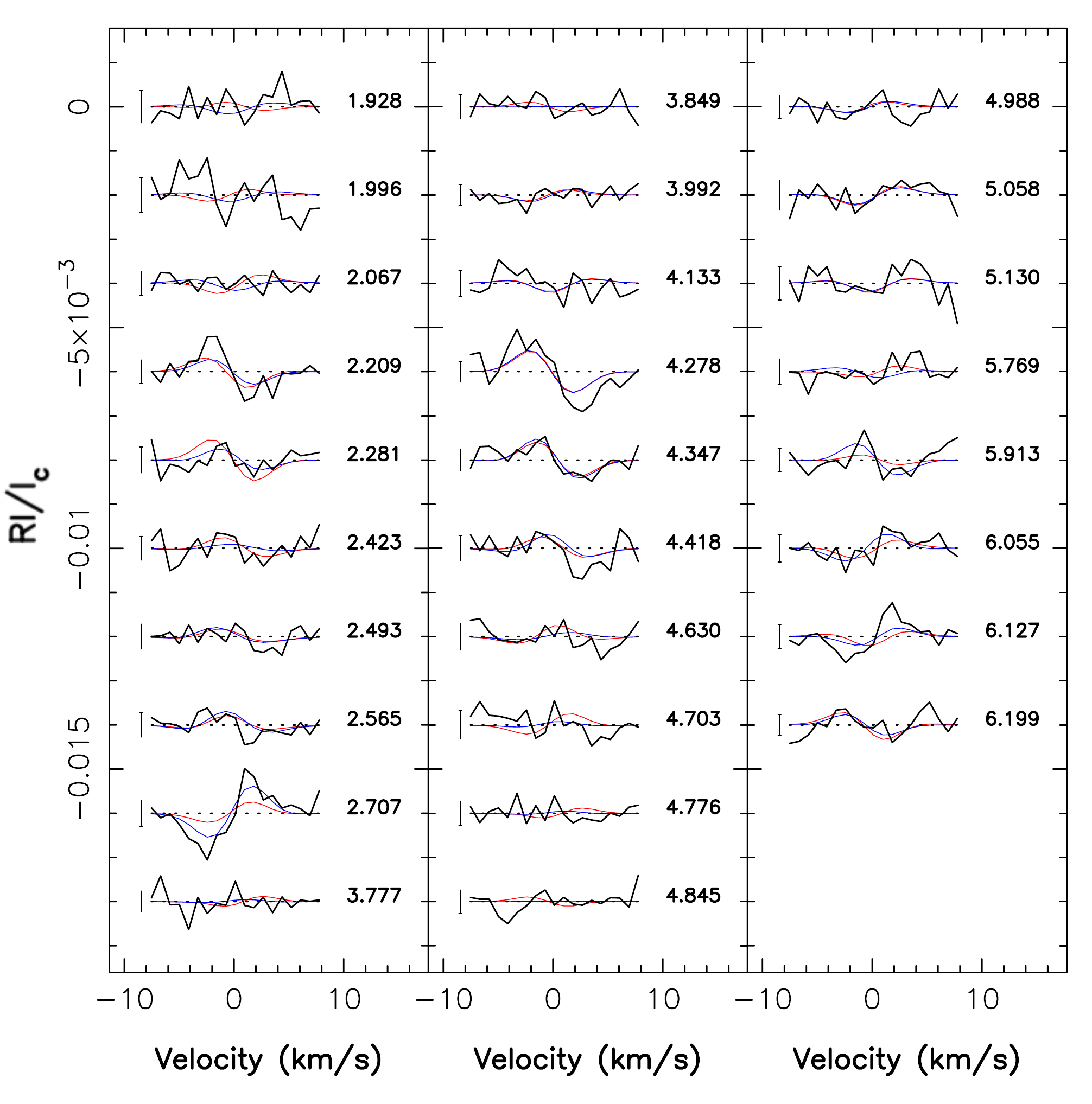}
\caption{
As Fig.~\ref{fig:res358}, for GJ~410. The blue fit represents the fit from epochs \#1 to \#3, the red fir represents the fit from the full data set.
}
\label{fig:res410}
\end{center}
\end{figure}
\begin{figure*}
\begin{center}
\includegraphics[scale = 0.40]{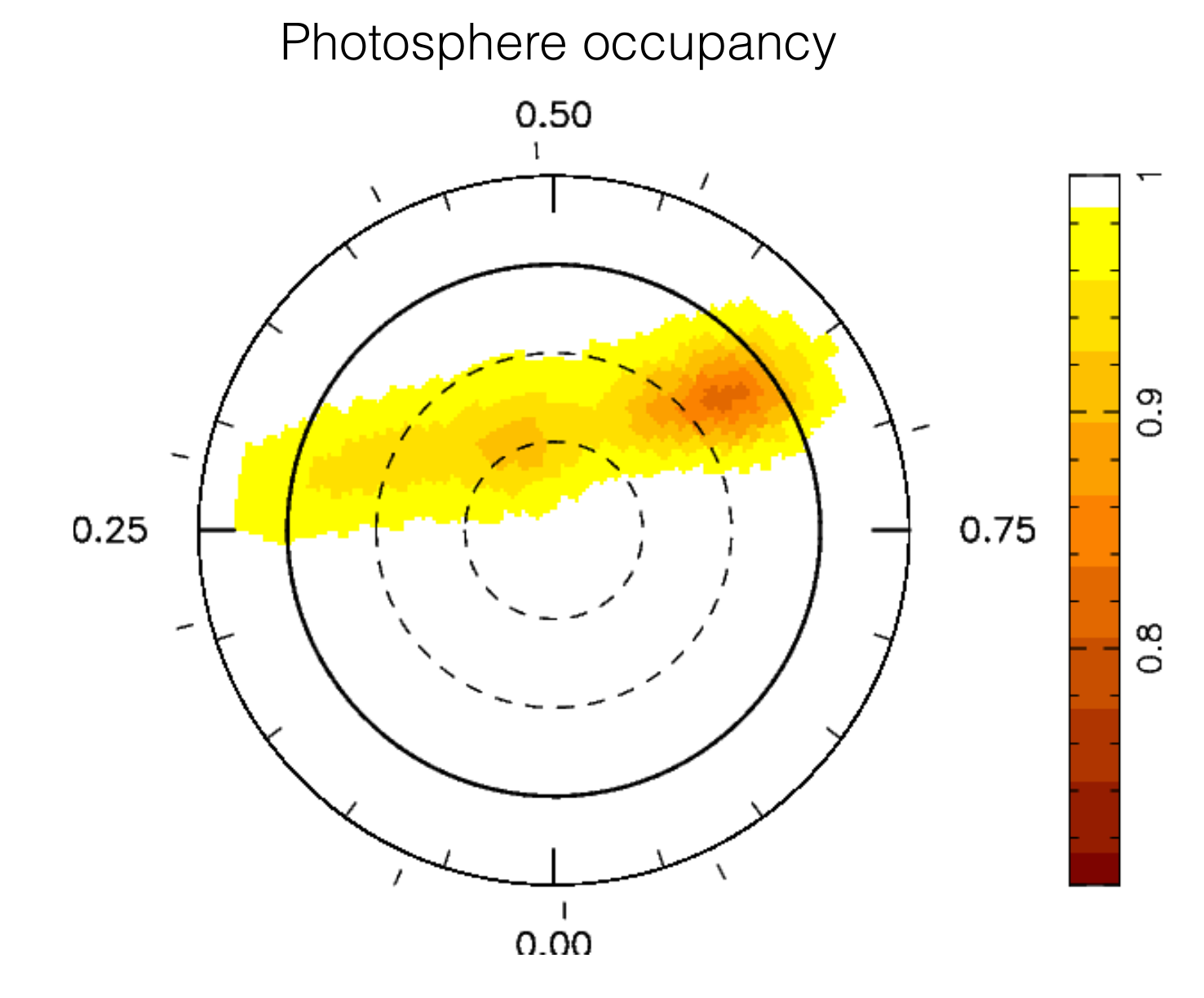}
\includegraphics[scale = 0.40]{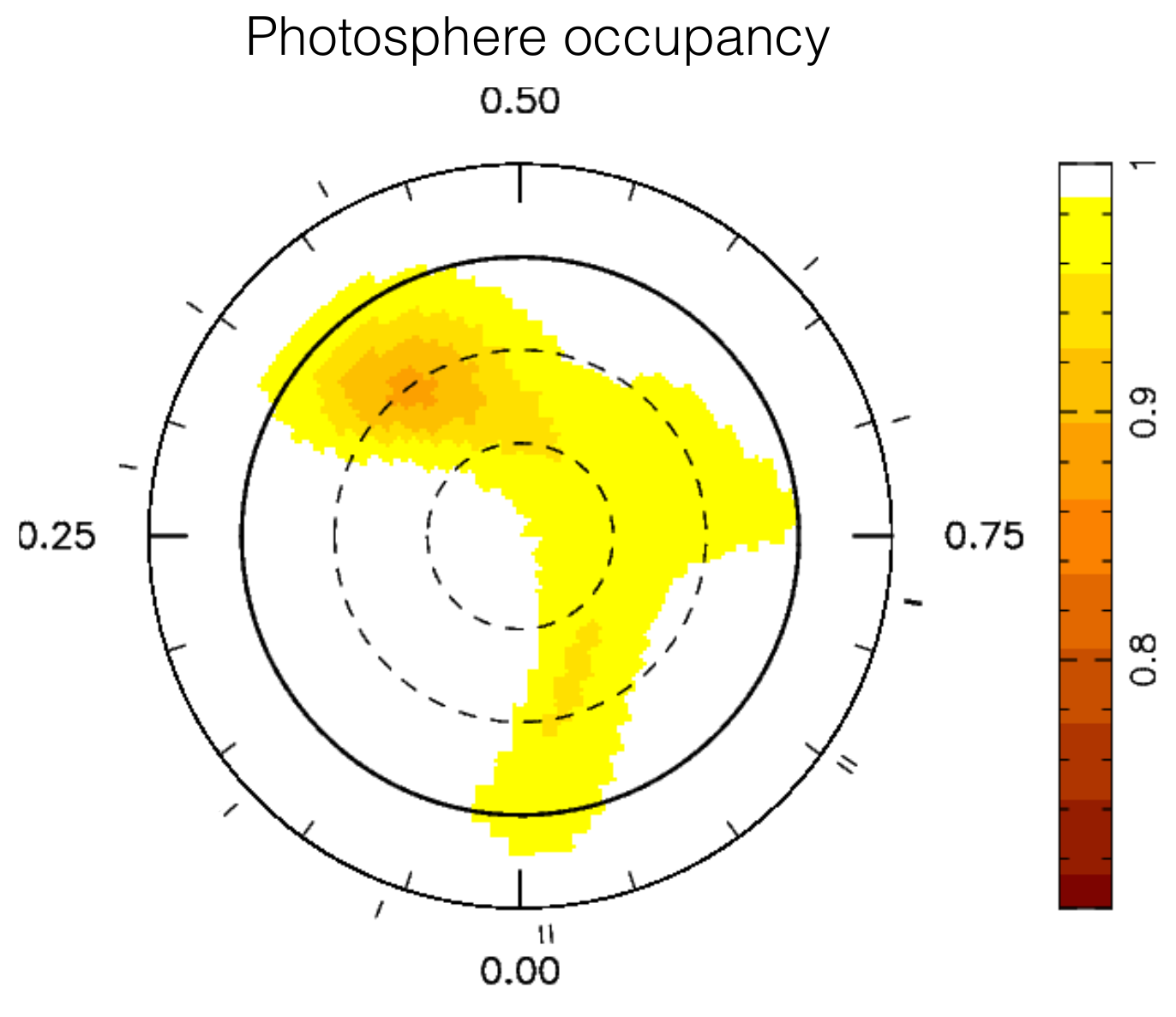}
\includegraphics[scale = 0.40]{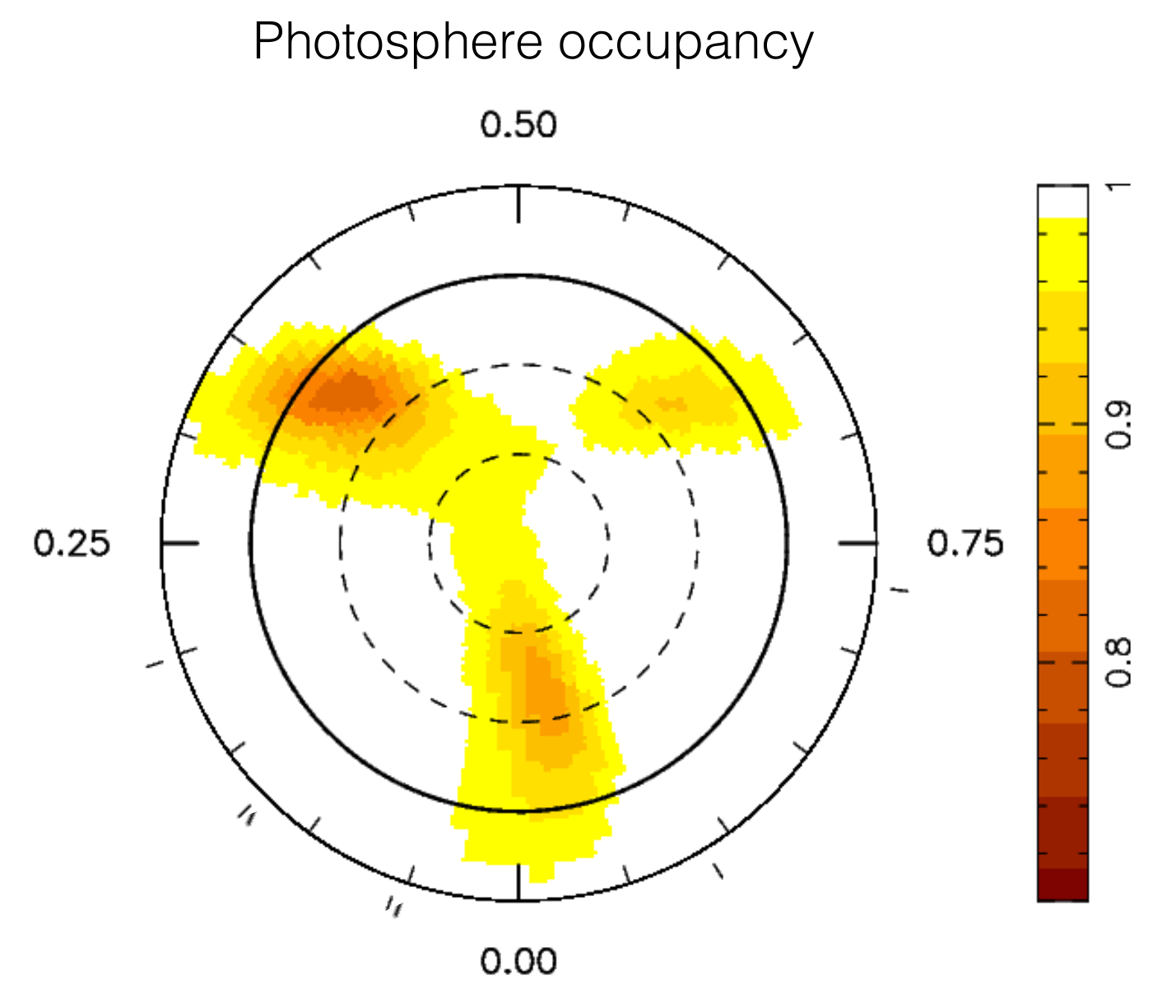}
\caption{
Maps of the filling factor of the photosphere of GJ~410 (white means that there is only quiet photosphere, brown means there is only spot in the cell) at the three epochs (from (1) to (3), from left to right).}
\label{fig:mapcq410}
\end{center}
\end{figure*}

For GJ~410, we collected observations over three months (\emph{i.e}, six stellar rotations). We note that all the $RI$ spectra do not identically repeat from one rotation cycle to the next (see, \emph{e.g.}, phases 2.7 and 3.7, or phases 4.9 and 5.9 in Fig.~\ref{fig:res410}). For this star, one of the most active of the studied sample, 
we first carried out a reconstruction for the whole data set (see Fig.~\ref{fig:carteall_410}). In a second step, we divided the data set into three sequential subsets to take into account the evolution of spot coverage on the stellar surface, respectively corresponding to rotation cycles 1.928-2.707 (epoch \#1, 9 observations), 3.777-4.988 (epoch \#2, 12 obs) and 5.058-6.199 (epoch \#3, 7 obs). 
The results are given Fig.~\ref{fig:mapcq410}.

Dividing the data into multiple subsets allows us to improve the fit to the data, with a final $\chi^2_r$ decreased from 2.0 (for the complete set) to 1.2-1.7 (for the individual subsets). 
The reconstructed maps show that in epoch \#1 a main dark spot at low latitude around phase 0.6 is visible at the stellar surface, with a fainter spot at phase 0.25 and 30\degr\ of latitude. This secondary spot grows and strengthens in epoch \#2 and \#3, and a new spot appears at phase 0.95 from epoch \#2 onwards. 

Moreover, the quality of the RV filtering increases within each of our subsets. The modeled RV curves we derive match the observed ones at a $\chi^2_r$ level of 1.0-1.9, to be compared with 4.1 when processing the whole data set.  
We conclude that in the case of GJ~410, the main variability observed in RV data likely comes from short-lived spots, inducing an evolution in the shape of the RV curve on a timescale of only ~2 rotation cycles.\\

 \begin{table*}
  \begin{tabular}{cccc||ccc||ccc|ccc}
\cline{2-13}
   & \multicolumn{3}{c||}{  Raw RV} &  \multicolumn{3}{c||}{Multiple } & \multicolumn{5}{c}{  DI modeling} \\
   & \multicolumn{3}{c||}{ data} &  \multicolumn{3}{c||}{ sine-fit} & \multicolumn{5}{c}{} \\
   & \multicolumn{3}{c||}{ } &  \multicolumn{3}{c||}{ (fund. + 1 or 2 harm.) } & \multicolumn{3}{c|}{RV curves} & \multicolumn{2}{c}{LSD profiles}\\
\cline{2-13}
   & $\sigma_0$ & rms$_0$ &$\chi^2_{r,0}$  & rms$_1$    & $\chi^2_{r,1}$ & FAP  &  rms$_2$    & $\chi^2_{r,2}$ & FAP & $\chi^2_{r,i}$ & $\chi^2_{r,f}$ & FAP \\             
   &  (\ms)  & (\ms)  && (\ms)  & &(\%) & (\ms)    &&(\%)  & & & (\%)\\             
\hline
\multicolumn{1}{c|}{epoch \#1} & 3.43  & 9.64 & 8.40 & 3.43 & 1.0 & $\leqslant$ 0.01 & 4.67  & 1.92 & $\leqslant$ 0.01  & 2.8 & 1.7 & $\leqslant$ 0.01\\
\multicolumn{1}{c|}{epoch \#2} &  3.22 & 7.14 & 6.29 & 3.83 & 1.03 & $\leqslant$ 0.01 & 3.95 & 1.66 & $\leqslant$ 0.01  &2.5 & 1.6 &$\leqslant$ 0.01\\
\multicolumn{1}{c|}{epoch \#3} &  3.06 & 5.94 & 3.91 & 3.06 & 1.0 & $\leqslant$ 0.01  &3.19 & 1.02 & 17 & 2.2 & 1.2&$\leqslant$ 0.01\\
\multicolumn{1}{c|}{full set} &  3.28 & 8.84 &  7.85 & 6.55 & 3.96 & 0.04 & 6.78 & 4.14 & 2.8& 2.9 & 2.0 & $\leqslant$ 0.01\\
\hline
\end{tabular}
\caption{ Same as Table~\ref{tab:fitvr2} for the three observation epochs of GJ~410. For the observation epochs (1) and (3) a single harmonics is sufficient to reach residual RVs lower than 3~\ms. The RV jitter can by entirely modeled with rotational modulation (rms$_{J, r} \sim$ 0\ms, et rms$_{J, tot} \sim$ rms$_{J, m}$), and we choose $\chi^2_{r,1}$ = 1.0 and  rms$_1$ = $\sigma_0$. 
}
\label{tab:fitvr410}
\end{table*}

\subsubsection{GJ 205} 
For GJ~205, the amplitude of the $RI$ spectra is low ($\leqslant$~0.05\%, see Fig.~\ref{fig:res479205}). The DI reconstruction leads to an equivalent spotted area $\epsilon$ of $\sim$0.9\% that allows to decrease the $\chi^2_r$ fit to the profiles from 4.3 to 3.8 only. The reconstructed features exhibit faint spot clusters, located at hight and mid-latitude, however, this reconstruction is not reliable given the FAP of 75\% associated with the Stokes $I$ LSD fit.
The RV jitter is not efficiently filtered (FAP~$\sim$~98\%). These results validate that there is likely no signals at \pstar\ and that DR might strongly affect the dark spot location at the surface of GJ~205 and thus the RV activity jitter of the star. Further work, taking explicitly into accound differential rotation, is thus needed for this star. This would require in particular a high quality spectropolarimetric data set from which differential rotation can be reliably estimated.\\

\subsubsection{Discussion}
\label{sec:discussion}
The efficiency of the RV filtering depends on the relative importance of the rotationally modulated RV component with respect to the random component. The importance of each component is reminded in Table~\ref{tab:jitterf1}.

For the lowest mass star of this sample, GJ~358, the rotational modulated component J$_{\rm m}$ of the RV jitter have been divided by 2.8 (and J$_{\rm tot}$ by 2.2).
For the earliest M-dwarfs (GJ~205 and GJ~410), neither the DI modeling nor the multiple sine fit succeed at obtaining a decent match to the observed RV jitter (high FAP), because of a higher level of intrinsic variability of the RV curve. 
In the particular case of GJ410, we observe that this higher level of intrinsic variability is directly related to the short spot lifetimes (1-2 rotation cycles), as evidenced by the significant improvement in the efficiency of the DI filtering when considering shorter time intervals (see Table~\ref{tab:jitterf1}).
Contrary to a simple multiple sine fit, the use of the imaging techniques allows one to (i) to better constrain the origin of the activity jitter (dark spots and rotational modulation, DR or short spot lifetime), and (ii) to obtain a self-consistent physically-motivated, though still simple, description of the activity jitter rather than to perform a blind filtering of the RV data.
\begin{table}
\begin{center}
  \begin{tabular}{c|ccc|c|c|c|c}
   & J$_{\rm tot}$ & J$_{\rm m}$ & J$_{\rm r}$  & J$_{\rm m}$ with DI & A$_1$ &  A$_2$ & A$_3$\\   
 \cline{2-5}
 \multicolumn{6}{c}{~~~~~(\ms)} \\          
\hline
GJ~205 &  3.21 & 1.75 &  2.68  & 1.53 & 1.2& 1.1& 2.1\\
\hline
GJ~410 &  8.21 & 5.67 &  5.93 & 5.82 & 1.4& 1.3& 3.1 \\
epoch \#1 &  9.01 & 9.01 & -  & 8.43 & -&3.8 & 3.9\\
epoch \#2 &  6.37 & 6.02 &  2.07 & 5.95 & 3.1 &2.8& 6.6\\
epoch \#3 &  5.10 & 5.10& -  & 5.04 & -&5.2&6.5\\
\hline
GJ~479 &  4.89 & 3.83 & 3.04& 3.54 & 1.7 & 1.5&2.6\\
\hline
GJ~358 &  4.31 & 4.10 &  1.33& 3.83 & 3.2 & 2.2 & 2.8\\
\hline
\end{tabular}
\caption{
Same as Table~\ref{tab:jmjr}, with three additional columns: column 5 gives J$_{\rm m}$, the rms of the RV data modelled using the DI imaging, column 7 gives A$_2$, quantifying how we can reduce the activity jitter thanks to DI imaging, and column 8 gives A$_3$ = $\hbox{J}_{\rm m}$/ $\sqrt{ \hbox{J}^{2}_{\rm m} - \hbox{J}^{2}_{\rm {m,DI}}}$, denoting the factor of decrease of the J$_m$ component. The dash indicates that data can be reproduced down to the noise level, \emph{i.e.}, that the RV variations are due to rotational modulations only.
}
\label{tab:jitterf1}
\end{center}
\end{table}

Our model is based on the assumption that the dominant contribution to the total RV signal in the M dwarfs should be the effect of dark spots. This assumption mainly relies on Sun-like stars studies, and on the low temperature of M dwarfs. However, we have to note that the current DI model does not yet allow us to faithfully reproduce the full amplitude of FWHM of the four studied stars. The phase of the variations are fitted, but the peak-to-peak amplitude is always underestimated in each case. 
This caveat may reflect inadequate assumptions/approximations in our modeling and will be further explored in forthcoming papers. 
The next step will be to add more physical realism in our model (\emph{e.g.,} use a more realistic line profile $I_{\rm s}$ to characterize the spotted regions) to improve the modeling of the effects of the activity jitter in M dwarfs.


\section{Summary and perspectives}
\label{sec:dis}
\begin{figure}
\includegraphics[scale=0.35,angle=0]{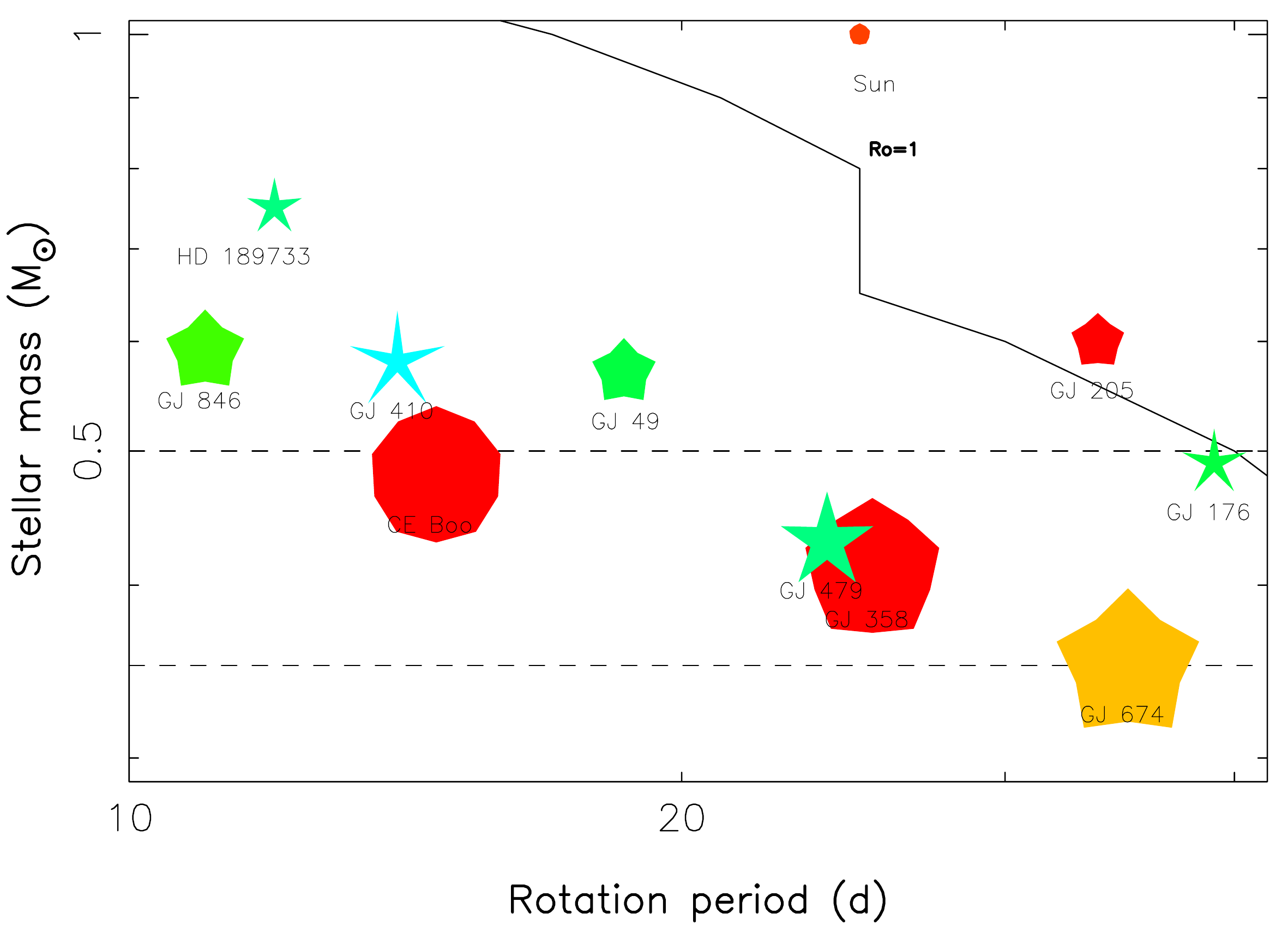} 
\caption{Properties of the magnetic topologies of our sample of five M dwarfs as a function of rotation period and stellar mass. Larger symbols indicate larger magnetic fields while symbol shapes depict the different degrees of axisymmetry of the reconstructed magnetic field (from decagons for purely axisymmetric fields to sharp stars for purely non-axisymmetric fields). Colors illustrate the field configuration (dark blue for purely toroidal fields, dark red for purely poloidal fields and intermediate colours for intermediate configurations). The solid line represents the contour of constant Rossby number $R_o$ = 1. The dotted line correspond to the 0.5 and 0.35 \msun\ thresholds. The Sun, GJ~49 and CE Boo \citep{Donati08c}, HD~189733 \citep{Fares10} and GJ~674 \& GJ~176 presented in detail in a forthcoming paper are shown for comparison.
}
\label{fig:config_magn}
\end{figure}

The magnetic analysis gives access to the large-scale magnetic field map of the observed weakly-active M dwarfs, as well as to a reliable and accurate estimate of \pstar.
Fig.~\ref{fig:config_magn} summarizes the magnetic properties of our sample. These data allow us to add new observations in the \mstar\  - \pstar\ diagram, covering a mostly unexplored domain so far. 
The magnetic fields detected for the early-M dwarfs exhibit strengths of a few tens of G, and are lower by a factor 5 than those of more active and rapidly rotating mid-M dwarfs \citep{Morin08b}. 
We note that for the stars with a stellar mass larger than 0.5~\msun, the toroidal component is significant, except for GJ~205 whose large-scale magnetic field is dominated by a poloidal component. 
GJ~205 is the only observed star with a Rossby number $R_o$\footnote{$R_o$ is defined as $R_o=\pstar/\tau_c$, where $\tau_c$ is the convective turnover time derived by \citet{Kiraga07} from the rotation-activity relation in X-rays} higher than 1 (as the Sun). 
This is in agreement with the trends previously reported in \cite{Donati09}, where stars with $R_o$~>~1 tend to exhibit weak poloidal fields mostly aligned with the rotation axis.

For \mstar~<~0.5~\msun, the large-scale magnetic properties are diverse, with some stars featuring mainly poloidal and axisymmetric fields (GJ~358, GJ~674) and some others exhibiting more complex topologies (GJ~479, GJ~176). In particular, we note that 2 stars of our sample feature different types of fields while sharing the same location in the \mstar\ vs \pstar\ plane.
This is reminiscent of the bi-stable behavior of dynamo processes, as previously pointed out by, \emph{e.g.}, \cite{Morin11} in the case of active very low-mass dwarfs.
The theoretical models \citep[e.g.,][]{Gastine13} foresee a bistability around $R_o$~=~0.1, with a transition between fields with a simple dipolar topology ($R_o$~<~0.1) and fields with a complex topology ($R_o$~>~1). 
Our observations suggest that dynamo bi-stability may indeed be present at different places of the \mstar\ vs \pstar\ diagram than previously identified by \cite{Morin11} and whose relation with theoretical predictions is yet to be checked in more details.  
More spectropolarimetric observations of M dwarfs in this range of mass and rotation periods are necessary to investigate this result in more details. \\

To find an Earth-like planet (in terms of size, mass and effective stellar flux) thanks to the RV method, moderately active M dwarfs appear to be natural targets with their reduced effective temperature and their low-mass. However we still need to model and filter out the RV activity jitter to reveal these plausible low-mass planets RV signatures. 
To characterize the activity jitter of cool low-mass stars, we used the studies done for Sun-like stars \citep[\emph{e.g.,}][]{Dumusque14}, taking into account their reduced photospheric temperature.
We then assumed the dominant contributor to the activity-modulated RV signal that plagues RV data is the rotational modulation caused by dark spot at the stellar surface (in agreement with theoretical studies as \citealt{Beeck15}).
With this hypothesis, we were able to develop a technique, based on a tomographic imaging (ZDI), to model the spot distribution at the surface of the four weakly active early M dwarfs we observed.

The sampled stars being slow rotators (\vsini~$\leqslant$~2~\kms), the observed spectral line width reflects directly intrinsic profiles rather than the Doppler broadening. To overcome this issue, we adapted the Doppler Imaging technique to reconstruct the profile residuals instead of the observed profiles themselves (see Sec.~\ref{sec:di}).
Thanks to this approach, we are not dependent anymore on our ability at achieving a detailed description of the local profile, and the code is focussed on the profile distortion modeling only.
Besides, this method relies on the knowledge of the rotational period \pstar, parameter previously estimated from the magnetic analysis.

The novel imaging method we devised is found to be reasonably successful at reconstructing the spot distribution at the surface of the early-type slowly-rotating stars that we studied. From this map and its associated set of spectra, we model the RV activity jitter whose period is commensurate to \pstar, \emph{i.e.}, J$_{\rm m}$ component only. For our early M dwarf sample, we found that the spots cover up to 2\% of the total stellar surface (in agreement with previous estimates for rapid rotators, see, \emph{e.g.}, \citealt{Morin08a}). The rotationally modulated RV component deduced from the brightness maps allows to reduce the observed RV jitter by a factor of 2-3, and the observed rotationally modulated component by a factor 3-6. 
The efficiency increases with decreasing stellar mass.
For the earliest M dwarfs, we speculate that the high level of intrinsic variability likely caused by short-lived spots having lifetimes of <~2 rotation cycles limits the efficiency of the modeling. A modeling of such effect is possible but requests specific observational strategy, with a high cadency sampling during more than 3 rotational cycles.
Although relying on a simple assumption, this method gives promising results, and allows us to get a better insight on the origin of the activity RV jitter of early M dwarfs. 
This study of slowly-rotating early-M dwarfs complements the analysis already done for late M dwarfs \citep[\emph{e.g.,}][]{Barnes11, Barnes14, Barnes15}.

\begin{figure*}
\begin{center}
\includegraphics[scale = 0.4]{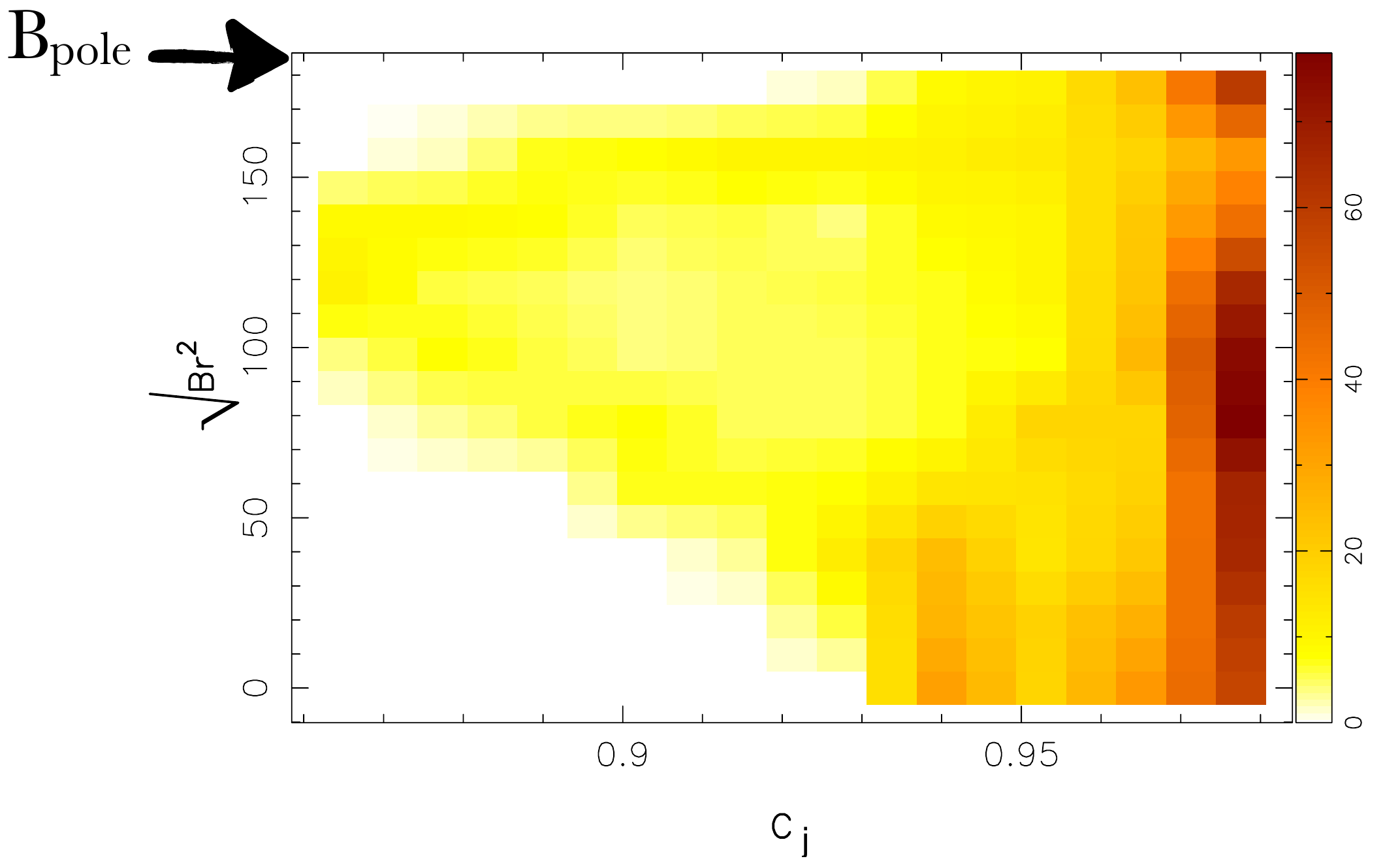}
\includegraphics[scale = 0.4]{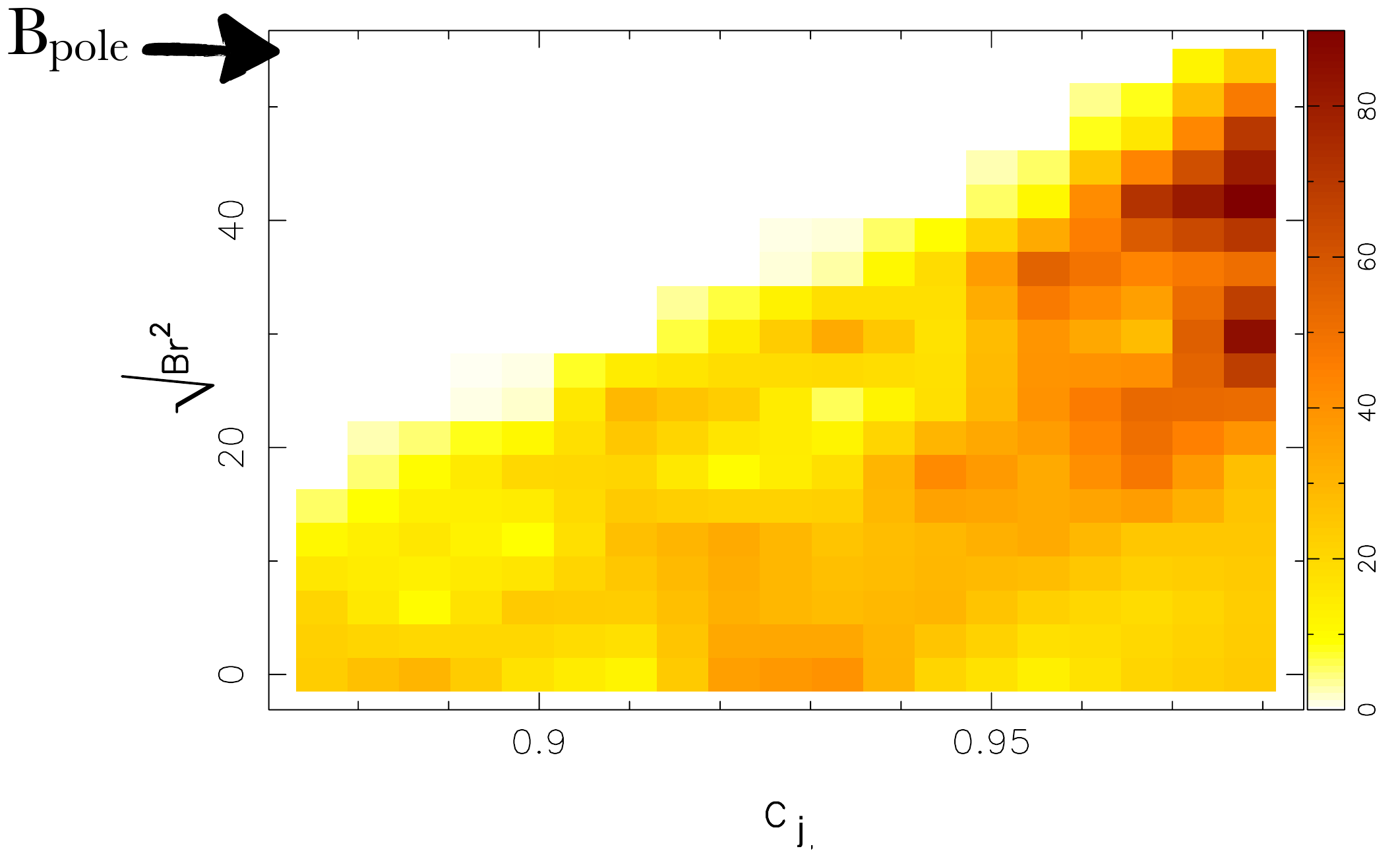}
\includegraphics[scale = 0.4]{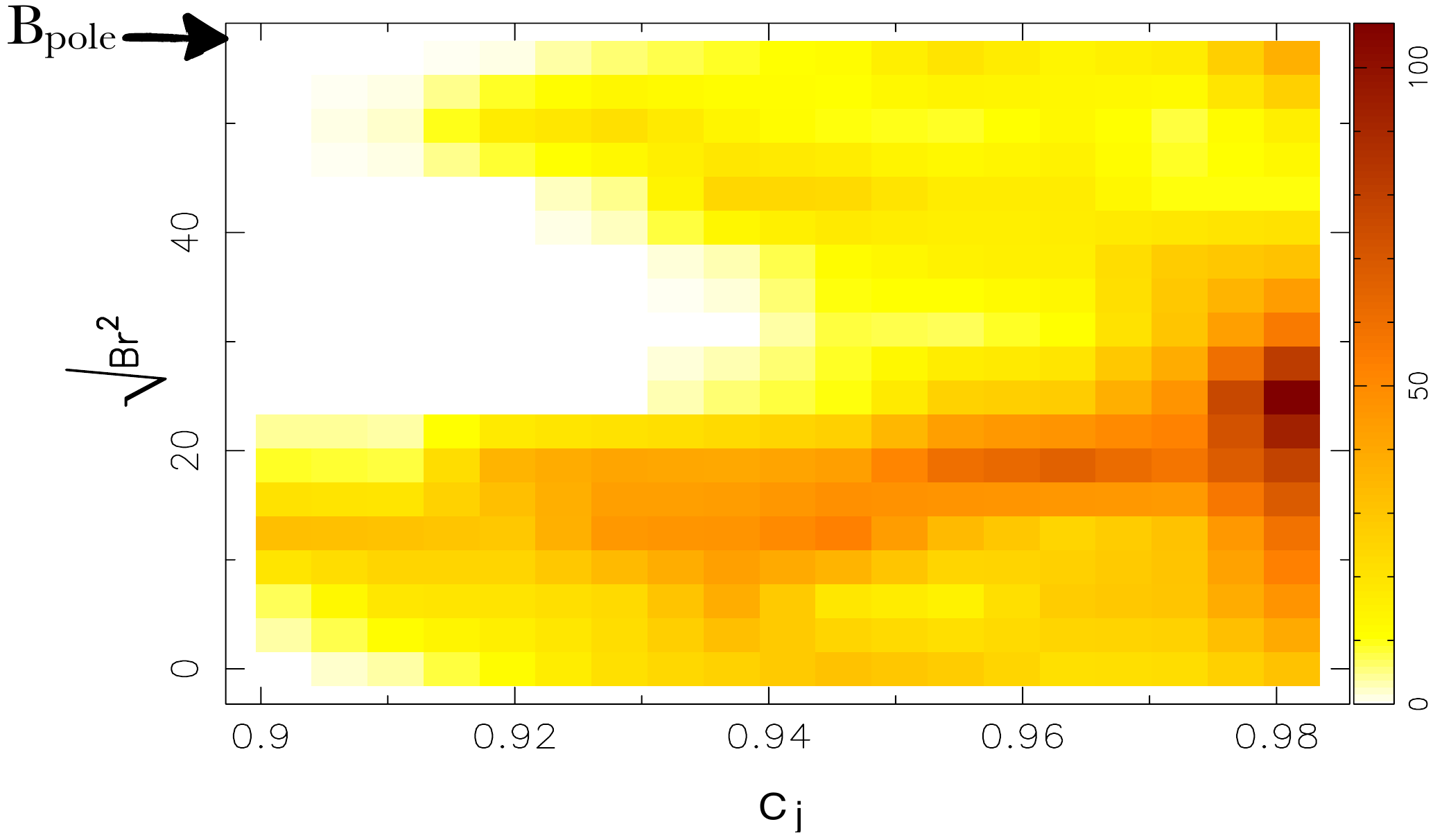}
\caption{
The x-axis depicts the $C_j$, denoting the proportion of photosphere inside each cell, and the y-axis the absolute value of the radial component of the large-scale magnetic field. The colorscale represents the number of cells affected by both a radial magnetic field and a dark spot.
The arrow indicates the absolute value of the radial polar magnetic field of the star. 
\textit{Top left} :  GJ~358
\textit{Top right} :  GJ~479
\textit{Bottom} :  GJ~410.
}
\label{fig:brcq}
\end{center}
\end{figure*}

To investigate if spot distributions relate to large-scale magnetic field topologies, we compare both the brightness and magnetic filed maps that we obtained (see  Fig.~\ref{fig:brcq}). It seems that the darkest spots concentrate either close to the magnetic poles (GJ~358 whose magnetic field is mainly poloidal), or to the magnetic equator (GJ~410 and GJ~479 whose magnetic field is significantly toroidal) - we exclude GJ~205 in this study, given the weak reliability of the spot distribution map. 
This suggests that the large-scale magnetic field may indeed be controlling where surface spots tend to preferentially appear at the stellar surface, as it does in the particular case of the Sun. 
This tendency needs to be examined in a larger sample with new spectropolarimetric surveys of moderately active M dwarfs.
Besides, given that surface spot distributions are significantly impacting our ability at detecting Earth-like planets \citep[see, \emph{e.g.,} simulations done by][]{Andersen15}, it will be worthwhile to extent this study to later-type M dwarfs to improve our knowledge of their spot patterns. For example, the Doppler Imaging is a powerful tool to investigate whether spot patterns change and, \emph{e.g.,} evolve towards more even distributions of smaller features, when going from partially to fully convective stars.

Further improvements can to be implemented to obtain a more accurate filtering of the RV curves of M dwarfs (\emph{e.g.}, by incorporating the temporal evolution of spots within the imaging process).
Moreover, adapting our method to stars of other spectral types (G and K), for which the activity jitter is no longer dominated by the spot brightness contrast  but by plages and the suppression of convective blueshift (\emph{e.g.}, \citealt{Haywood14}), is another obvious avenue worth exploring. 
Finally, a complementary study is in preparation to present the performances of this technique for M-dwarfs hosting a planet.

To disentangle stellar from planetary signals, a powerful analysis  should be to carry out observations at both optical and IR wavelengths, particularly for M dwarfs emitting a large fraction of their flux in the IR. Several studies showed that the RV jitter will be divided by at least a factor of 2 due to the lower contrast between the dark spot regions and the quiet photosphere (\citet{Marchwinski15} \& \citet{Reiners10, Rodler11}, respectively for Sun-like stars \& late M dwarfs). 
In this context, the new generation of high resolution/precision velocimeters working in the nIR domain (\emph{e.g.,} CARMENES\footnote{Calar Alto high-Resolution search for M dwarfs with Exoearths with Near-infrared and optical Echelle Spectrographs, a high-precision velocimeter at the 3.5m telescope at the Calar Alto Observatory.}, SPIRou\footnote{ SpectroPolarimeter in the near InfraRed, a spectropolarimeter/high-precision velocimeter for the 3.6m Canada-France-Hawaii Telescope. It will operate at near-infrared wavelengths (first light in 2017).}, CRIRES+\footnote{upgrade of the CRyogenic InfraRed Echelle Spectrograph at ESO/VLT (first light in 2018).}) present a tremendous interest. However, characterizing, modeling and filtering out the RV activity jitter of M dwarfs remain mandatory steps for all future velocimetric studies aiming at detecting small Earth-mass rocky planets. They allow to define the best adapted observational strategies, taking into account the specificities of the M dwarf activity that hampers RV measurements.
Moreover, while the brightness contrast decreases in the IR, the impact of small scale-magnetic field on RVs strengthens through Zeeman effect.
Therefore, the method we presented will be particularly adapted for SPIRou, which will be both a high-precision velocimeter and a spectropolarimeter. Spectropolarimetric surveys in nIR will give new options for filtering RV curves from the activity jitter using tomographic techniques like ZDI, and will efficiently further enhance the sensitivity to low-mass planets, as well as to the magnetic stellar activity RV signal itself.

\section*{Acknowledgements}
This paper is based on observations obtained at the 3.6m telescope in La Silla, Chile, operated by ESO, the European Southern Observatory. We thank the La Silla staff for their valuable help throughout our observing runs and F.Bouchy for his help in the coordination of observing runs. We also thank the TBL staff for their help during data collection of GJ~410, GJ~205 and GJ~846. Finally we thank X.Bonfils for his M mask that we used to compute RVs. JM was sponsored by a postdoctoral fellowship of the Alexander von Humboldt foundation in G\"ottingen.
The authors acknowledge the PNPS/INSU (Programme national de Physique Stellaire) to fund this project. XD acknowledges funding from the LabEx OSUG@2020. 
Finally thank you yo the anonymous referee for the useful comments.



\bibliographystyle{mnras}
\bibliography{allbiblio}

\begin{thebibliography}{}
\makeatletter
\relax
\def\mn@urlcharsother{\let\do\@makeother \do\$\do\&\do\#\do\^\do\_\do\%\do\~}
\def\mn@doi{\begingroup\mn@urlcharsother \@ifnextchar [ {\mn@doi@}
  {\mn@doi@[]}}
\def\mn@doi@[#1]#2{\def\@tempa{#1}\ifx\@tempa\@empty \href
  {http://dx.doi.org/#2} {doi:#2}\else \href {http://dx.doi.org/#2} {#1}\fi
  \endgroup}
\def\mn@eprint#1#2{\mn@eprint@#1:#2::\@nil}
\def\mn@eprint@arXiv#1{\href {http://arxiv.org/abs/#1} {{\tt arXiv:#1}}}
\def\mn@eprint@dblp#1{\href {http://dblp.uni-trier.de/rec/bibtex/#1.xml}
  {dblp:#1}}
\def\mn@eprint@#1:#2:#3:#4\@nil{\def\@tempa {#1}\def\@tempb {#2}\def\@tempc
  {#3}\ifx \@tempc \@empty \let \@tempc \@tempb \let \@tempb \@tempa \fi \ifx
  \@tempb \@empty \def\@tempb {arXiv}\fi \@ifundefined
  {mn@eprint@\@tempb}{\@tempb:\@tempc}{\expandafter \expandafter \csname
  mn@eprint@\@tempb\endcsname \expandafter{\@tempc}}}

\bibitem[\protect\citeauthoryear{{Aigrain}, {Pont}  \& {Zucker}}{{Aigrain}
  et~al.}{2012}]{Aigrain12}
{Aigrain} S.,  {Pont} F.,   {Zucker} S.,  2012, \mnras, 419

\bibitem[\protect\citeauthoryear{{Andersen} \& {Korhonen}}{{Andersen} \&
  {Korhonen}}{2015}]{Andersen15}
{Andersen} J.~M.,  {Korhonen} H.,  2015, \mn@doi [\mnras]
  {10.1093/mnras/stu2731}, \href
  {http://cdsads.u-strasbg.fr/abs/2015MNRAS.448.3053A} {448, 3053}

\bibitem[\protect\citeauthoryear{{Baraffe}, {Homeier}, {Allard}  \&
  {Chabrier}}{{Baraffe} et~al.}{2015}]{Baraffe15}
{Baraffe} I.,  {Homeier} D.,  {Allard} F.,   {Chabrier} G.,  2015, \mn@doi
  [\aap] {10.1051/0004-6361/201425481}, \href
  {http://cdsads.u-strasbg.fr/abs/2015A\%26A...577A..42B} {577, A42}

\bibitem[\protect\citeauthoryear{{Barnes}, {Jeffers}  \& {Jones}}{{Barnes}
  et~al.}{2011}]{Barnes11}
{Barnes} J.,  {Jeffers} S.,   {Jones} H.,  2011, \mn@doi [\mnras]
  {10.1111/j.1365-2966.2010.17979.x}, 412, 1599

\bibitem[\protect\citeauthoryear{{Barnes} et~al.,}{{Barnes}
  et~al.}{2014}]{Barnes14}
{Barnes} J.~R.,  et~al., 2014, \mn@doi [\mnras] {10.1093/mnras/stu172}, \href
  {http://cdsads.u-strasbg.fr/abs/2014MNRAS.439.3094B} {439, 3094}

\bibitem[\protect\citeauthoryear{{Barnes}, {Jeffers}, {Jones}, {Pavlenko},
  {Jenkins}, {Haswell}  \& {Lohr}}{{Barnes} et~al.}{2015}]{Barnes15}
{Barnes} J.~R.,  {Jeffers} S.~V.,  {Jones} H.~R.~A.,  {Pavlenko} Y.~V.,
  {Jenkins} J.~S.,  {Haswell} C.~A.,   {Lohr} M.~E.,  2015, \mn@doi [\apj]
  {10.1088/0004-637X/812/1/42}, \href
  {http://adsabs.harvard.edu/abs/2015ApJ...812...42B} {812, 42}

\bibitem[\protect\citeauthoryear{{Beeck}, {Sch{\"u}ssler}, {Cameron}  \&
  {Reiners}}{{Beeck} et~al.}{2015}]{Beeck15}
{Beeck} B.,  {Sch{\"u}ssler} M.,  {Cameron} R.~H.,   {Reiners} A.,  2015,
  \mn@doi [\aap] {10.1051/0004-6361/201525788}, \href
  {http://cdsads.u-strasbg.fr/abs/2015A%26A...581A..42B} {581, A42}

\bibitem[\protect\citeauthoryear{{Boisse} et~al.,}{{Boisse}
  et~al.}{2009}]{Boisse09}
{Boisse} I.,  et~al., 2009, \mn@doi [\aap] {10.1051/0004-6361:200810648}, \href
  {http://cdsads.u-strasbg.fr/abs/2009A\%26A...495..959B} {495, 959}

\bibitem[\protect\citeauthoryear{{Boisse}, {Bouchy}, {H{\'e}brard}, {Bonfils},
  {Santos}  \& {Vauclair}}{{Boisse} et~al.}{2011}]{Boisse11}
{Boisse} I.,  {Bouchy} F.,  {H{\'e}brard} G.,  {Bonfils} X.,  {Santos} N.,
  {Vauclair} S.,  2011, \aap, 528

\bibitem[\protect\citeauthoryear{{Bonfils} et~al.,}{{Bonfils}
  et~al.}{2007}]{Bonfils07}
{Bonfils} X.,  et~al., 2007, \mn@doi [\aap] {10.1051/0004-6361:20077068}, 474,
  293

\bibitem[\protect\citeauthoryear{{Bonfils}, {Gillon}, {Udry}, {Armstrong}  \&
  {Bouchy}}{{Bonfils} et~al.}{2012}]{Bonfils12}
{Bonfils} X.,  {Gillon} M.,  {Udry} S.,  {Armstrong} D.,   {Bouchy} F.,  2012,
  \aap, 546, A27

\bibitem[\protect\citeauthoryear{{Bonfils}, {Delfosse}, {Udry}, {Forveille},
  {Mayor}, {Perrier}  \& {Bouchy}}{{Bonfils} et~al.}{2013}]{Bonfils13}
{Bonfils} X.,  {Delfosse} X.,  {Udry} S.,  {Forveille} T.,  {Mayor} M.,
  {Perrier} C.,   {Bouchy} F.,  2013, \aap, 549, A109

\bibitem[\protect\citeauthoryear{{Borgniet}, {Meunier}  \&
  {Lagrange}}{{Borgniet} et~al.}{2015}]{Borgniet15}
{Borgniet} S.,  {Meunier} N.,   {Lagrange} A.-M.,  2015, preprint, \href
  {http://cdsads.u-strasbg.fr/abs/2015arXiv150507361B} {} (\mn@eprint {arXiv}
  {1505.07361})

\bibitem[\protect\citeauthoryear{{Brown}, {Donati}, {Rees}  \& {Semel}}{{Brown}
  et~al.}{1991}]{Brown91}
{Brown} S.~F.,  {Donati} J.-F.,  {Rees} D.~E.,   {Semel} M.,  1991, \aap, \href
  {http://adsabs.harvard.edu/abs/1991A\%26A...250..463B} {250, 463}

\bibitem[\protect\citeauthoryear{{Collier Cameron}}{{Collier
  Cameron}}{1992}]{Cameron92}
{Collier Cameron} A.,  1992, in {Byrne} P.~B.,  {Mullan} D.~J.,  eds,  Lecture
  Notes in Physics, Berlin Springer Verlag Vol. 397, Surface Inhomogeneities on
  Late-Type Stars. p.~33, \mn@doi{10.1007/3-540-55310-X_131}

\bibitem[\protect\citeauthoryear{{Delfosse}, {Forveille}, {S{\'e}gransan},
  {Beuzit}, {Udry}, {Perrier}  \& {Mayor}}{{Delfosse}
  et~al.}{2000}]{Delfosse00}
{Delfosse} X.,  {Forveille} T.,  {S{\'e}gransan} D.,  {Beuzit} J.-L.,  {Udry}
  S.,  {Perrier} C.,   {Mayor} M.,  2000, \aap, 364, 217

\bibitem[\protect\citeauthoryear{{Desort}, {Lagrange}, {Galland}, {Udry}  \&
  {Mayor}}{{Desort} et~al.}{2007}]{Desort07}
{Desort} M.,  {Lagrange} A.,  {Galland} F.,  {Udry} S.,   {Mayor} M.,  2007,
  \aap, 473, 983

\bibitem[\protect\citeauthoryear{{Donati}}{{Donati}}{2001}]{Donati01}
{Donati} J.-F.,  2001, in {Boffin} H.~M.~J.,  {Steeghs} D.,   {Cuypers} J.,
  eds,  Lecture Notes in Physics, Berlin Springer Verlag Vol. 573,
  Astrotomography, Indirect Imaging Methods in Observational Astronomy. p.~207

\bibitem[\protect\citeauthoryear{{Donati}}{{Donati}}{2003}]{Donati03}
{Donati} J.-F.,  2003. p.~41

\bibitem[\protect\citeauthoryear{{Donati} \& {Landstreet}}{{Donati} \&
  {Landstreet}}{2009}]{Donati09}
{Donati} J.,  {Landstreet} J.~D.,  2009, \mn@doi [\araa]
  {10.1146/annurev-astro-082708-101833}, \href
  {http://adsabs.harvard.edu/abs/2009ARA%26A..47..333D} {47, 333}

\bibitem[\protect\citeauthoryear{{Donati}, {Semel}, {Carter}, {Rees}  \&
  {Collier Cameron}}{{Donati} et~al.}{1997}]{Donati97}
{Donati} J.-F.,  {Semel} M.,  {Carter} B.~D.,  {Rees} D.~E.,   {Collier
  Cameron} A.,  1997, \mnras, \href
  {http://cdsads.u-strasbg.fr/abs/1997MNRAS.291..658D} {291, 658}

\bibitem[\protect\citeauthoryear{{Donati}, {Collier Cameron}  \&
  {Petit}}{{Donati} et~al.}{2003}]{Donati03b}
{Donati} J.-F.,  {Collier Cameron} A.,   {Petit} P.,  2003, \mnras, 345, 1187

\bibitem[\protect\citeauthoryear{{Donati}, {Forveille}, {Cameron}, {Barnes},
  {Delfosse}, {Jardine}  \& {Valenti}}{{Donati} et~al.}{2006a}]{Donati06a}
{Donati} J.-F.,  {Forveille} T.,  {Cameron} A.~C.,  {Barnes} J.~R.,  {Delfosse}
  X.,  {Jardine} M.~M.,   {Valenti} J.~A.,  2006a, \mn@doi [Science]
  {10.1126/science.1121102}, \href
  {http://adsabs.harvard.edu/abs/2006Sci...311..633D} {311, 633}

\bibitem[\protect\citeauthoryear{{Donati} et~al.,}{{Donati}
  et~al.}{2006b}]{Donati06b}
{Donati} J.-F.,  et~al., 2006b, \mn@doi [\mnras]
  {10.1111/j.1365-2966.2006.10558.x}, \href
  {http://adsabs.harvard.edu/abs/2006MNRAS.370..629D} {370, 629}

\bibitem[\protect\citeauthoryear{{Donati} et~al.,}{{Donati}
  et~al.}{2008}]{Donati08c}
{Donati} J.-F.,  et~al., 2008, \mn@doi [\mnras]
  {10.1111/j.1365-2966.2008.13799.x}, \href
  {http://adsabs.harvard.edu/abs/2008MNRAS.390..545D} {390, 545}

\bibitem[\protect\citeauthoryear{{Dressing} \& {Charbonneau}}{{Dressing} \&
  {Charbonneau}}{2015}]{Dressing15}
{Dressing} C.~D.,  {Charbonneau} D.,  2015, \mn@doi [\apj]
  {10.1088/0004-637X/807/1/45}, \href
  {http://cdsads.u-strasbg.fr/abs/2015ApJ...807...45D} {807, 45}

\bibitem[\protect\citeauthoryear{{Dumusque}, {Boisse}  \& {Santos}}{{Dumusque}
  et~al.}{2014}]{Dumusque14}
{Dumusque} X.,  {Boisse} I.,   {Santos} N.,  2014, \mn@doi [\apj]
  {10.1088/0004-637X/796/2/132}, 796

\bibitem[\protect\citeauthoryear{{Fares} et~al.,}{{Fares}
  et~al.}{2010}]{Fares10}
{Fares} R.,  et~al., 2010, \mn@doi [\mnras] {10.1111/j.1365-2966.2010.16715.x},
  \href {http://cdsads.u-strasbg.fr/abs/2010MNRAS.406..409F} {406, 409}

\bibitem[\protect\citeauthoryear{{Forveille} et~al.,}{{Forveille}
  et~al.}{2009}]{Forveille09}
{Forveille} T.,  et~al., 2009, \mn@doi [\aap] {10.1051/0004-6361:200810557},
  493, 645

\bibitem[\protect\citeauthoryear{{Gastine}, {Morin}, {Duarte}, {Reiners},
  {Christensen}  \& {Wicht}}{{Gastine} et~al.}{2013}]{Gastine13}
{Gastine} T.,  {Morin} J.,  {Duarte} L.,  {Reiners} A.,  {Christensen} U.~R.,
  {Wicht} J.,  2013, \mn@doi [\aap] {10.1051/0004-6361/201220317}, \href
  {http://cdsads.u-strasbg.fr/abs/2013A%26A...549L...5G} {549, L5}

\bibitem[\protect\citeauthoryear{{Gomes Da Silva}, {Santos}, {Bonfils}  \&
  {Delfosse}}{{Gomes Da Silva} et~al.}{2011}]{DaSilva11}
{Gomes Da Silva} J.,  {Santos} N.,  {Bonfils} X.,   {Delfosse} X.,  2011, \aap,
  534, A30

\bibitem[\protect\citeauthoryear{{Gray}}{{Gray}}{1982}]{Gray82}
{Gray} D.~F.,  1982, \mn@doi [\apj] {10.1086/159818}, \href
  {http://adsabs.harvard.edu/abs/2010MNRAS.407.2269M} {255, 200}

\bibitem[\protect\citeauthoryear{{Haywood}, {Collier-Cameron}  \&
  {Quloz}}{{Haywood} et~al.}{2014}]{Haywood14}
{Haywood} R.,  {Collier-Cameron} A.,   {Quloz} D.,  2014, \mnras, 443

\bibitem[\protect\citeauthoryear{{H\'ebrard}, {Donati}, {Delfosse}, {Morin}  \&
  {Boisse}}{{H\'ebrard} et~al.}{2014}]{Hebrard14}
{H\'ebrard} E.,  {Donati} J.,  {Delfosse} X.,  {Morin} J.,   {Boisse} I.,
  2014, \mnras, 443, 2899

\bibitem[\protect\citeauthoryear{{Kasting}, {Kopparapu}, {Ramirez}  \&
  {Harman}}{{Kasting} et~al.}{2014}]{Kasting14}
{Kasting} J.~F.,  {Kopparapu} R.,  {Ramirez} R.~M.,   {Harman} C.~E.,  2014,
  \mn@doi [Proceedings of the National Academy of Science]
  {10.1073/pnas.1309107110}, \href
  {http://cdsads.u-strasbg.fr/abs/2014PNAS..11112641K} {111, 12641}

\bibitem[\protect\citeauthoryear{{Kiraga} \& {Stepien}}{{Kiraga} \&
  {Stepien}}{2007}]{Kiraga07}
{Kiraga} M.,  {Stepien} K.,  2007, Acta Astronomica, \href
  {http://adsabs.harvard.edu/abs/2007AcA....57..149K} {57, 149}

\bibitem[\protect\citeauthoryear{{Koen}, {Kilkenny}, {van Wyk}  \&
  {Marang}}{{Koen} et~al.}{2010}]{Koen10}
{Koen} C.,  {Kilkenny} F.,  {van Wyk} F.,   {Marang} F.,  2010, \mnras, 403,
  1949

\bibitem[\protect\citeauthoryear{{Kurucz}}{{Kurucz}}{1993}]{Kurucz93}
{Kurucz} R.,  1993, CDROM \#~13 (ATLAS9 atmospheric models) and \#~18 (ATLAS9
  and SYNTHE routines, spectral line database).
Smithsonian Astrophysical Observatory, Washington D.C.

\bibitem[\protect\citeauthoryear{{Leggett}, {Allard}, {Geballe}, {Hauschild}
  \& {Schweitzer}}{{Leggett} et~al.}{2001}]{Leggett01}
{Leggett} S.,  {Allard} F.,  {Geballe} T.,  {Hauschild} P.,   {Schweitzer} A.,
  2001, \apj, 548, 908

\bibitem[\protect\citeauthoryear{{Lomb}}{{Lomb}}{1976}]{Lomb76}
{Lomb} N.~R.,  1976, \mn@doi [\apss] {10.1007/BF00648343}, \href
  {http://cdsads.u-strasbg.fr/abs/1976Ap\%26SS..39..447L} {39, 447}

\bibitem[\protect\citeauthoryear{{Marchwinski}, {Mahadevan}, {Robertson},
  {Ramsey}  \& {Harder}}{{Marchwinski} et~al.}{2015}]{Marchwinski15}
{Marchwinski} R.~C.,  {Mahadevan} S.,  {Robertson} P.,  {Ramsey} L.,   {Harder}
  J.,  2015, \mn@doi [\apj] {10.1088/0004-637X/798/1/63}, \href
  {http://cdsads.u-strasbg.fr/abs/2015ApJ...798...63M} {798, 63}

\bibitem[\protect\citeauthoryear{{Mayor} et~al.,}{{Mayor}
  et~al.}{2003}]{Mayor03}
{Mayor} M.,  et~al., 2003, The Messenger, \href
  {http://cdsads.u-strasbg.fr/abs/2003Msngr.114...20M} {114, 20}

\bibitem[\protect\citeauthoryear{{Melo} et~al.,}{{Melo} et~al.}{2007}]{Melo07}
{Melo} C.,  et~al., 2007, \mn@doi [\aap] {10.1051/0004-6361:20066845}, \href
  {http://cdsads.u-strasbg.fr/abs/2007A\%26A...467..721M} {467, 721}

\bibitem[\protect\citeauthoryear{{Meunier}, {Desort}  \& {Lagrange}}{{Meunier}
  et~al.}{2010}]{Meunier10}
{Meunier} N.,  {Desort} M.,   {Lagrange} A.-M.,  2010, \mn@doi [\aap]
  {10.1051/0004.6361/200913551}, 512, A39

\bibitem[\protect\citeauthoryear{{Morin} et~al.,}{{Morin}
  et~al.}{2008a}]{Morin08a}
{Morin} J.,  et~al., 2008a, \mnras, 384, 77

\bibitem[\protect\citeauthoryear{{Morin} et~al.,}{{Morin}
  et~al.}{2008b}]{Morin08b}
{Morin} J.,  et~al., 2008b, \mn@doi [\mnras]
  {10.1111/j.1365-2966.2008.13809.x}, \href
  {http://adsabs.harvard.edu/abs/2008MNRAS.390..567M} {390, 567}

\bibitem[\protect\citeauthoryear{{Morin}, {Donati}, {Petit}, {Delfosse},
  {Forveille}  \& {Jardine}}{{Morin} et~al.}{2010}]{Morin10}
{Morin} J.,  {Donati} J.,  {Petit} P.,  {Delfosse} X.,  {Forveille} T.,
  {Jardine} M.~M.,  2010, \mn@doi [\mnras] {10.1111/j.1365-2966.2010.17101.x},
  \href {http://adsabs.harvard.edu/abs/2010MNRAS.407.2269M} {407, 2269}

\bibitem[\protect\citeauthoryear{{Morin}, {Dormy}, {Schrinner}  \&
  {Donati}}{{Morin} et~al.}{2011}]{Morin11}
{Morin} J.,  {Dormy} E.,  {Schrinner} M.,   {Donati} J.-F.,  2011, \mn@doi
  [\mnras] {10.1111/j.1745-3933.2011.01159.x}, \href
  {http://cdsads.u-strasbg.fr/abs/2011MNRAS.418L.133M} {418, L133}

\bibitem[\protect\citeauthoryear{{Moutou} et~al.,}{{Moutou}
  et~al.}{2007}]{Moutou07}
{Moutou} C.,  et~al., 2007, \mn@doi [\aap] {10.1051/0004-6361:20077795}, \href
  {http://adsabs.harvard.edu/abs/2007A\%26A...473..651M} {473, 651}

\bibitem[\protect\citeauthoryear{{Queloz}, {Henry}, {Sivan}, {Baliunas},
  {Beuzit}  \& {Donahue}}{{Queloz} et~al.}{2001}]{Queloz01}
{Queloz} D.,  {Henry} G.,  {Sivan} J.,  {Baliunas} S.,  {Beuzit} J.,
  {Donahue} R.,  2001, \aap, 379

\bibitem[\protect\citeauthoryear{{Reiners}, {Bean}, {Hubert}, {Dreizler}  \&
  {Seifahrt}}{{Reiners} et~al.}{2010}]{Reiners10}
{Reiners} A.,  {Bean} J.,  {Hubert} K.,  {Dreizler} S.,   {Seifahrt} A.,  2010,
  \apj, 710, 432

\bibitem[\protect\citeauthoryear{{Robertson}, {Mahadevan}, {Endl}  \&
  {Roy}}{{Robertson} et~al.}{2014}]{Robertson14}
{Robertson} P.,  {Mahadevan} S.,  {Endl} M.,   {Roy} A.,  2014, \mn@doi
  [Science] {10.1126/science.1253253}, \href
  {http://cdsads.u-strasbg.fr/abs/2014Sci...345..440R} {345, 440}

\bibitem[\protect\citeauthoryear{{Rodler}, {Del Burgo}, {Witte}, {Helling},
  {Hauschildt}, {Mart{\'{\i}}n}, {{\'A}lvarez}  \& {Deshpande}}{{Rodler}
  et~al.}{2011}]{Rodler11}
{Rodler} F.,  {Del Burgo} C.,  {Witte} S.,  {Helling} C.,  {Hauschildt} P.~H.,
  {Mart{\'{\i}}n} E.~L.,  {{\'A}lvarez} C.,   {Deshpande} R.,  2011, \mn@doi
  [\aap] {10.1051/0004-6361/201015490}, \href
  {http://cdsads.u-strasbg.fr/abs/2011A%26A...532A..31R} {532, A31}

\bibitem[\protect\citeauthoryear{{Scargle}}{{Scargle}}{1982}]{Scargle82}
{Scargle} J.~D.,  1982, \mn@doi [\apj] {10.1086/160554}, \href
  {http://cdsads.u-strasbg.fr/abs/1982ApJ...263..835S} {263, 835}

\bibitem[\protect\citeauthoryear{{Snik}, {Kochukhov}  \& {Piskunov}}{{Snik}
  et~al.}{2011}]{Snik11}
{Snik} F.,  {Kochukhov} O.,   {Piskunov} N.,  2011, \mn@doi [ASP Conference
  Series] {2011ASPC..437..237S}, 437, 237

\bibitem[\protect\citeauthoryear{{Zechmeister} \& {K{\"u}rster}}{{Zechmeister}
  \& {K{\"u}rster}}{2009}]{Zechmeister09}
{Zechmeister} M.,  {K{\"u}rster} M.,  2009, \mn@doi [\aap]
  {10.1051/0004-6361:200811296}, \href
  {http://cdsads.u-strasbg.fr/abs/2009A%26A...496..577Z} {496, 577}

\makeatother
\end{thebibliography}



\appendix

\section{Rotation periods derived from \bl, RV, FWHM and \hal.}
\begin{table}
\begin{center}
\begin{tabular}{cccccccc}
\hline
Star  & $P_{\rm rot}^{\bl}$   &    $P_{\rm rot}^{\rm RV}$ & $P_{\rm rot}^{\rm FWHM}$     & $P_{\rm rot}^{\hal}$    \\
  & (d)   &    (d)  &             (d)      & (d)    \\
\hline
GJ~205& 33.63$\pm$0.37 & 39.70$\pm$0.85  & 41.9$\pm$1.9  & 33.46$\pm$0.80  \\
GJ~410& 13.83$\pm$0.10  & 14.20$\pm$0.10  & 14.76$\pm$0.20  & 15.15$\pm$0.30   \\
GJ~479& 24.04$\pm$0.75  & 23.2$\pm$1.9  & 25.48$\pm$0.81  & 22.94$\pm$0.60   \\
GJ~358& 25.37$\pm$0.32  & 24.47$\pm$0.60  & 25.49$\pm$0.42  & 23.8$\pm$2.7   \\
\hline
\end{tabular}
\caption{ Rotation periods derived from \bl, RV, FWHM and \hal\ measurements and the estimated error-bars at 1$\sigma$ for the 4 stars of the sample.
}
\label{tab:all_period}
\end{center}
\end{table}

\section{Observations journal}
\label{ann:journal}
Observations journal for the four M-dwarfs observed from October 2013 to August 2014 with HARPS-Pol@LaSilla and NARVAL@TBL.

\begin{table*}
\begin{center}
\begin{tabular}{cccccccc}
\hline
Cycle  & Date   &   BJD &             instrument     & S/N  &  B  &RV  \\
            &                  & (+ 2 456 000) &                  &          &(G) &(\kms) \\
\hline
0.024 & 30apr14 & 778.5870 & HARPS-Pol & 133 & 20.54 $\pm$ 7.03 & 8.17e-03 $\pm$ 1.56e-03\\
0.069 & 01may14 & 779.6500 & HARPS-Pol & 121 & 3.36 $\pm$ 7.82 & 4.08e-03 $\pm$ 1.74e-03\\
0.151 & 03may14 & 781.6390 & HARPS-Pol & 108 & 8.61 $\pm$ 9.02 & 5.16e-03 $\pm$ 1.97e-03\\
0.190 & 04may14 & 782.5590 & HARPS-Pol & 135 & 5.46 $\pm$ 7.04 & 5.53e-03 $\pm$ 1.43e-03\\
0.231 & 05may14 & 783.5600 & HARPS-Pol & 102 & 24.94 $\pm$ 9.97 & -6.72e-03 $\pm$ 2.09e-03\\
0.272 & 06may14 & 784.5480 & HARPS-Pol & 120 & 18.45 $\pm$ 8.25 & 5.52e-03 $\pm$ 1.75e-03\\
0.356 & 08may14 & 786.5590 & HARPS-Pol & 146 & 21.14 $\pm$ 6.50 & 1.32e-02 $\pm$ 1.43e-03\\
1.314 & 31may14 & 809.5840 & HARPS-Pol & 99 & 32.36 $\pm$ 9.77 & 7.16e-04 $\pm$ 2.20e-03\\
1.397 & 02jun14 & 811.5850 & HARPS-Pol & 120 & 16.67 $\pm$ 7.65 & 9.20e-04 $\pm$ 1.69e-03\\
1.523 & 05jun14 & 814.6220 & HARPS-Pol & 99 & -16.66 $\pm$ 9.66 & -4.61e-03 $\pm$ 2.14e-03\\
1.607 & 07jun14 & 816.6220 & HARPS-Pol & 95 & -34.92 $\pm$ 10.45 & -4.82e-03 $\pm$ 2.25e-03\\
1.648 & 08jun14 & 817.6150 & HARPS-Pol & 85 & -16.75 $\pm$ 11.69 & -4.24e-03 $\pm$ 2.31e-03\\
1.689 & 09jun14 & 818.6140 & HARPS-Pol & 75 & -16.33 $\pm$ 13.67 & -1.93e-03 $\pm$ 2.87e-03\\
1.729 & 10jun14 & 819.5700 & HARPS-Pol & 97 & -24.67 $\pm$ 10.03 & -5.63e-03 $\pm$ 2.18e-03\\
1.894 & 14jun14 & 823.5230 & HARPS-Pol & 89 & 1.53 $\pm$ 11.60 & 3.14e-03 $\pm$ 2.39e-03\\
1.935 & 15jun14 & 824.5230 & HARPS-Pol & 106 & 15.89 $\pm$ 9.20 & 5.63e-03 $\pm$ 1.96e-03\\
2.016 & 17jun14 & 826.4650 & HARPS-Pol & 113 & 14.38 $\pm$ 8.71 & 2.82e-03 $\pm$ 1.83e-03\\
2.099 & 19jun14 & 828.4620 & HARPS-Pol & 129 & 22.29 $\pm$ 7.44 & 2.27e-03 $\pm$ 1.60e-03\\
2.144 & 20jun14 & 829.5330 & HARPS-Pol & 118 & 10.17 $\pm$ 7.95 & 6.55e-03 $\pm$ 1.76e-03\\
2.184 & 21jun14 & 830.5110 & HARPS-Pol & 114 & 19.65 $\pm$ 8.41 & 4.76e-03 $\pm$ 1.82e-03\\
2.352 & 25jun14 & 834.5360 & HARPS-Pol & 101 & 40.06 $\pm$ 9.60 & 3.41e-03 $\pm$ 2.12e-03\\
2.643 & 02jul14 & 841.5440 & HARPS-Pol & 63 & 3.14 $\pm$ 17.37 & 2.85e-03 $\pm$ 3.50e-03\\
2.684 & 03jul14 & 842.5310 & HARPS-Pol & 101 & -18.62 $\pm$ 9.72 & -5.48e-03 $\pm$ 2.08e-03\\
\hline
\end{tabular}
\caption{Journal of observations for GJ~479. Columns 1 and 5, respectively, list, the rotational cycle (computed with the rotation period mentioned in Table~\ref{tab:starparam} according to ephemeris given by Eq~\ref{eq:eph}, the date of the beginning of the night, the Barycentric Julian Date, the observation site, the peak S/N (per 0.85~\kms\ velocity bin). Column 6-7 give, respectively, \bl\ and RV values.
}
\label{tab:obs479}
\end{center}
\end{table*}

\begin{table*}
\begin{center}
\begin{tabular}{cccccccc}
\hline
Cycle  & Date   &   BJD &             instrument     & S/N  &  B  &RV  \\
            &                  & (+ 2 456 000) &                  &          &(G) &(\kms) \\
\hline
0.000 & 17jan14 & 675.7090 & HARPS-Pol & 109 & -51.09 $\pm$ 12.08 & -6.39e-03 $\pm$ 1.93e-03\\
0.078 & 19jan14 & 677.6780 & HARPS-Pol & 96 & -54.90 $\pm$ 14.22 & -1.23e-03 $\pm$ 2.21e-03\\
0.158 & 21jan14 & 679.7130 & HARPS-Pol & 86 & -99.91 $\pm$ 16.07 & -8.14e-04 $\pm$ 2.47e-03\\
0.235 & 23jan14 & 681.6580 & HARPS-Pol & 93 & -94.82 $\pm$ 14.74 & -3.57e-03 $\pm$ 2.26e-03\\
0.477 & 29jan14 & 687.7800 & HARPS-Pol & 88 & -49.06 $\pm$ 15.49 & -3.10e-03 $\pm$ 2.43e-03\\
0.553 & 31jan14 & 689.7110 & HARPS-Pol & 123 & -15.37 $\pm$ 10.66 & 5.98e-03 $\pm$ 1.68e-03\\
0.631 & 02feb14 & 691.6650 & HARPS-Pol & 121 & -1.68 $\pm$ 10.99 & 8.18e-03 $\pm$ 1.73e-03\\
0.789 & 06feb14 & 695.6510 & HARPS-Pol & 123 & 18.37 $\pm$ 10.49 & 4.64e-03 $\pm$ 1.71e-03\\
0.869 & 08feb14 & 697.6850 & HARPS-Pol & 113 & 10.17 $\pm$ 11.80 & -3.22e-03 $\pm$ 1.88e-03\\
1.468 & 23feb14 & 712.8430 & HARPS-Pol & 112 & -53.01 $\pm$ 11.43 & 2.75e-03 $\pm$ 1.85e-03\\
1.547 & 25feb14 & 714.8380 & HARPS-Pol & 76 & -13.01 $\pm$ 18.93 & 9.27e-03 $\pm$ 2.86e-03\\
1.626 & 27feb14 & 716.8240 & HARPS-Pol & 107 & -8.13 $\pm$ 12.01 & 4.92e-03 $\pm$ 1.99e-03\\
1.704 & 01mar14 & 718.8090 & HARPS-Pol & 86 & 7.23 $\pm$ 15.59 & -7.48e-04 $\pm$ 2.51e-03\\
1.743 & 02mar14 & 719.8050 & HARPS-Pol & 119 & 12.56 $\pm$ 10.58 & -2.58e-03 $\pm$ 1.78e-03\\
1.821 & 04mar14 & 721.7630 & HARPS-Pol & 124 & -9.67 $\pm$ 9.97 & -1.89e-03 $\pm$ 1.70e-03\\
1.899 & 06mar14 & 723.7550 & HARPS-Pol & 109 & -29.50 $\pm$ 11.70 & -4.77e-04 $\pm$ 1.96e-03\\
2.014 & 09mar14 & 726.6640 & HARPS-Pol & 106 & -49.88 $\pm$ 12.37 & -5.61e-03 $\pm$ 2.02e-03\\
2.094 & 11mar14 & 728.6900 & HARPS-Pol & 133 & -81.82 $\pm$ 9.12 & -9.77e-04 $\pm$ 1.56e-03\\
2.171 & 13mar14 & 730.6400 & HARPS-Pol & 86 & -83.74 $\pm$ 15.55 & 1.07e-03 $\pm$ 2.49e-03\\
2.645 & 25mar14 & 742.6240 & HARPS-Pol & 101 & -22.29 $\pm$ 12.71 & 1.19e-02 $\pm$ 2.12e-03\\
2.723 & 27mar14 & 744.5980 & HARPS-Pol & 130 & 12.28 $\pm$ 9.45 & -1.70e-03 $\pm$ 1.60e-03\\
2.804 & 29mar14 & 746.6290 & HARPS-Pol & 107 & 7.83 $\pm$ 11.74 & -1.37e-03 $\pm$ 1.97e-03\\
2.880 & 31mar14 & 748.5630 & HARPS-Pol & 70 & 16.42 $\pm$ 19.71 & -1.72e-03 $\pm$ 3.16e-03\\
\hline
\end{tabular}
\caption{
Same as Table~\ref{tab:obs479} for GJ~358.
}
\label{tab:obs358}
\end{center}
\end{table*}

\begin{table*}
\begin{center}
\begin{tabular}{cccccccc}
\hline
Cycle  & Date   &   BJD &             instrument     & S/N  &  B  &RV  \\
            &                  & (+ 2 456 000) &                  &          &(G) &(\kms) \\
\hline
0.000 & 03oct13 & 569.8850 & HARPS-Pol & 177 & 8.29 $\pm$ 2.25 & -1.05e-05 $\pm$ 1.23e-03\\
0.030 & 04oct13 & 570.9030 & HARPS-Pol & 196 & 11.79 $\pm$ 2.00 & -4.79e-03 $\pm$ 1.12e-03\\
0.178 & 09oct13 & 575.8680 & HARPS-Pol & 228 & 3.30 $\pm$ 1.70 & -6.62e-04 $\pm$ 8.86e-04\\
0.208 & 10oct13 & 576.8910 & HARPS-Pol & 219 & 5.16 $\pm$ 1.80 & 9.26e-04 $\pm$ 9.94e-04\\
0.444 & 18oct13 & 584.8280 & HARPS-Pol & 188 & 11.05 $\pm$ 2.13 & 1.51e-03 $\pm$ 1.17e-03\\
0.505 & 20oct13 & 586.8820 & HARPS-Pol & 194 & 4.40 $\pm$ 2.04 & 2.49e-03 $\pm$ 1.14e-03\\
0.682 & 26oct13 & 592.8720 & HARPS-Pol & 210 & -5.45 $\pm$ 1.84 & -7.92e-04 $\pm$ 1.04e-03\\
0.741 & 28oct13 & 594.8510 & HARPS-Pol & 171 & -2.36 $\pm$ 2.33 & 7.87e-04 $\pm$ 1.28e-03\\
0.918 & 03nov13 & 600.8070 & HARPS-Pol & 180 & 6.55 $\pm$ 2.20 & 6.62e-03 $\pm$ 1.22e-03\\
0.977 & 05nov13 & 602.8160 & HARPS-Pol & 189 & 11.98 $\pm$ 2.09 & 1.44e-03 $\pm$ 1.07e-03\\
1.036 & 07nov13 & 604.7880 & HARPS-Pol & 197 & 11.14 $\pm$ 2.00 & 3.71e-03 $\pm$ 1.12e-03\\
1.274 & 15nov13 & 612.7990 & HARPS-Pol & 170 & 2.45 $\pm$ 2.36 & 1.69e-03 $\pm$ 1.30e-03\\
1.332 & 17nov13 & 614.7790 & HARPS-Pol & 203 & 8.53 $\pm$ 1.94 & -1.66e-03 $\pm$ 1.09e-03\\
1.571 & 25nov13 & 622.8050 & HARPS-Pol & 209 & 2.80 $\pm$ 1.88 & -4.04e-03 $\pm$ 9.73e-04\\
1.623 & 27nov13 & 624.5610 & NARVAL & 313 & -1.60 $\pm$ 1.31 & - \\
1.630 & 27nov13 & 624.8090 & HARPS-Pol & 174 & -2.16 $\pm$ 2.32 & 2.68e-03 $\pm$ 1.15e-03\\
1.688 & 29nov13 & 626.7720 & HARPS-Pol & 185 & -4.40 $\pm$ 2.13 & -1.92e-03 $\pm$ 1.18e-03\\
1.745 & 01dec13 & 628.7000 & HARPS-Pol & 138 & -5.52 $\pm$ 3.04 & 2.07e-04 $\pm$ 1.57e-03\\
1.864 & 05dec13 & 632.6810 & HARPS-Pol & 171 & 5.87 $\pm$ 2.39 & 1.84e-03 $\pm$ 1.26e-03\\
2.006 & 10dec13 & 637.4830 & NARVAL & 399 & 8.10 $\pm$ 0.97 & - \\
2.065 & 12dec13 & 639.4620 & NARVAL & 454 & 6.75 $\pm$ 0.83 & - \\
2.186 & 16dec13 & 643.5400 & NARVAL & 308 & 4.78 $\pm$ 1.31 & - \\
3.135 & 17jan14 & 675.5440 & HARPS-Pol & 172 & 4.19 $\pm$ 2.42 & 3.61e-03 $\pm$ 1.26e-03\\
3.197 & 19jan14 & 677.6080 & HARPS-Pol & 148 & 3.78 $\pm$ 2.76 & 7.27e-03 $\pm$ 1.59e-03\\
3.315 & 23jan14 & 681.5910 & HARPS-Pol & 133 & 8.39 $\pm$ 3.11 & 8.42e-03 $\pm$ 1.52e-03\\
3.523 & 30jan14 & 688.5870 & HARPS-Pol & 171 & 5.48 $\pm$ 2.37 & -4.19e-03 $\pm$ 1.27e-03\\
\hline
\end{tabular}
\caption{Same as Table~\ref{tab:obs479} for GJ~205.
}
\label{tab:obs205}
\end{center}
\end{table*}

\begin{table*}
\begin{center}
\begin{tabular}{ccccccccc}
\hline
Cycle  & Date   &   BJD &             instrument     & S/N  &  B & RV\\
            &                  & (+ 2 456 000) &                  &          &(G) & (\kms)\\
\hline
0.000 & 10sep13 & 546.4638 & NARVAL & 301 & 2.71 $\pm$ 1.43 &- \\
0.637 & 17sep13 & 553.4688 & NARVAL & 251 & 2.14 $\pm$ 1.80 &-\\
1.092 & 22sep13 & 558.4694 & NARVAL & 282 & -1.07 $\pm$ 1.57 &-\\
1.269 & 24sep13 & 560.4200 & NARVAL & 318 & -1.60 $\pm$ 1.36 &-\\
2.629 & 09oct13 & 575.3856 & NARVAL & 278 & 3.12 $\pm$ 1.57 &-\\
2.821 & 11oct13 & 577.4947 & NARVAL & 242 & 1.67 $\pm$ 1.82 &- \\
2.902 & 12oct13 & 578.3830 & NARVAL & 297 & 1.19 $\pm$ 1.41 &- \\
2.993 & 13oct13 & 579.3876 & NARVAL & 274 & -2.93 $\pm$ 1.55&- \\
3.442 & 18oct13 & 584.3262 & NARVAL & 209 & 4.21 $\pm$ 2.21 &- \\
4.622 & 31oct13 & 597.3060 & NARVAL & 189 & 2.37 $\pm$ 2.44 &- \\
7.075 & 27nov13 & 624.2851 & NARVAL & 194 & -6.40 $\pm$ 2.41 &- \\
8.252 & 10dec13 & 637.2351 & NARVAL & 280 & 1.27 $\pm$ 1.57&-\\
8.434 & 12dec13 & 639.2356 & NARVAL & 305 & 1.98 $\pm$ 1.42 &-\\
8.615 & 14dec13 & 641.2330 & NARVAL & 234 & 7.99 $\pm$ 1.92 &- \\
8.709 & 15dec13 & 642.2580 & NARVAL & 291 & 4.24 $\pm$ 1.52  &-\\
25.764 & 20jun14 & 829.8720 & HARPS-Pol & 91 & 17.14 $\pm$ 5.05 & 8.63e-03 $\pm$ 3.26e-03\\
25.950 & 22jun14 & 831.9170 & HARPS-Pol & 158 & 7.85 $\pm$ 2.62 & 1.08e-03 $\pm$ 1.78e-03\\
26.132 & 24jun14 & 833.9110 & HARPS-Pol & 140 & 8.96 $\pm$ 3.02 & -9.73e-05 $\pm$ 2.15e-03\\
26.314 & 26jun14 & 835.9120 & HARPS-Pol & 100 & 2.19 $\pm$ 4.55 & 2.84e-03 $\pm$ 3.03e-03\\
28.311 & 18jul14 & 857.8880 & HARPS-Pol & 111 & 6.47 $\pm$ 4.01 & 1.08e-02 $\pm$ 2.93e-03\\
28.398 & 19jul14 & 858.8430 & HARPS-Pol & 141 & -0.08 $\pm$ 2.96 & 8.27e-03 $\pm$ 2.00e-03\\
28.493 & 20jul14 & 859.8910 & HARPS-Pol & 129 & 2.69 $\pm$ 3.34 & 5.13e-03 $\pm$ 2.15e-03\\
28.761 & 23jul14 & 862.8310 & HARPS-Pol & 107 & 6.86 $\pm$ 7.38 & 8.41e-03 $\pm$ 2.78e-03\\
28.850 & 24jul14 & 863.8130 & HARPS-Pol & 119 & 10.93 $\pm$ 3.61 & 1.04e-02 $\pm$ 2.31e-03\\
28.940 & 25jul14 & 864.8020 & HARPS-Pol & 131 & 4.41 $\pm$ 3.28 & 9.71e-03 $\pm$ 1.98e-03\\
31.849 & 27aug14 & 896.8010 & HARPS-Pol & 102 & -0.01 $\pm$ 4.45 & 5.91e-03 $\pm$ 2.70e-03\\
\hline
\end{tabular}
\caption{Same as Table~\ref{tab:obs479} for GJ~846.
}
\label{tab:obs846}
\end{center}
\end{table*}

\begin{table*}
\begin{center}
\begin{tabular}{cccccccc}
\hline
Cycle  & Date   &   BJD &             instrument     & S/N  &  B  &RV  \\
            &                  & (+ 2 456 000) &                  &          &(G) &(\kms) \\
\hline
0.558& 09jan14 & 667.6948 & NARVAL & 282 & -8.94 $\pm$ 1.50 & - \\
0.696 & 11jan14 & 669.6198 & NARVAL & 255 & 5.28 $\pm$ 1.69 & - \\
1.000 & 15jan14 & 673.8840 & HARPS-Pol & 90 & 16.52 $\pm$ 5.62 & -1.97e-02 $\pm$ 3.85e-03\\
1.928 & 28jan14 & 686.8660 & HARPS-Pol & 84 & 27.53 $\pm$ 6.11 & -2.79e-03 $\pm$ 4.41e-03\\
1.996 & 29jan14 & 687.8170 & HARPS-Pol & 79 & 8.28 $\pm$ 6.59 & 4.17e-03 $\pm$ 4.34e-03\\
2.067 & 30jan14 & 688.8200 & HARPS-Pol & 108 & 4.11 $\pm$ 4.55 & -5.30e-04 $\pm$ 2.90e-03\\
2.209 & 01feb14 & 690.8100 & HARPS-Pol & 112 & 3.32 $\pm$ 4.35 & 1.31e-02 $\pm$ 3.06e-03\\
2.281 & 02feb14 & 691.8190 & HARPS-Pol & 103 & -1.09 $\pm$ 4.75 & 9.54e-04 $\pm$ 3.32e-03\\
2.423 & 04feb14 & 693.7960 & HARPS-Pol & 110 & -15.13 $\pm$ 4.41 & 2.79e-03 $\pm$ 3.08e-03\\
2.493 & 05feb14 & 694.7790 & HARPS-Pol & 106 & -13.66 $\pm$ 4.66 & 3.50e-03 $\pm$ 3.23e-03\\
2.565 & 06feb14 & 695.7870 & HARPS-Pol & 109 & -8.80 $\pm$ 4.45 & 6.28e-03 $\pm$ 3.13e-03\\
2.707 & 08feb14 & 697.7710 & HARPS-Pol & 100 & 10.94 $\pm$ 4.92 & -2.21e-02 $\pm$ 3.40e-03\\
3.777 & 23feb14 & 712.7580 & HARPS-Pol & 123 & 34.21 $\pm$ 3.85 & -2.86e-03 $\pm$ 2.59e-03\\
3.849 & 24feb14 & 713.7600 & HARPS-Pol & 107 & 30.37 $\pm$ 4.53 & 2.02e-03 $\pm$ 2.94e-03\\
3.992 & 26feb14 & 715.7630 & HARPS-Pol & 122 & 19.46 $\pm$ 3.86 & -3.82e-03 $\pm$ 2.82e-03\\
4.133 & 28feb14 & 717.7440 & HARPS-Pol & 101 & 3.45 $\pm$ 4.82 & 4.44e-03 $\pm$ 3.37e-03\\
4.278 & 02mar14 & 719.7690 & HARPS-Pol & 125 & -2.65 $\pm$ 3.82 & 2.06e-02 $\pm$ 2.52e-03\\
4.347 & 03mar14 & 720.7430 & HARPS-Pol & 118 & -7.61 $\pm$ 4.02 & 8.56e-03 $\pm$ 2.69e-03\\
4.418 & 04mar14 & 721.7260 & HARPS-Pol & 103 & -1.61 $\pm$ 4.73 & 8.66e-03 $\pm$ 3.61e-03\\
4.630 & 07mar14 & 724.7060 & HARPS-Pol & 98 & 9.48 $\pm$ 5.00 & 1.01e-03 $\pm$ 3.52e-03\\
4.703 & 08mar14 & 725.7160 & HARPS-Pol & 94 & 24.12 $\pm$ 5.32 & 2.93e-03 $\pm$ 3.62e-03\\
4.776 & 09mar14 & 726.7450 & HARPS-Pol & 108 & 16.25 $\pm$ 4.47 & 2.14e-03 $\pm$ 3.17e-03\\
4.845 & 10mar14 & 727.7130 & HARPS-Pol & 113 & 22.17 $\pm$ 4.30 & -2.84e-03 $\pm$ 3.05e-03\\
4.988 & 12mar14 & 729.7070 & HARPS-Pol & 112 & 9.37 $\pm$ 4.23 & -1.80e-04 $\pm$ 2.84e-03\\
5.058 & 13mar14 & 730.6860 & HARPS-Pol & 89 & 6.96 $\pm$ 5.69 & -6.63e-03 $\pm$ 3.84e-03\\
5.130 & 14mar14 & 731.6960 & HARPS-Pol & 83 & 7.69 $\pm$ 6.25 & -5.06e-03 $\pm$ 4.48e-03\\
5.769 & 23mar14 & 740.6520 & HARPS-Pol & 103 & 16.34 $\pm$ 4.84 & -5.91e-03 $\pm$ 3.06e-03\\
5.913 & 25mar14 & 742.6630 & HARPS-Pol & 113 & 11.95 $\pm$ 4.27 & 2.96e-03 $\pm$ 2.85e-03\\
6.036 & 27mar14 & 744.3780 & NARVAL & 169 & 11.67 $\pm$ 2.85 & - \\
6.055 & 27mar14 & 744.6480 & HARPS-Pol & 97 & 5.46 $\pm$ 5.06 & -8.96e-03 $\pm$ 3.52e-03\\
6.127 & 28mar14 & 745.6580 & HARPS-Pol & 110 & 6.51 $\pm$ 4.45 & -1.39e-02 $\pm$ 3.09e-03\\
6.199 & 29mar14 & 746.6660 & HARPS-Pol & 124 & -3.10 $\pm$ 3.82 & 1.81e-03 $\pm$ 2.76e-03\\
6.753 & 06apr14 & 754.4280 & NARVAL & 221 & 11.83 $\pm$ 2.09 & - \\
6.899 & 08apr14 & 756.4720 & NARVAL & 298 & 11.17 $\pm$ 1.43 & - \\
6.969 & 09apr14 & 757.4440 & NARVAL & 303 & 7.82 $\pm$ 1.42 & - \\
7.119 & 11apr14 & 759.5440 & NARVAL & 262 & 4.32 $\pm$ 1.73 & - \\
7.184 & 12apr14 & 760.4620 & NARVAL & 281 & 1.66 $\pm$ 1.53 & - \\
7.255 & 13apr14 & 761.4560 & NARVAL & 296 & -0.22 $\pm$ 1.47 & - \\
7.323 & 14apr14 & 762.4040 & NARVAL & 229 & -2.88 $\pm$ 1.91 & - \\
7.399 & 15apr14 & 763.4680 & NARVAL & 294 & -6.74 $\pm$ 1.47 & - \\
7.468 & 16apr14 & 764.4370 & NARVAL & 300 & -7.45 $\pm$ 1.45 & - \\
7.542 & 17apr14 & 765.4620 & NARVAL & 253 & -4.75 $\pm$ 1.76 & - \\
\hline
\end{tabular}
\caption{Same as Table~\ref{tab:obs479} for GJ~410.
}
\label{tab:obs410}
\end{center}
\end{table*}


\section{Supplementary \bl\ data}
\label{ann:suppl}
\begin{figure*}
\begin{center}
\includegraphics[scale=0.27,angle=0]{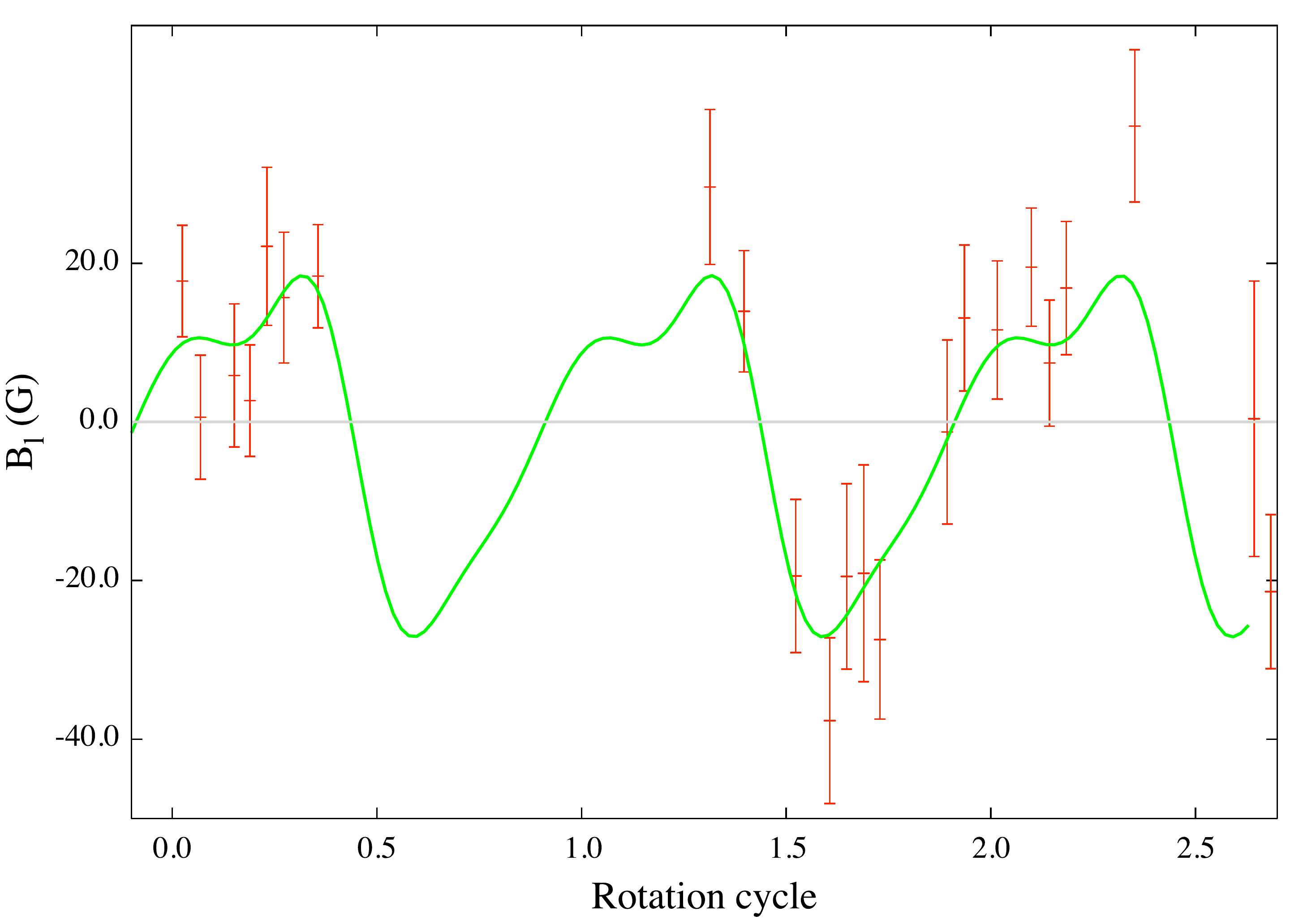} 
\includegraphics[scale=0.27,angle=0]{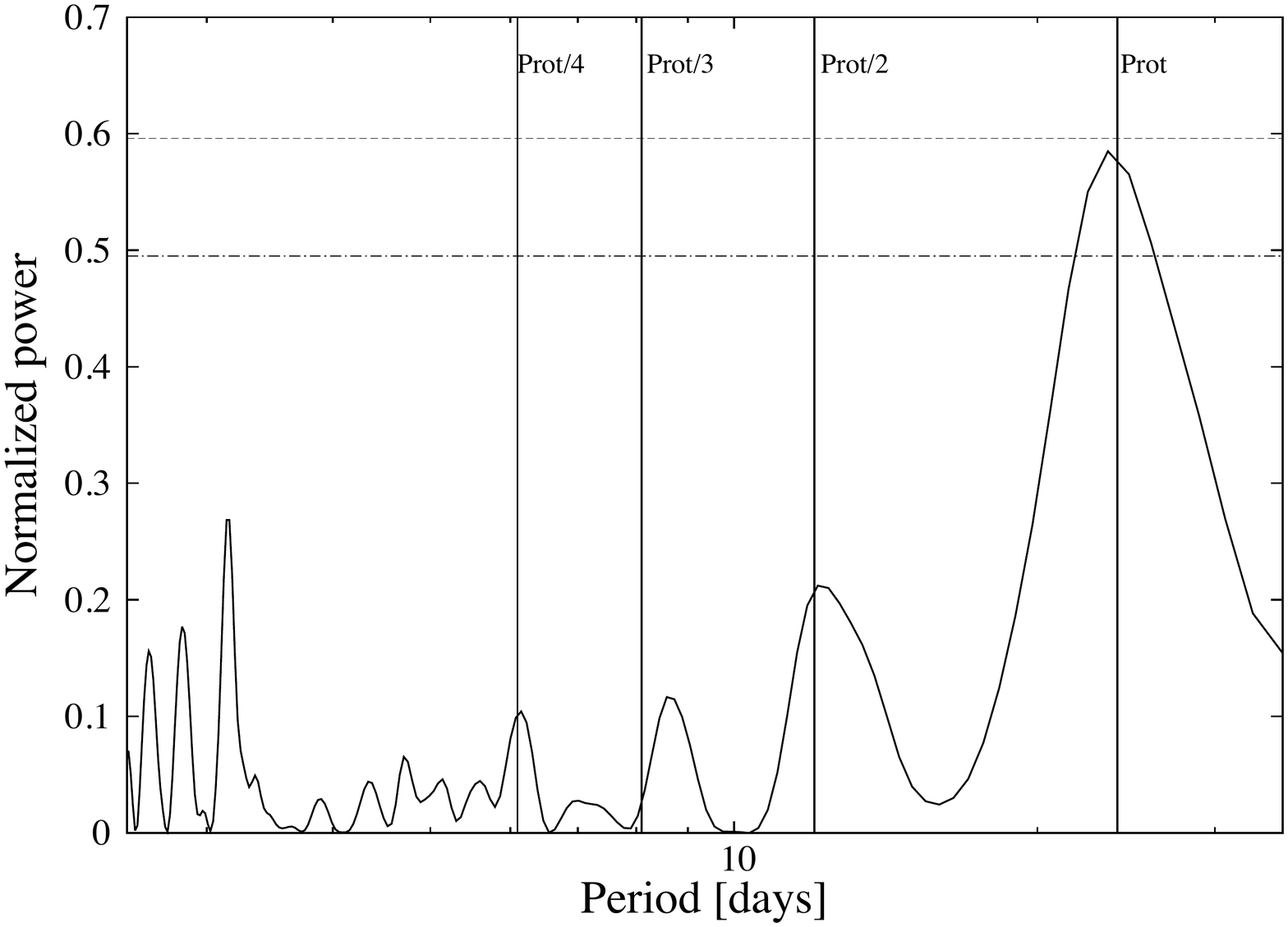} 
\caption{
Same as Fig.~\ref{fig:bl358} for GJ~479.}
\label{fig:bl479}
\end{center}
\end{figure*}

\begin{figure*}
\begin{center}
\includegraphics[scale=0.27,angle=0]{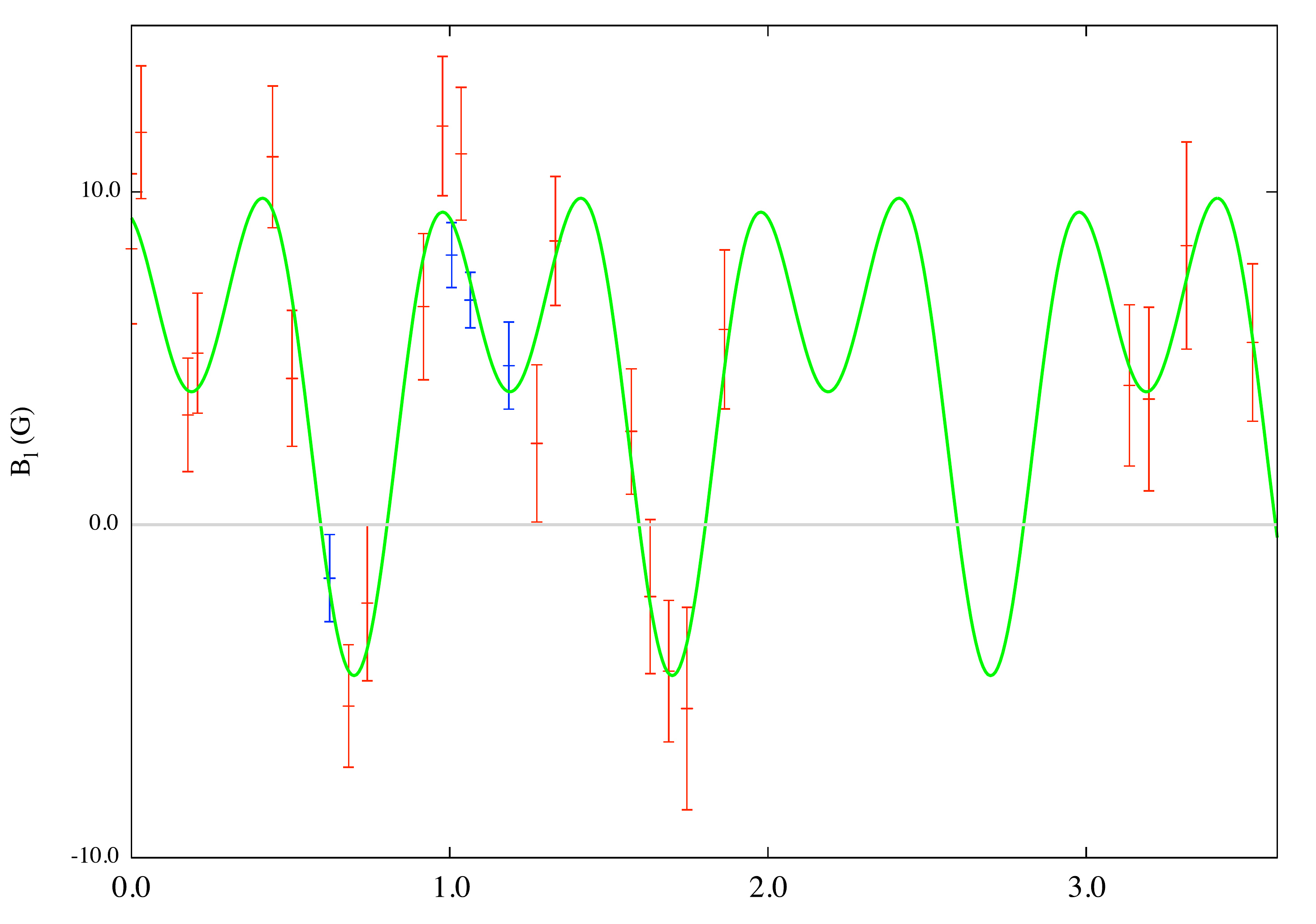} 
\includegraphics[scale=0.27,angle=0]{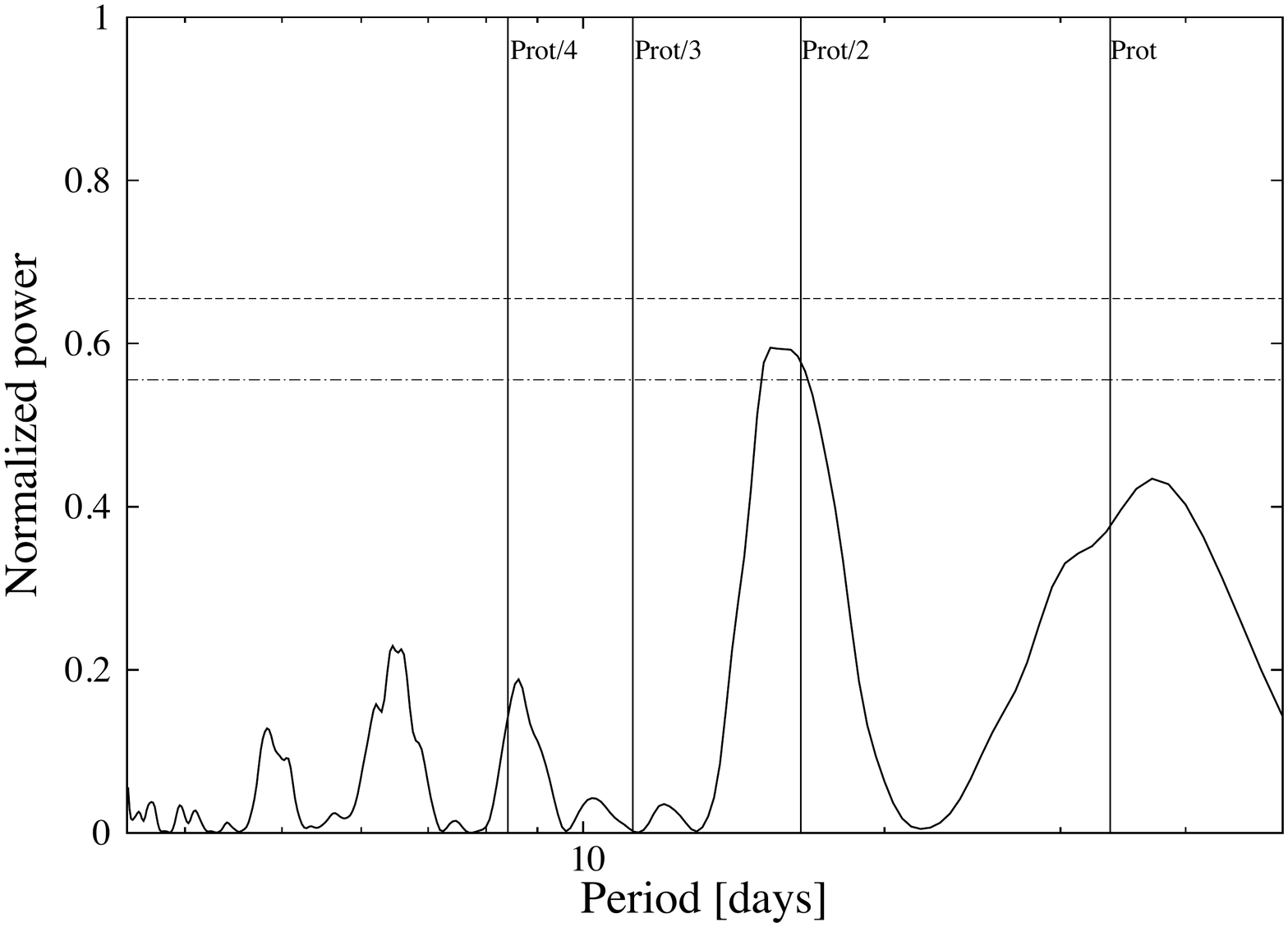} 
\caption{
Same as Fig.~\ref{fig:bl358} for GJ~205. HARPS-Pol data are in red, NARVAL data are blue.}
\label{fig:bl205}
\end{center}
\end{figure*}


\section{Supplementary Stokes $V$ LSD profiles and parent large-scale magnetic field maps}

\label{ann:suppl2}
\begin{figure*}
\includegraphics[scale=0.45,angle=0]{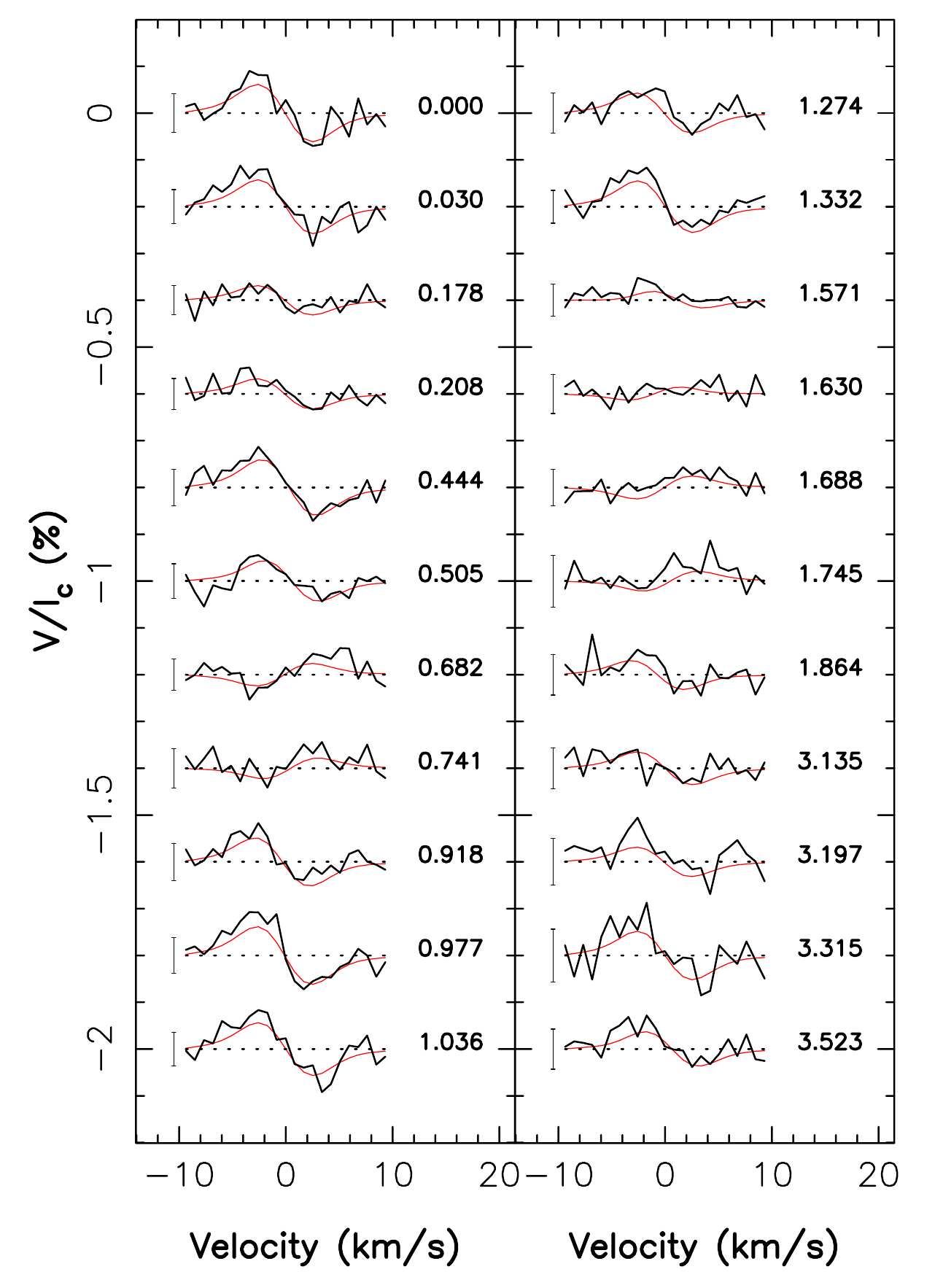} 
\includegraphics[scale=0.46,angle=0]{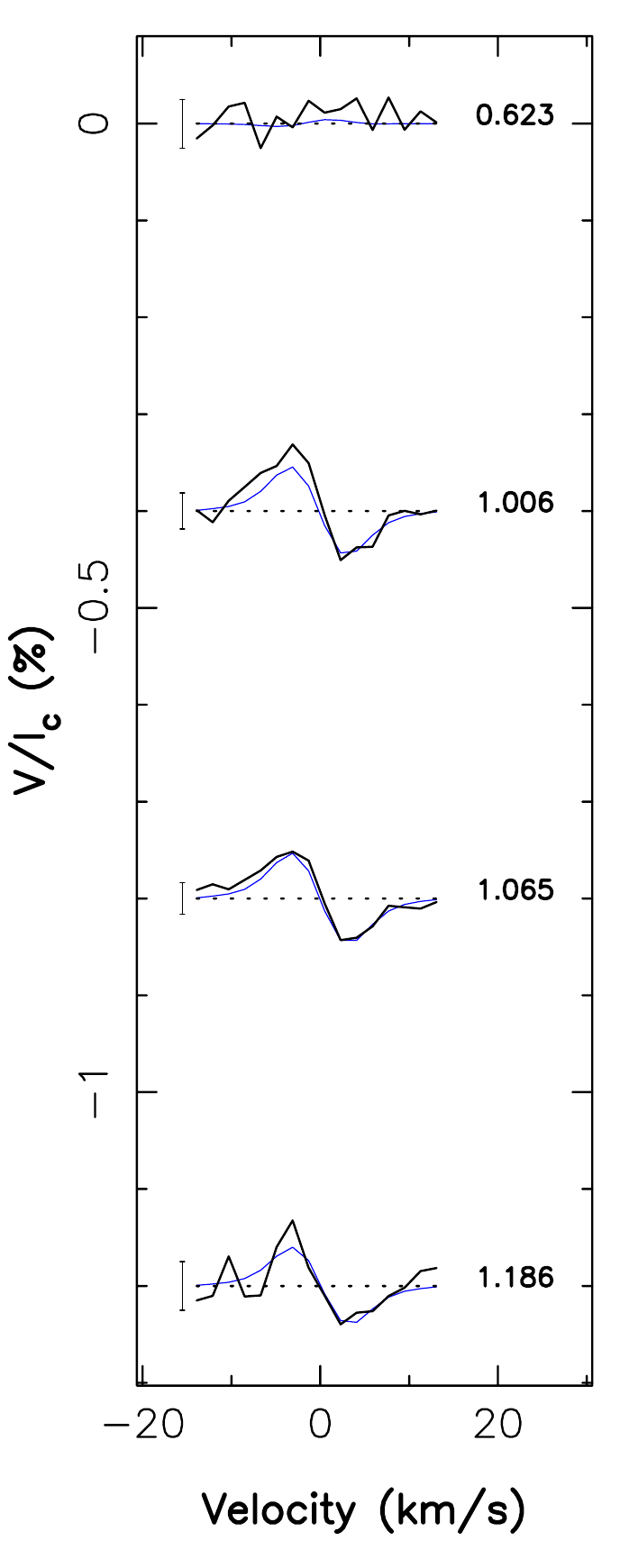} 
\includegraphics[scale=0.5,angle=0]{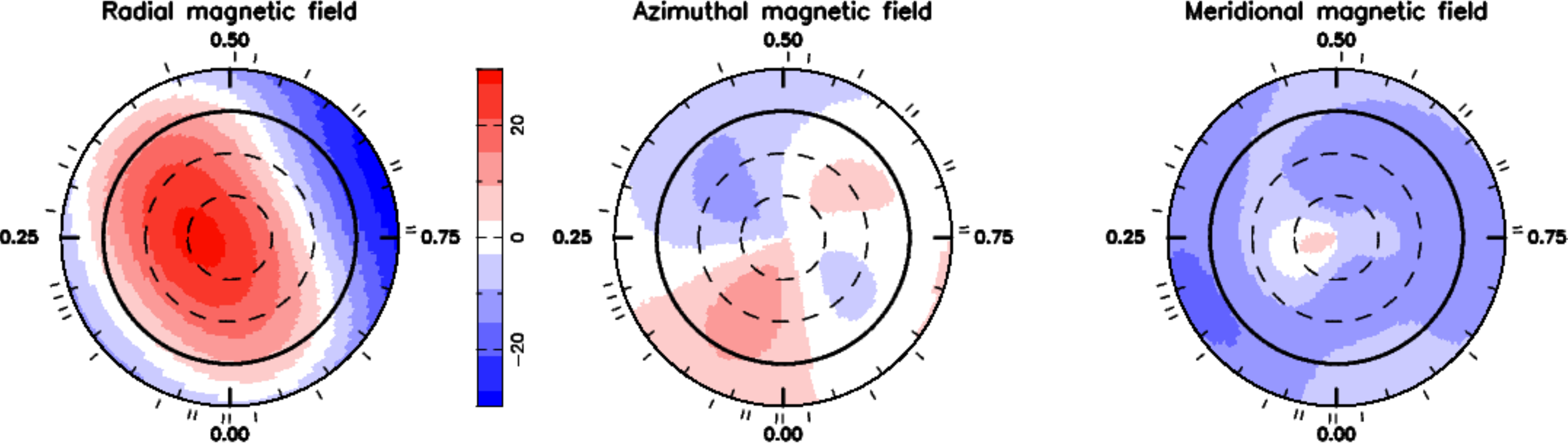} 
\caption{\textit{}
Same as Fig.~\ref{fig:map358} for GJ~205. LSD Stokes $V$ profiles in the top left and top right panels correspond to HARPS-Pol and NARVAL observations respectively.}
\label{fig:map205}
\end{figure*}

\begin{figure*}
\includegraphics[scale=0.45,angle=0]{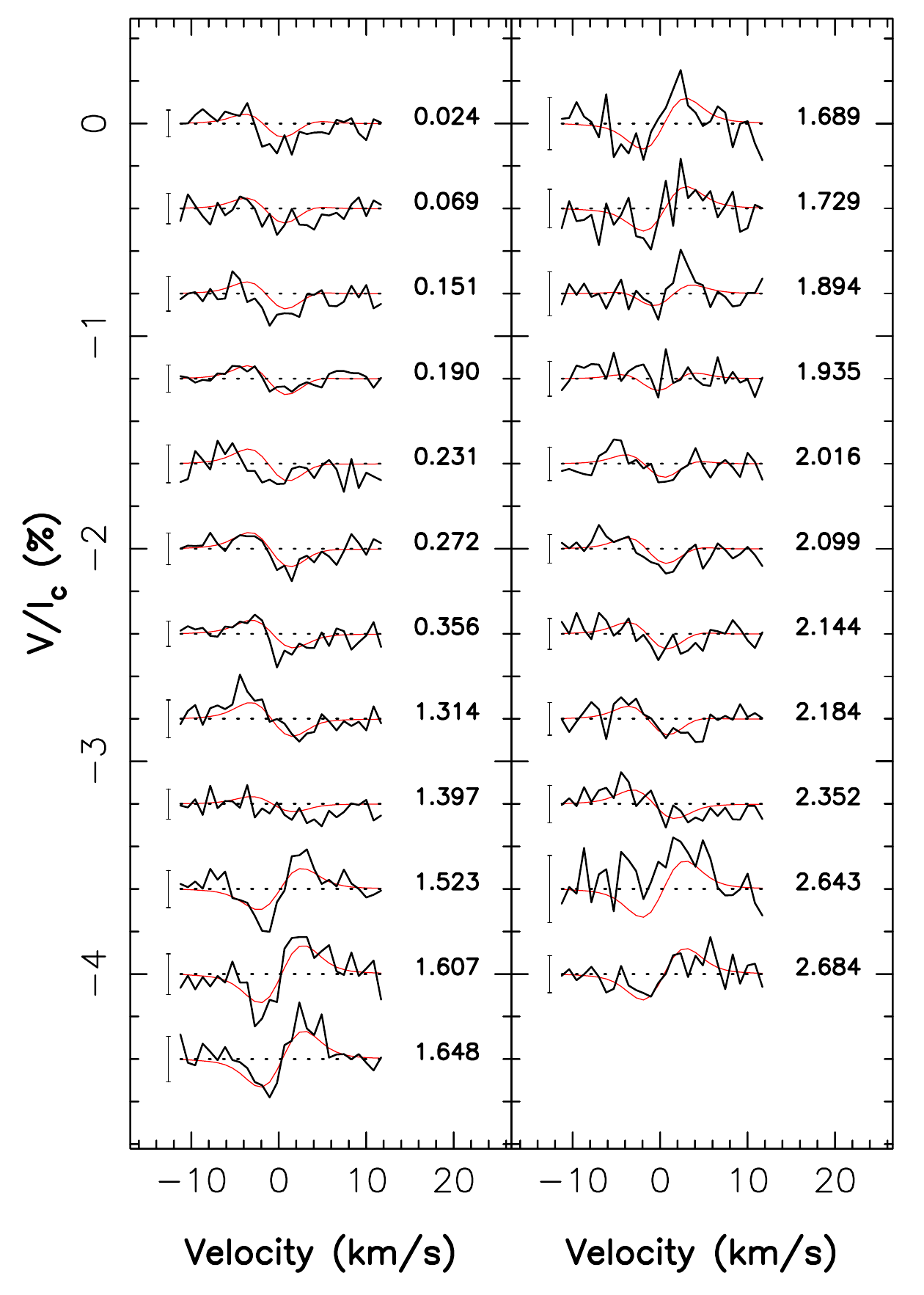} 
\includegraphics[scale=0.5,angle=0]{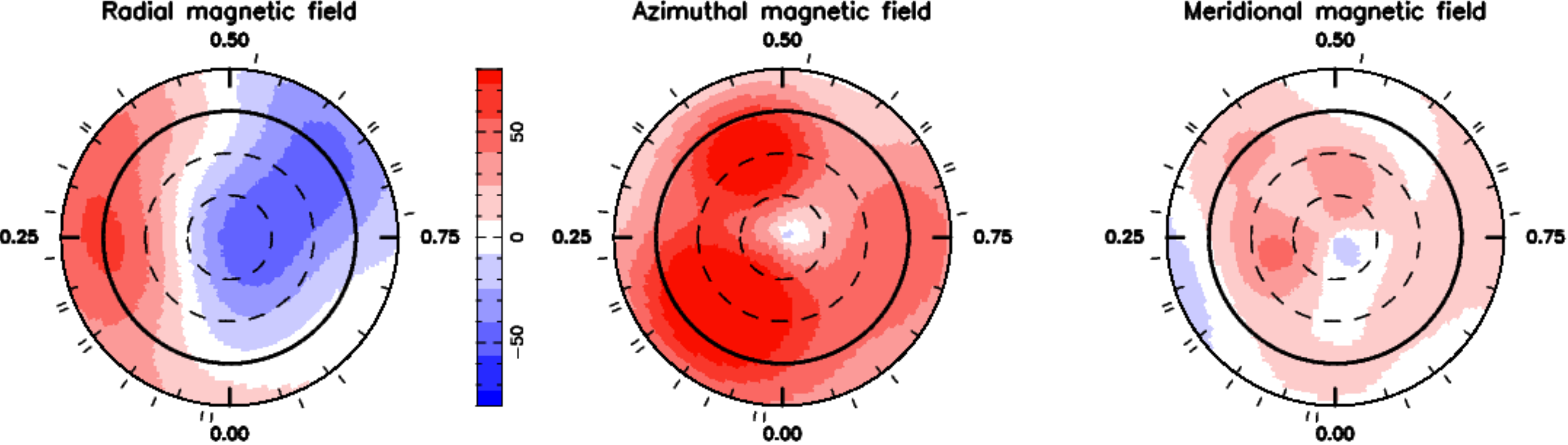} 
\caption{\textit{}
Same as Fig.~\ref{fig:map358} for GJ~479.}
\label{fig:map479}
\end{figure*}

\begin{figure*}
\includegraphics[scale=0.45,angle=0]{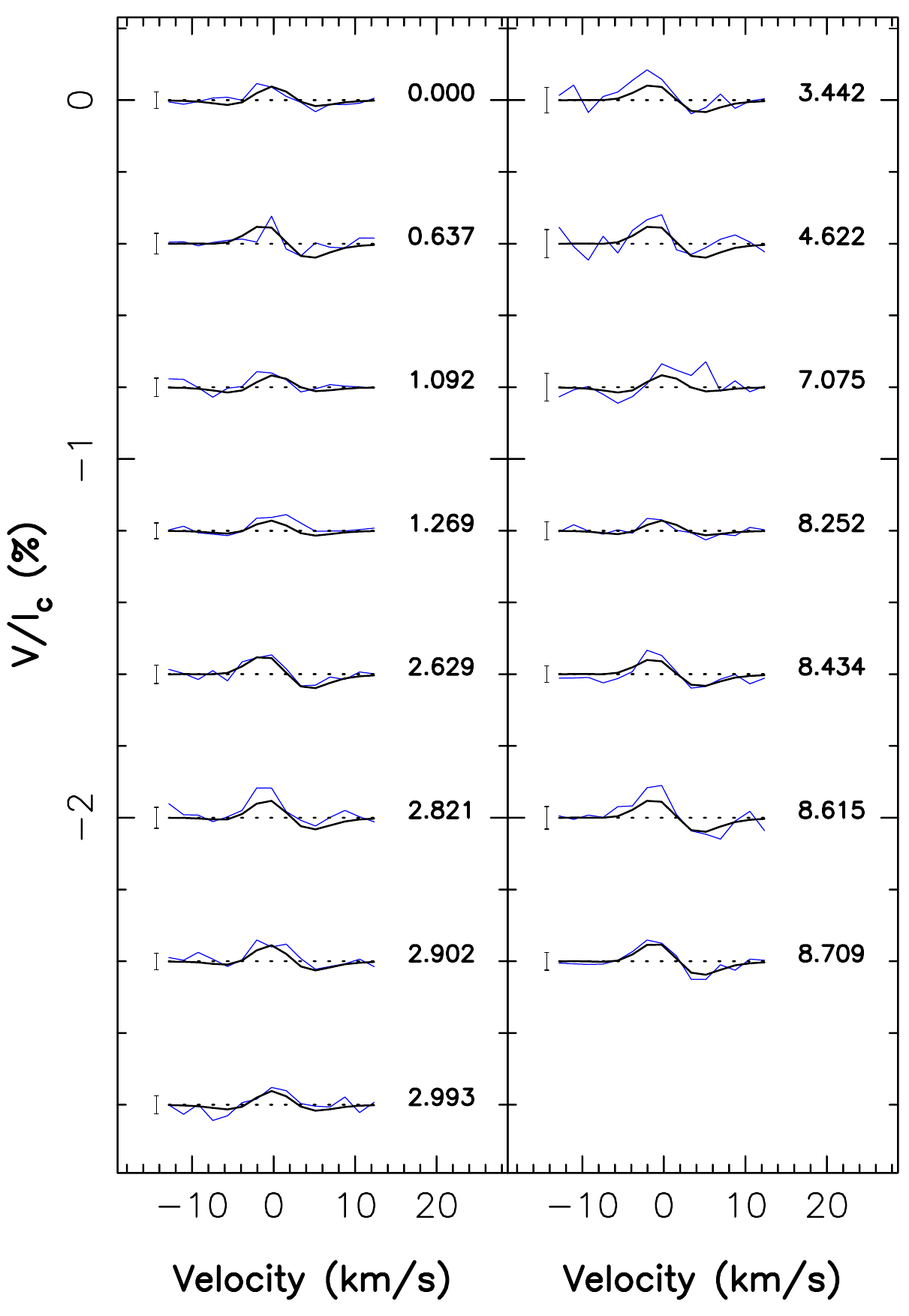} 
\includegraphics[scale=0.45,angle=0]{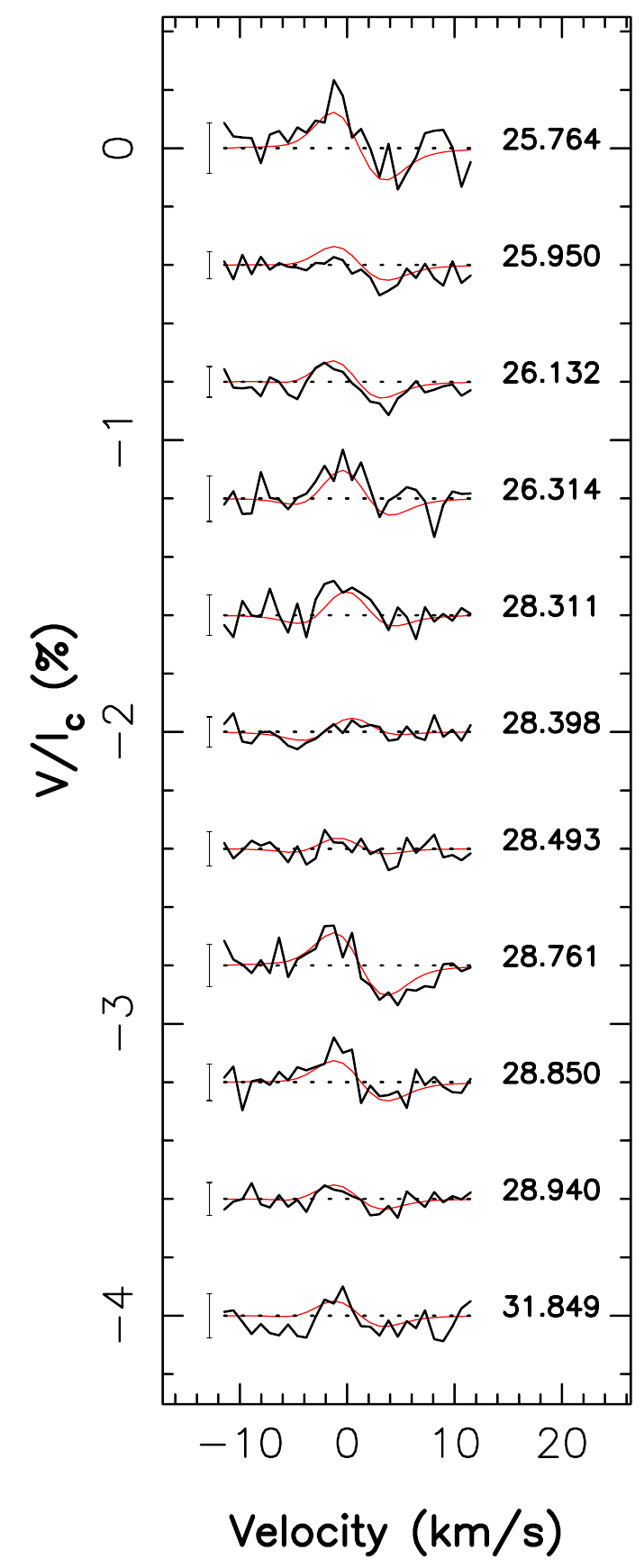} 
\includegraphics[scale=0.5,angle=0]{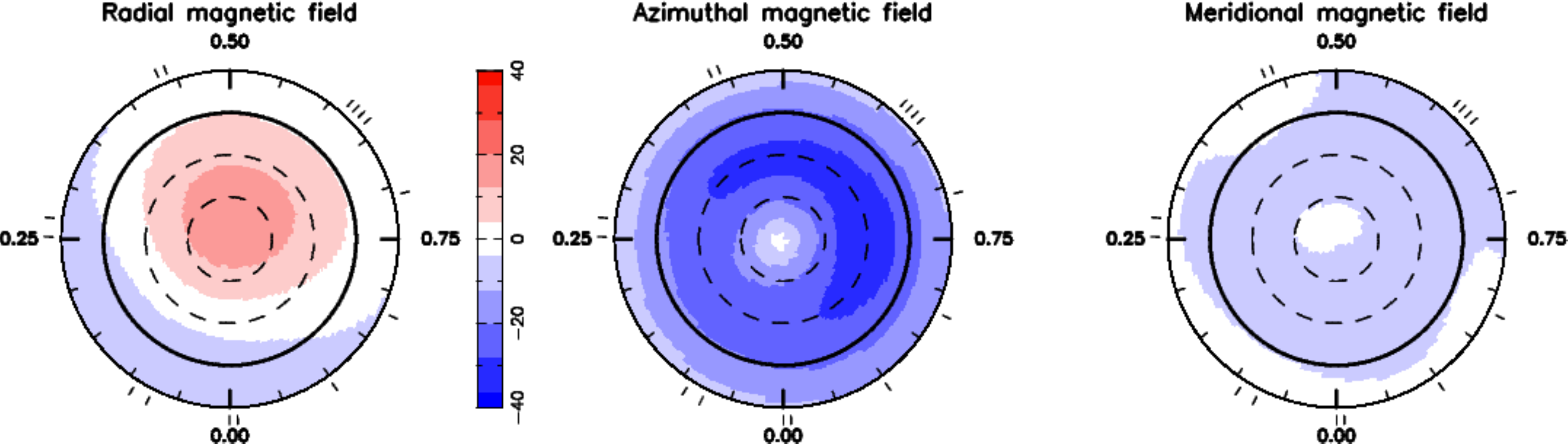} 
\includegraphics[scale=0.64,angle=0]{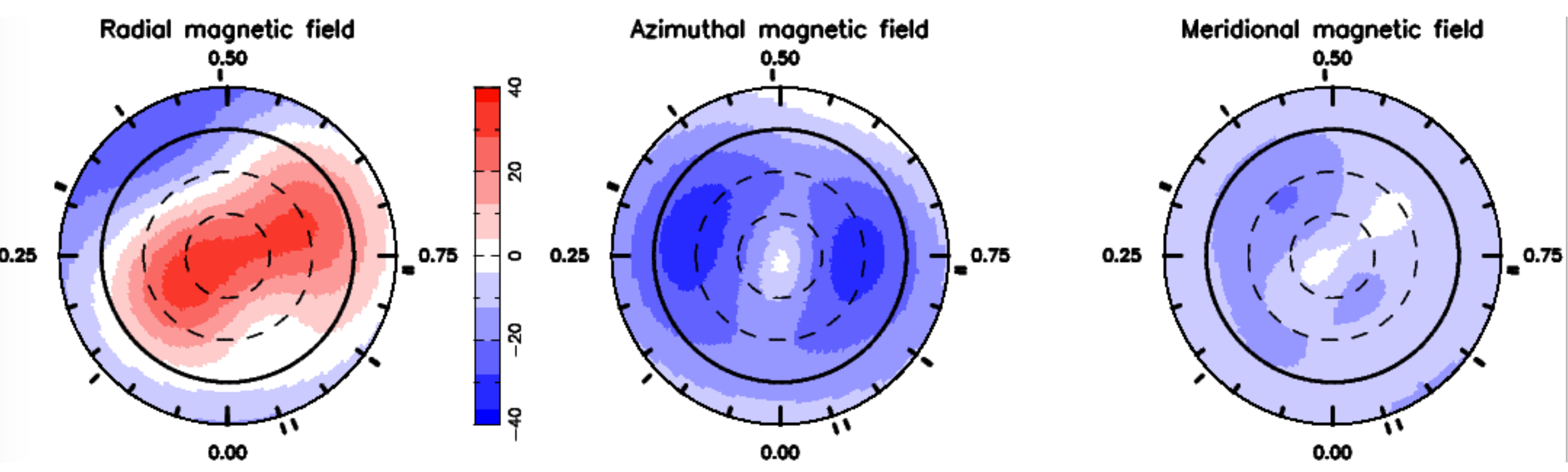} 
\caption{\textit{}
Same as Fig.~\ref{fig:map358} for GJ~846. LSD Stokes $V$ profiles in the top left and top right panels correspond toNARVAL and HARPS-Pol observations respectively.}
\label{fig:map846}
\end{figure*}


\section{Supplementary \bl, \vr, FWHM, \hal\ and \vs\ data}
\label{ann:suppl3}
\begin{figure*}
\includegraphics[scale=0.53,angle=0]{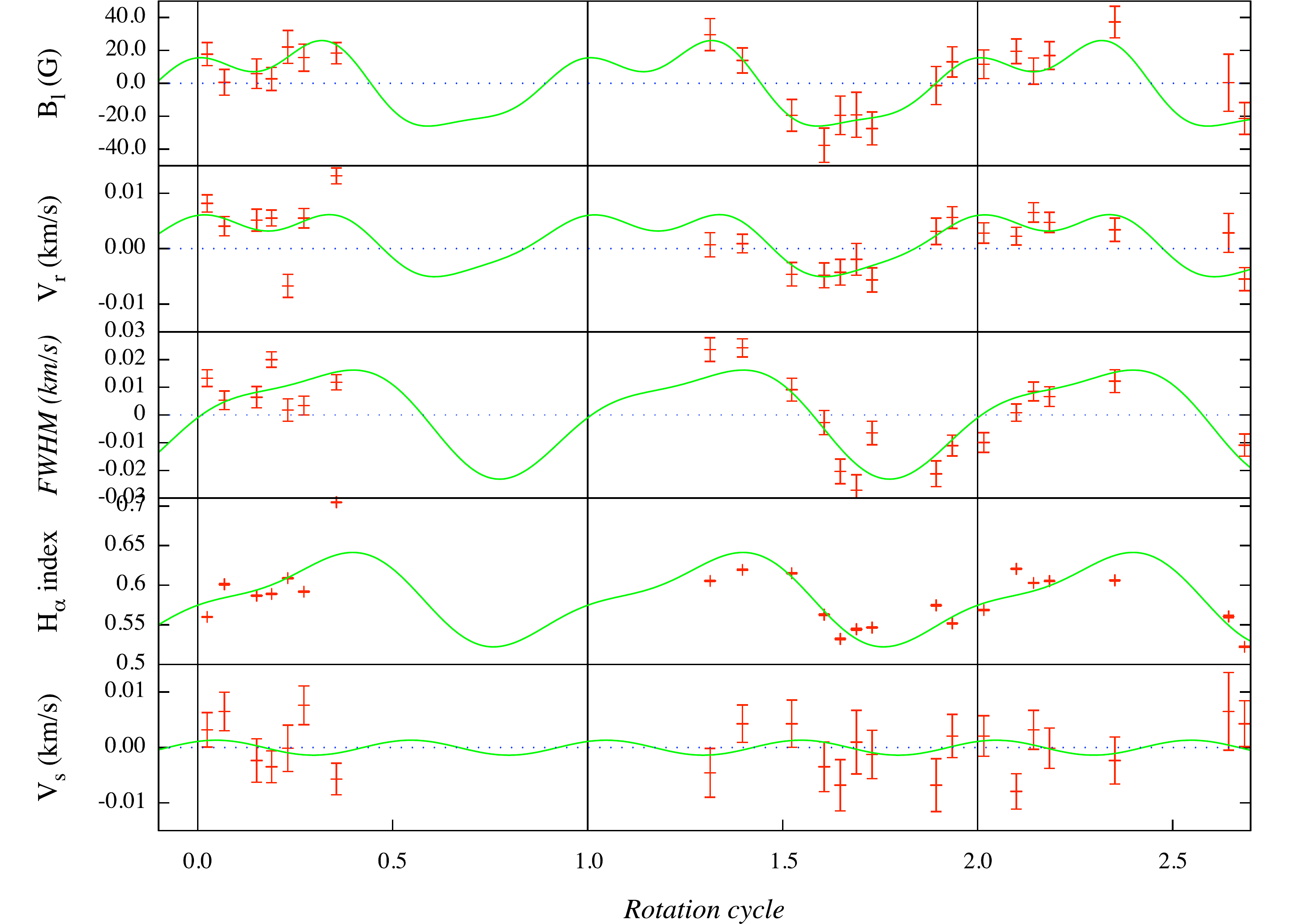} 
\includegraphics[scale=0.53,angle=0]{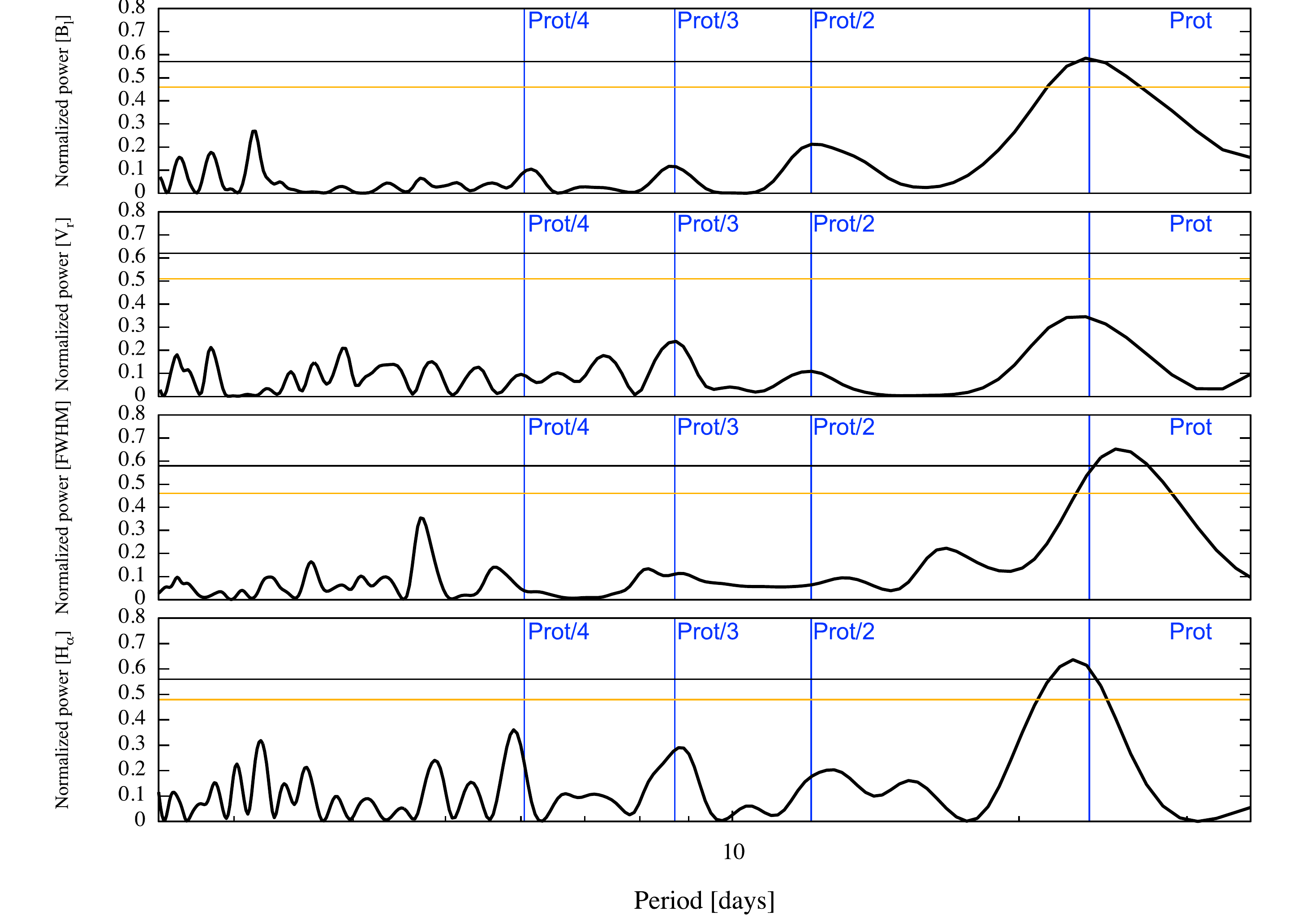} 
\caption{As Figure~\ref{fig:gj358} for GJ 479.}
\label{fig:gj479}
\end{figure*}

\begin{figure*}
\includegraphics[scale=0.53,angle=0]{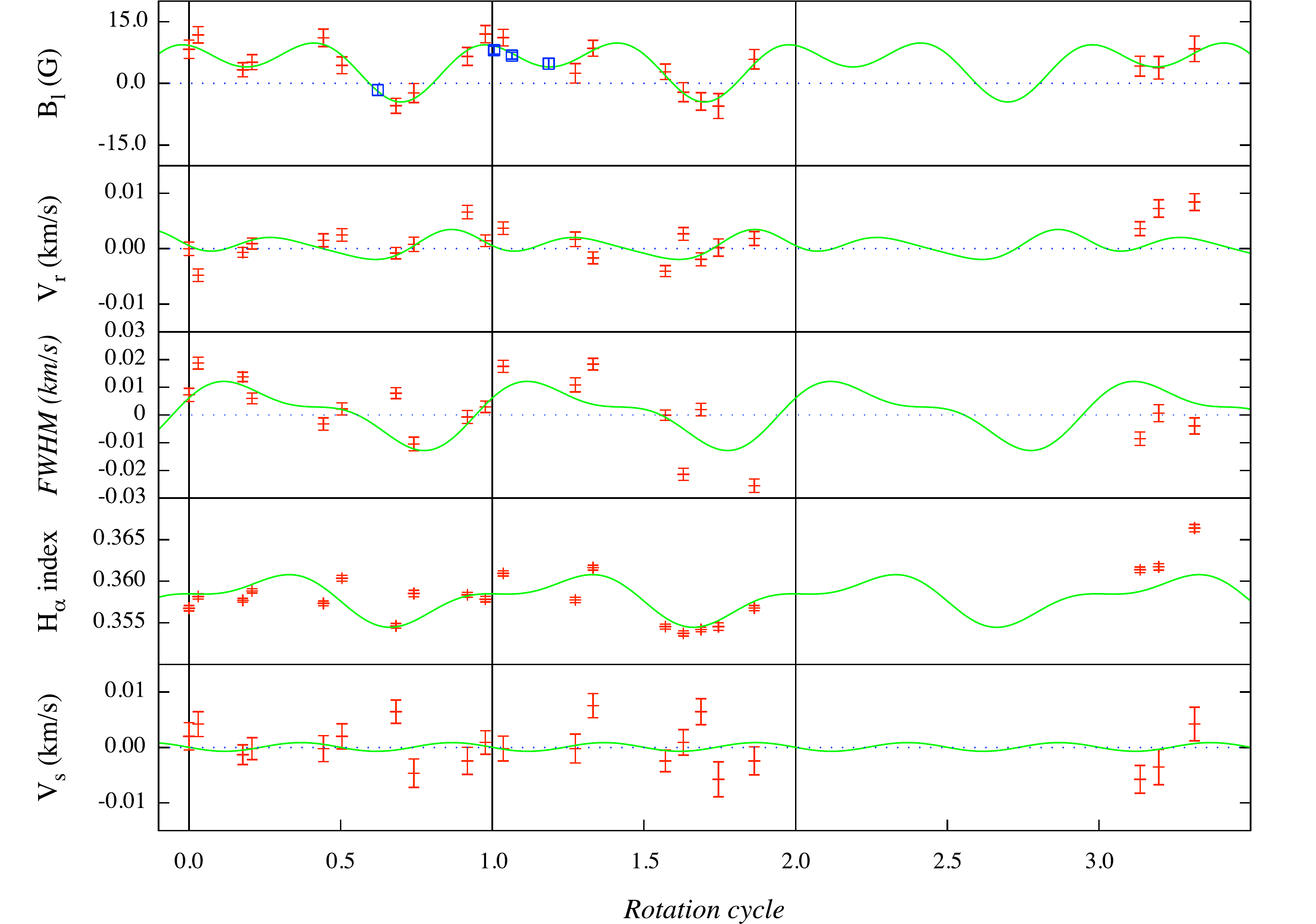} 
\includegraphics[scale=0.53,angle=0]{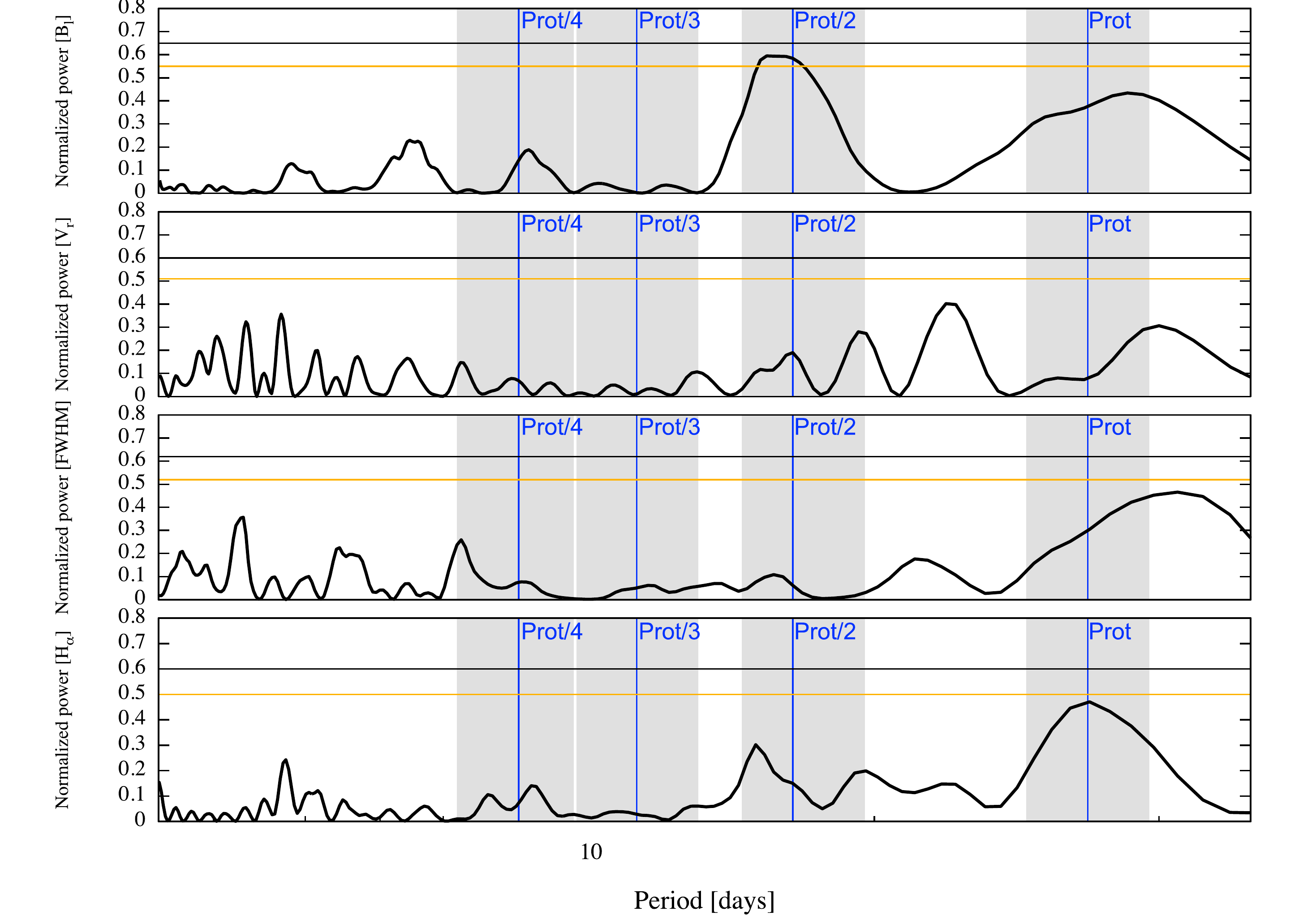} 
\caption{As Figure~\ref{fig:gj410} for GJ 205. Note that for this star, the DR is supposed only, and not measured from the data set.}
\label{fig:gj205}
\end{figure*}


\section{Complementary results of the simulations presented Sec.~5.2.}
\label{ann:suppl4}
\begin{figure*}
\begin{center}
\includegraphics[scale=0.4,angle=0]{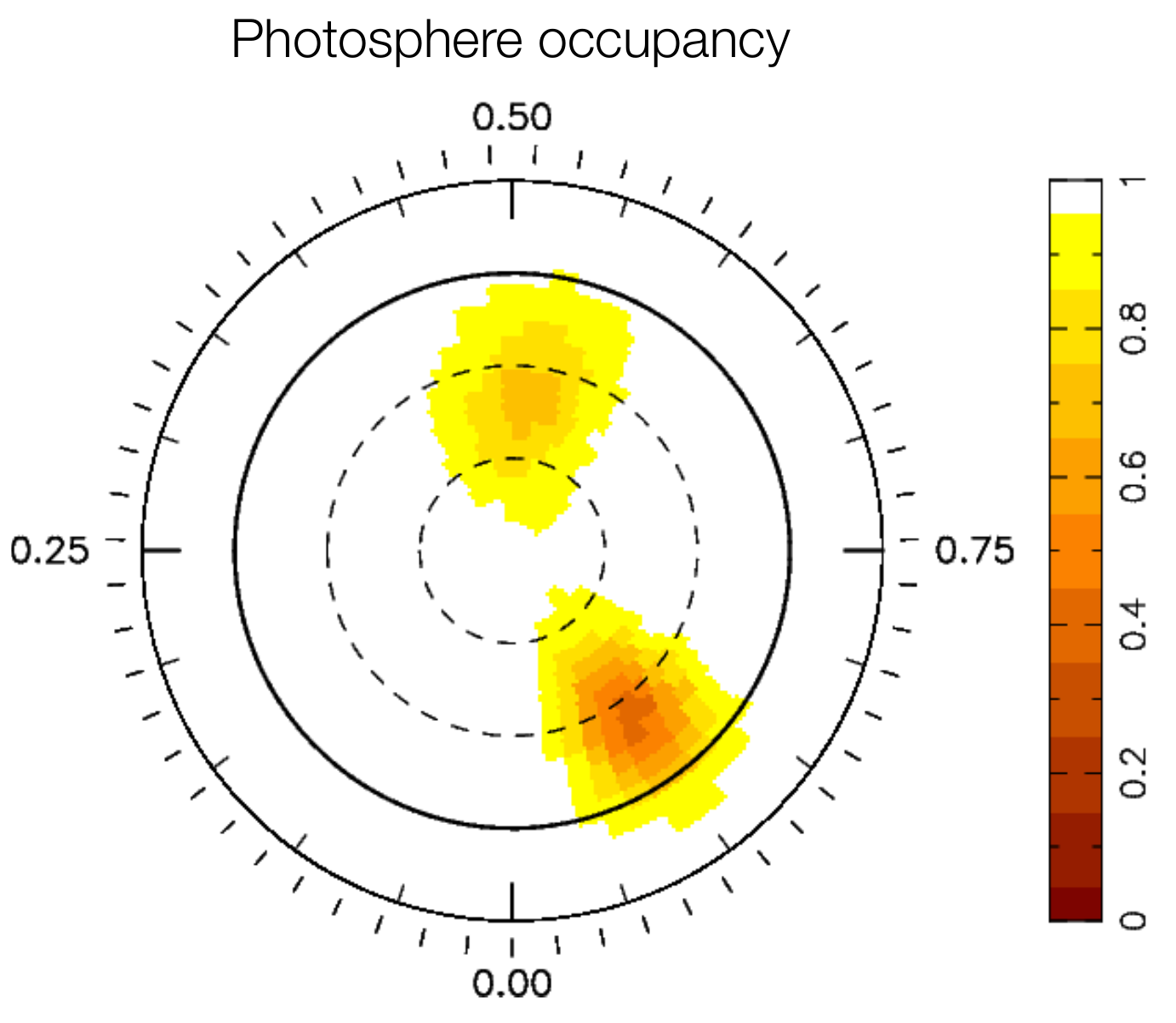}
\includegraphics[scale=0.2,angle=0]{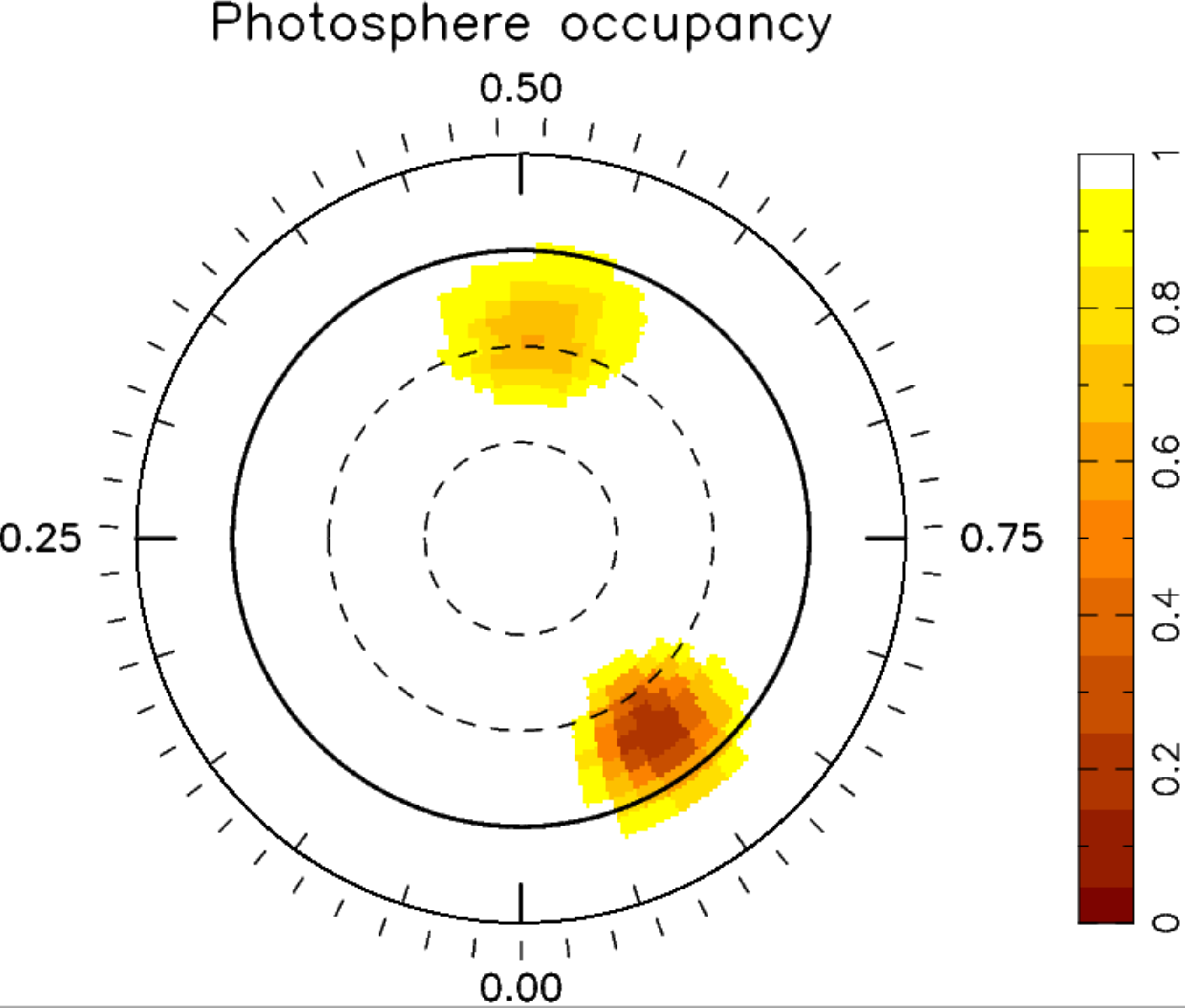}
\caption{ 
Reconstructed map obtained for a star with \vsini~=~1~\kms and $i$~=~60\degr\ with 2 spots covering 1.5\% of the stellar surface.
\textit{ Left}: Reconstructed map from $I$ with the sampling A,
\textit{Right}: Reconstructed map from $RI$ with the sampling A,
The colour-scale depicts the photosphere filling factor of each cell (white corresponding to a unspotted cell).
}
\label{fig:art3}
\end{center}
\end{figure*}


\section{Supplementary $RI$ spectra}
\label{ann:suppl5}
\begin{figure*}
\includegraphics[scale = 0.6]{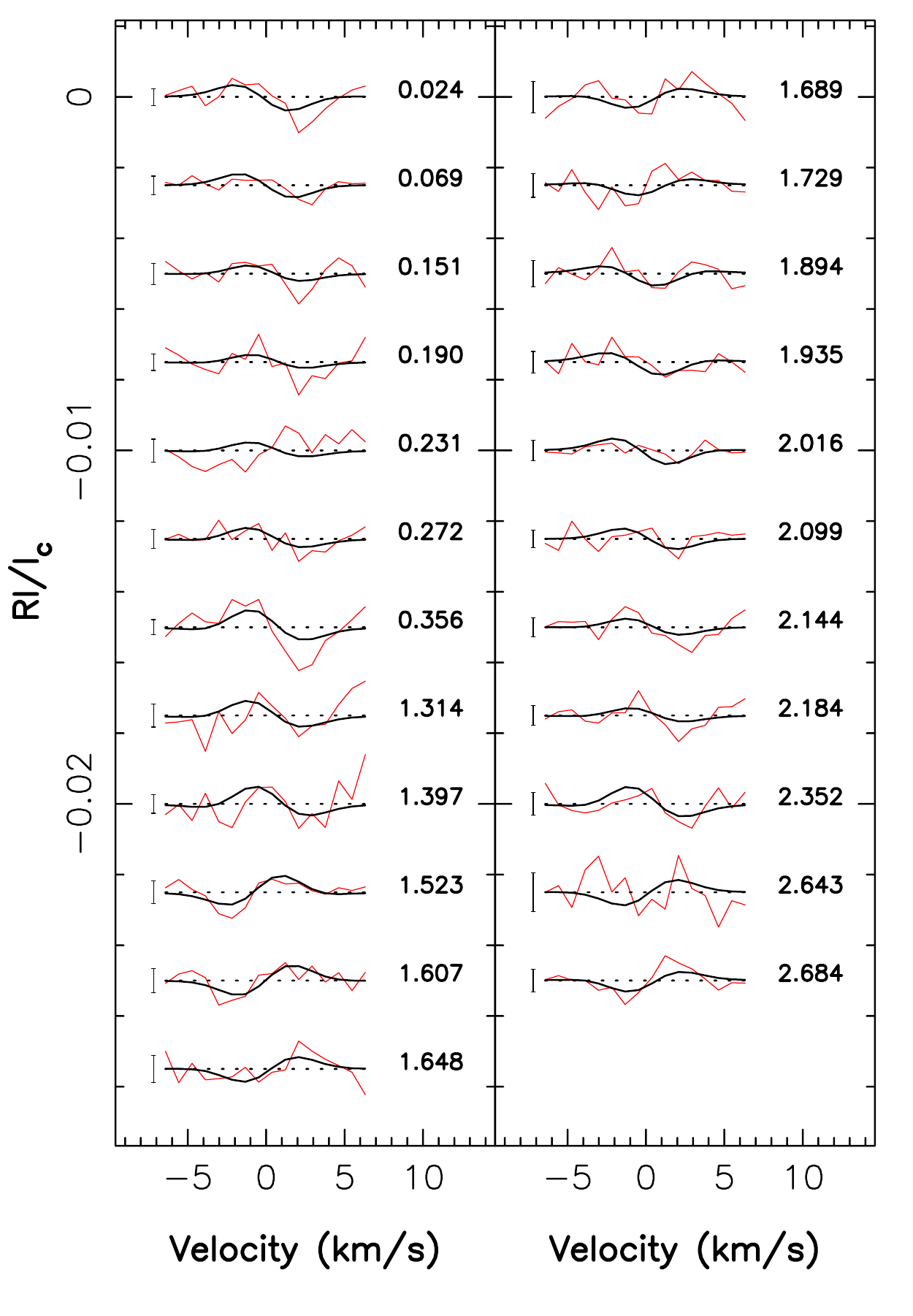}
\includegraphics[scale = 0.6]{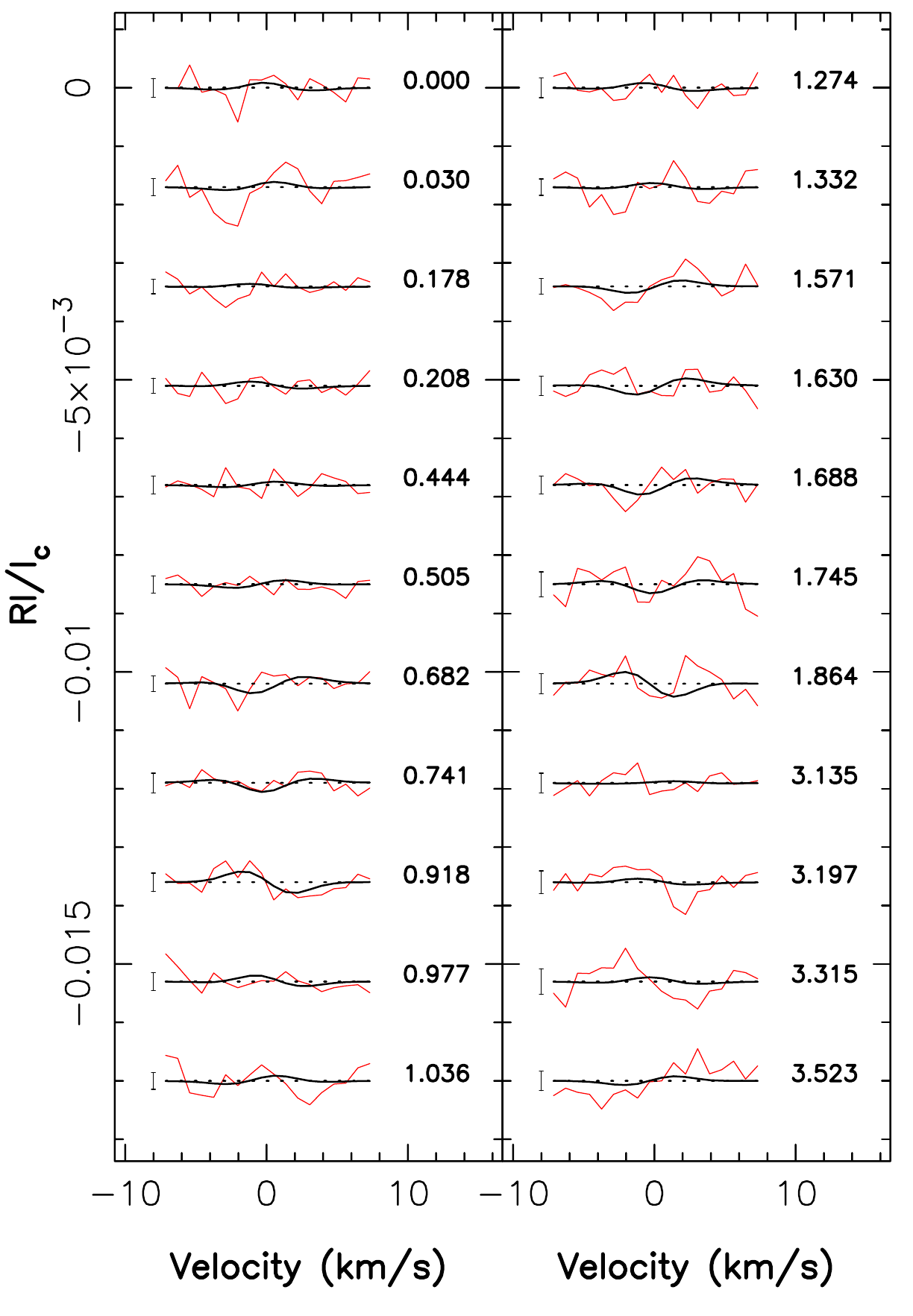}
\caption{
Same as Figure~\ref{fig:res358} for GJ~479 (left) and GJ~205 (right).
}
\label{fig:res479205}
\end{figure*}

\section{Supplementary \bl, \vr\ and FWHM reconstruction, and relative brightness maps}
\label{ann:suppl6}
\begin{figure*}
\begin{center}
\includegraphics[scale = 0.45]{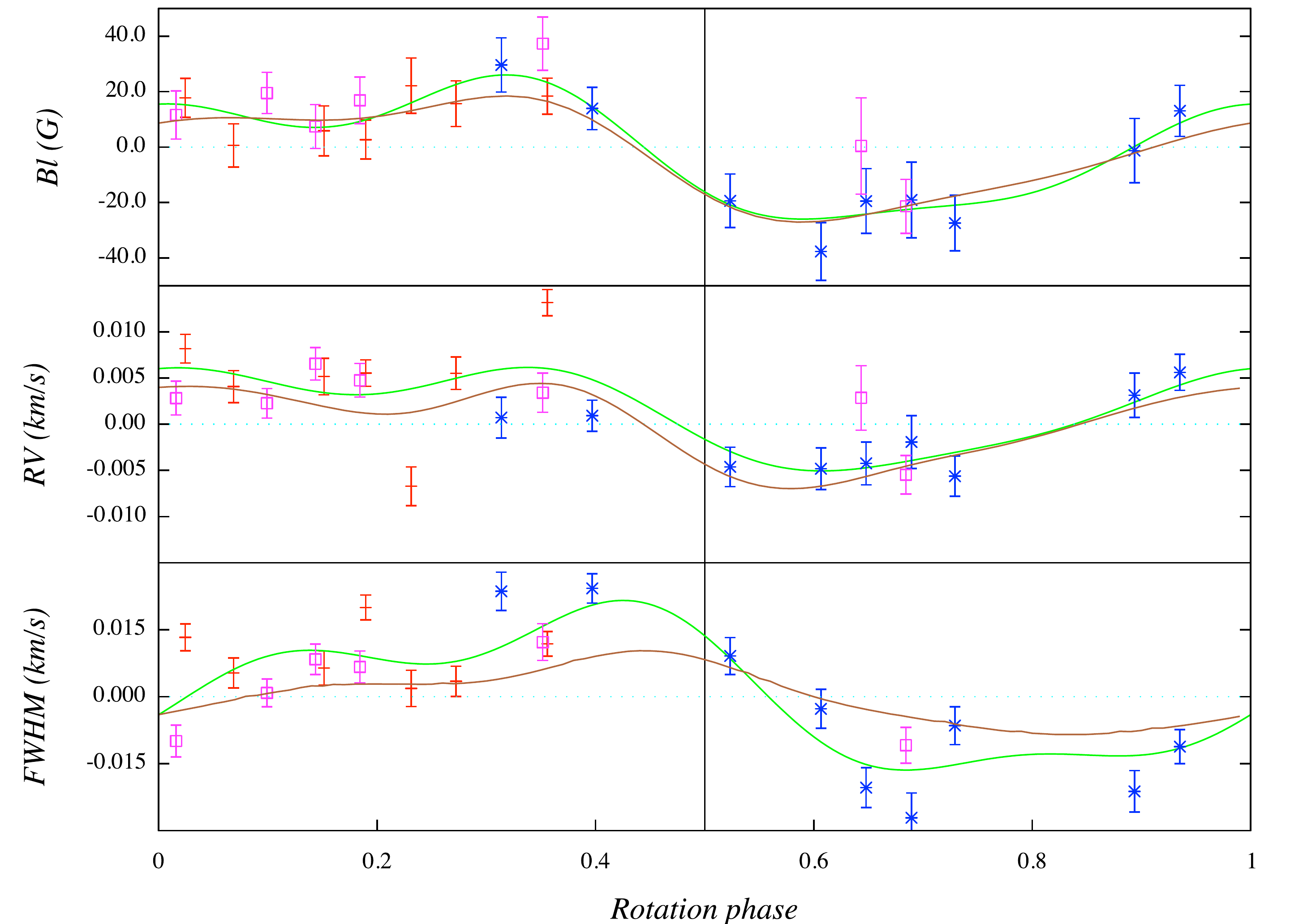}
\includegraphics[scale = 0.5]{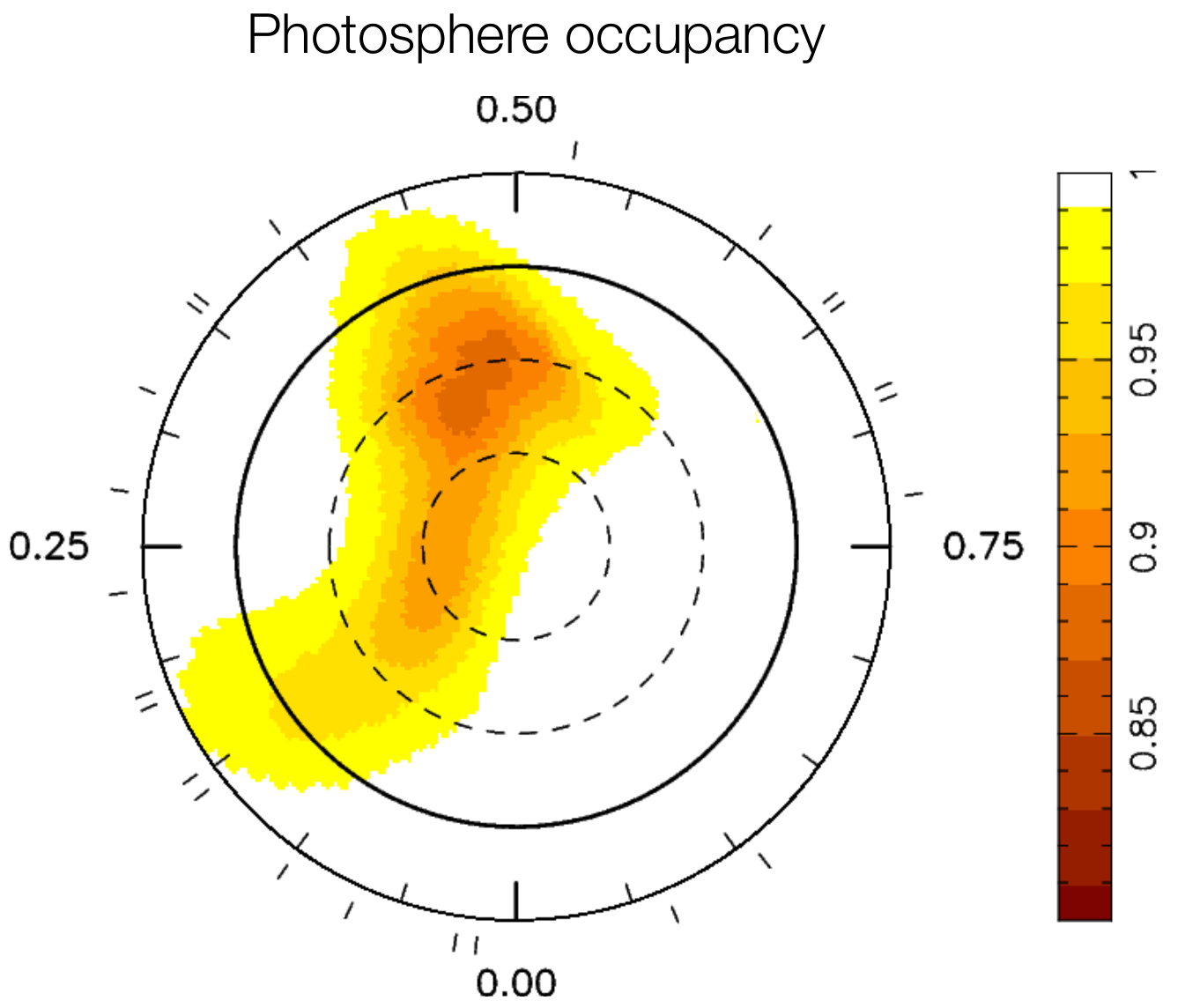}
\includegraphics[scale = 0.25]{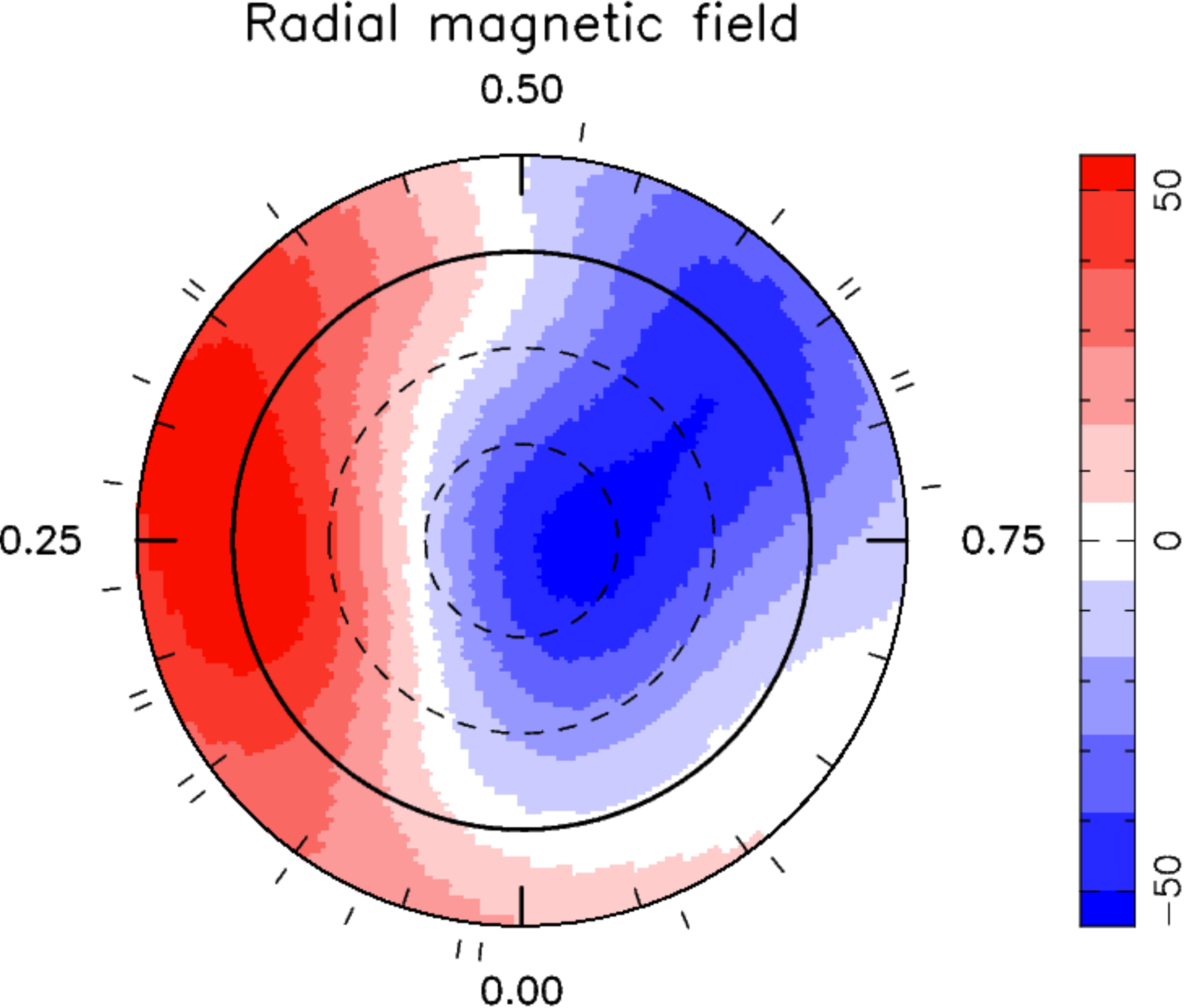}
\includegraphics[height = 6cm, width=16cm]{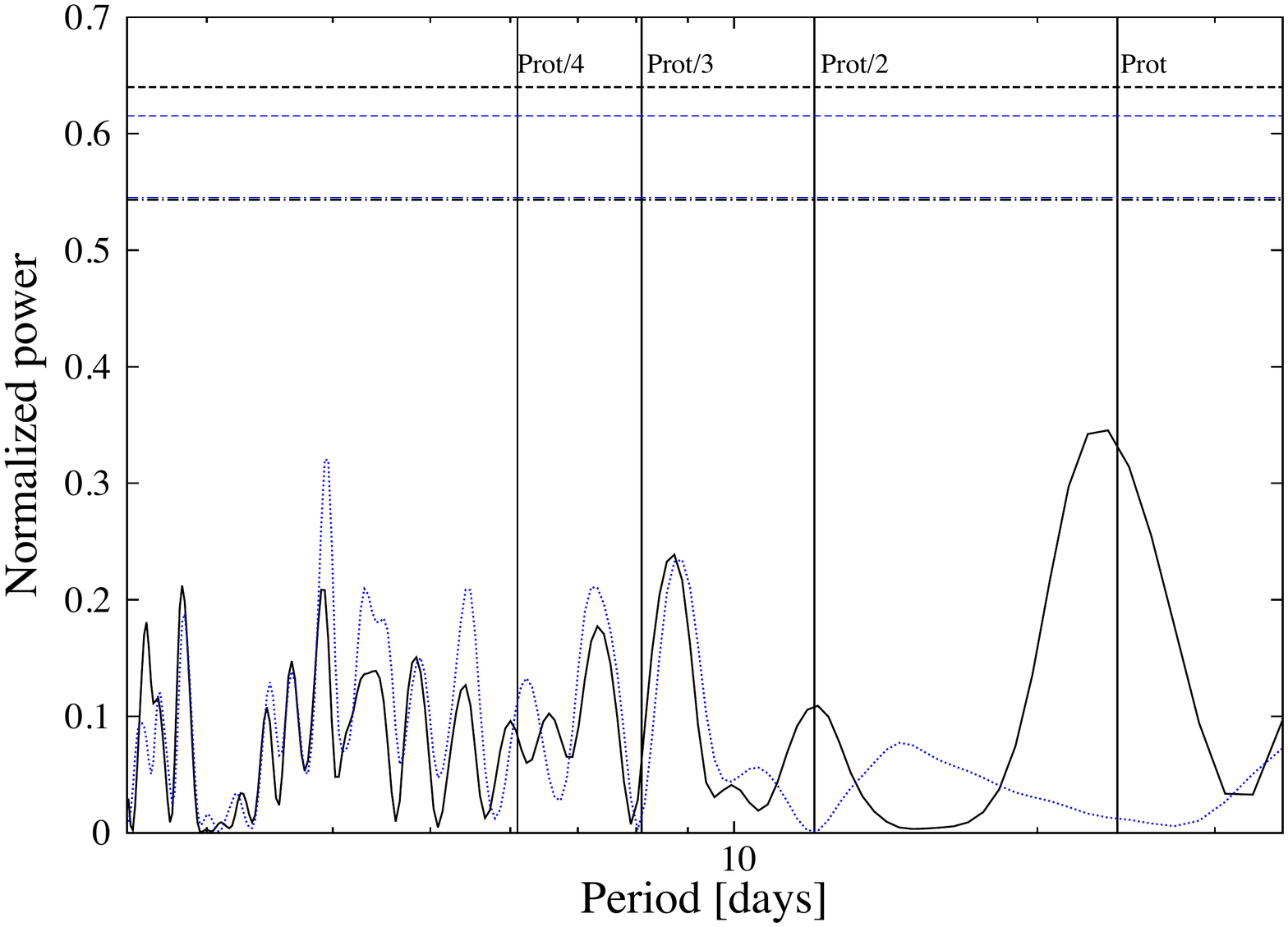}
\caption{
Same as Figure~\ref{fig:mapcq358} for GJ~479}
\label{fig:mapcq479}
\end{center}
\end{figure*}

\begin{figure*}
\begin{center}
\includegraphics[scale = 0.45]{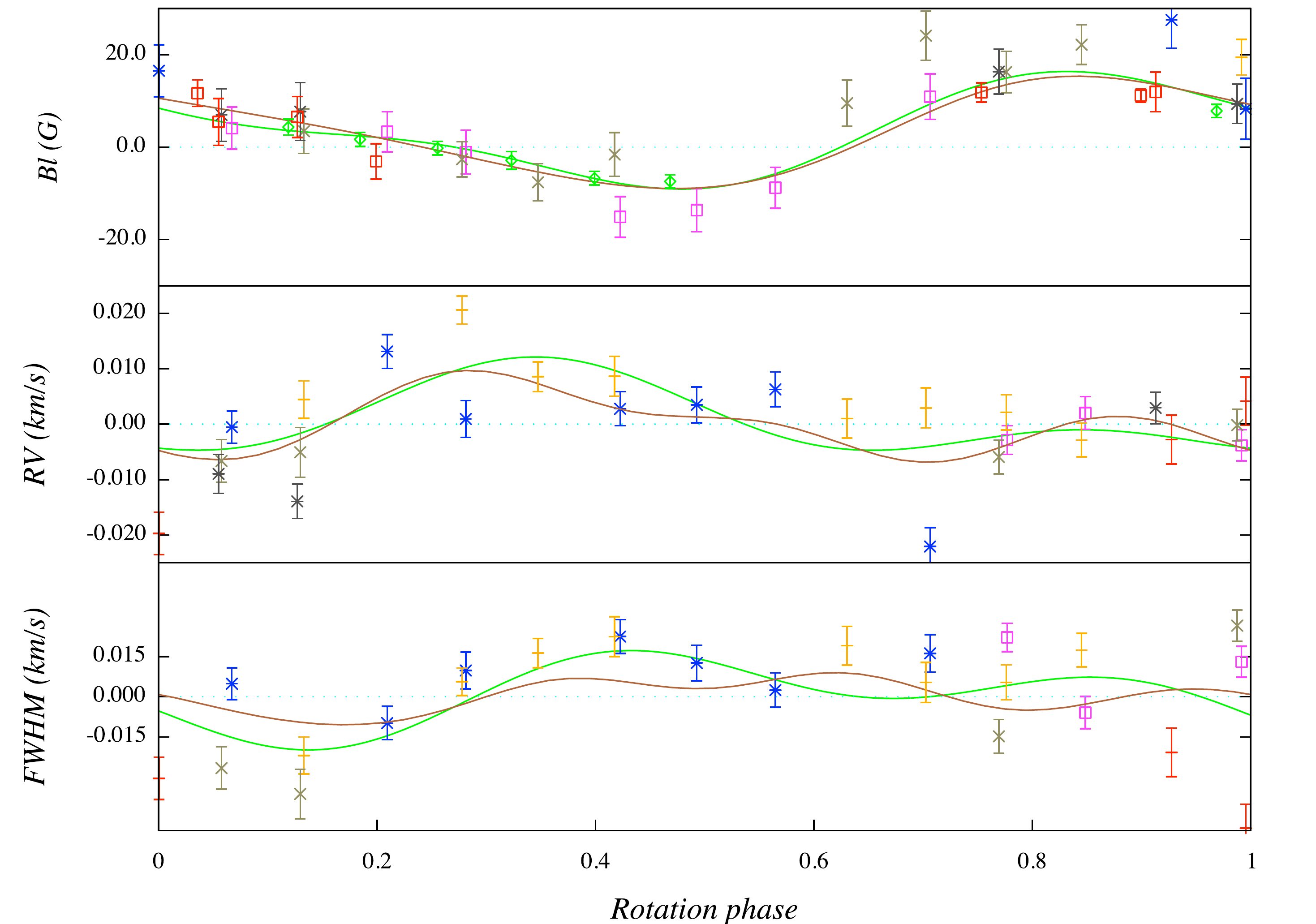}
\includegraphics[scale = 0.47]{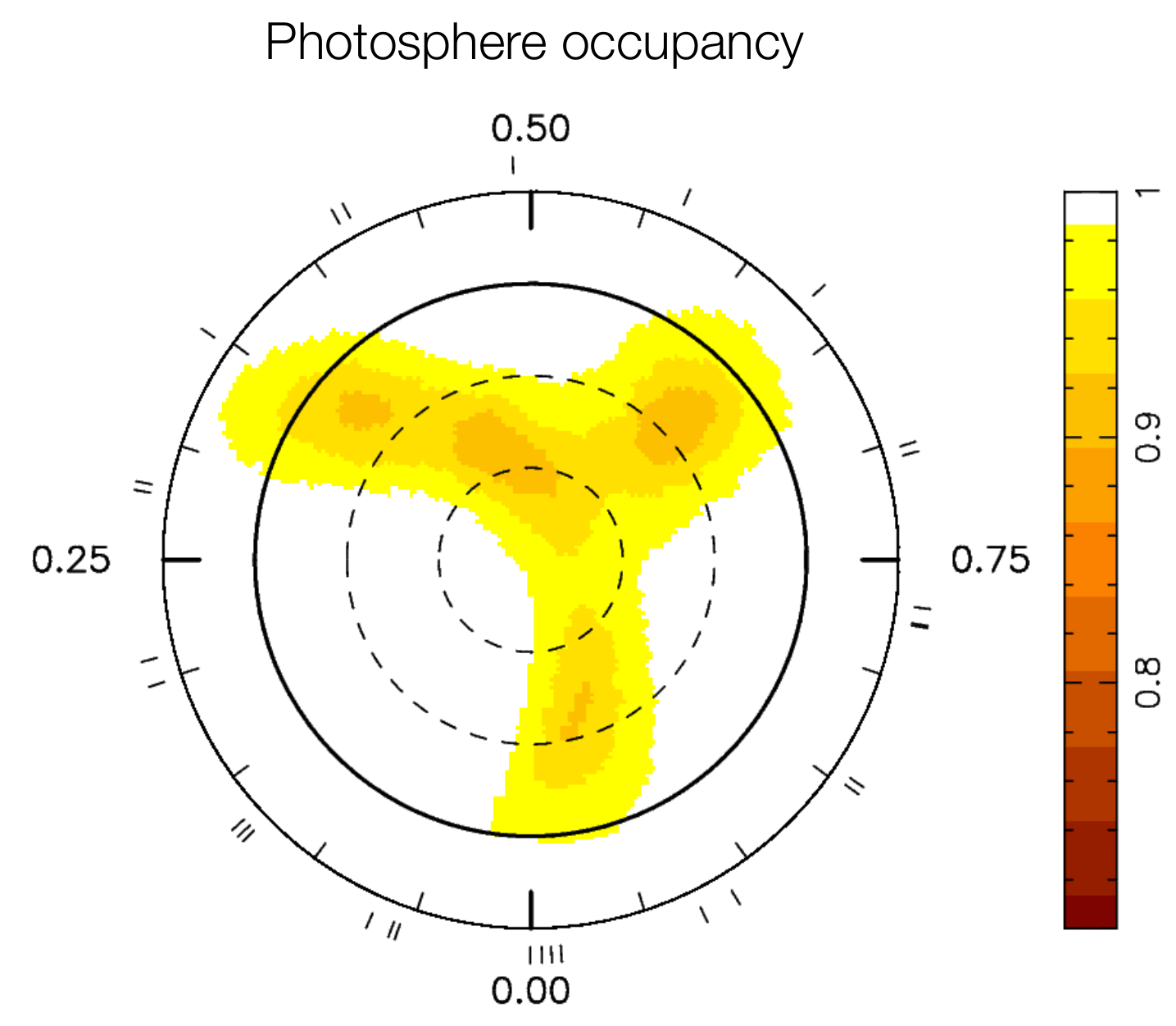}
\includegraphics[scale = 0.25]{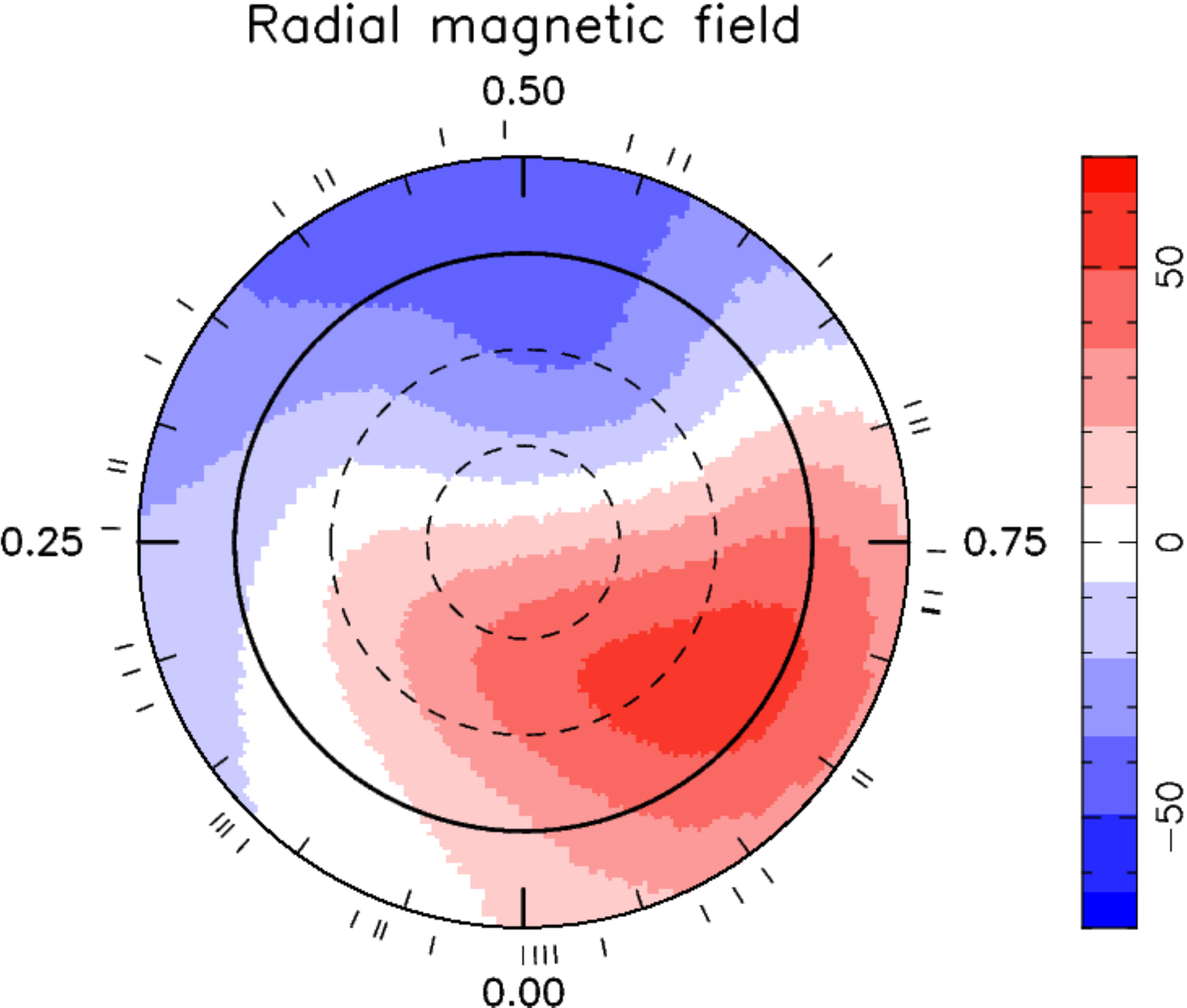}
\includegraphics[height = 6cm, width=16cm]{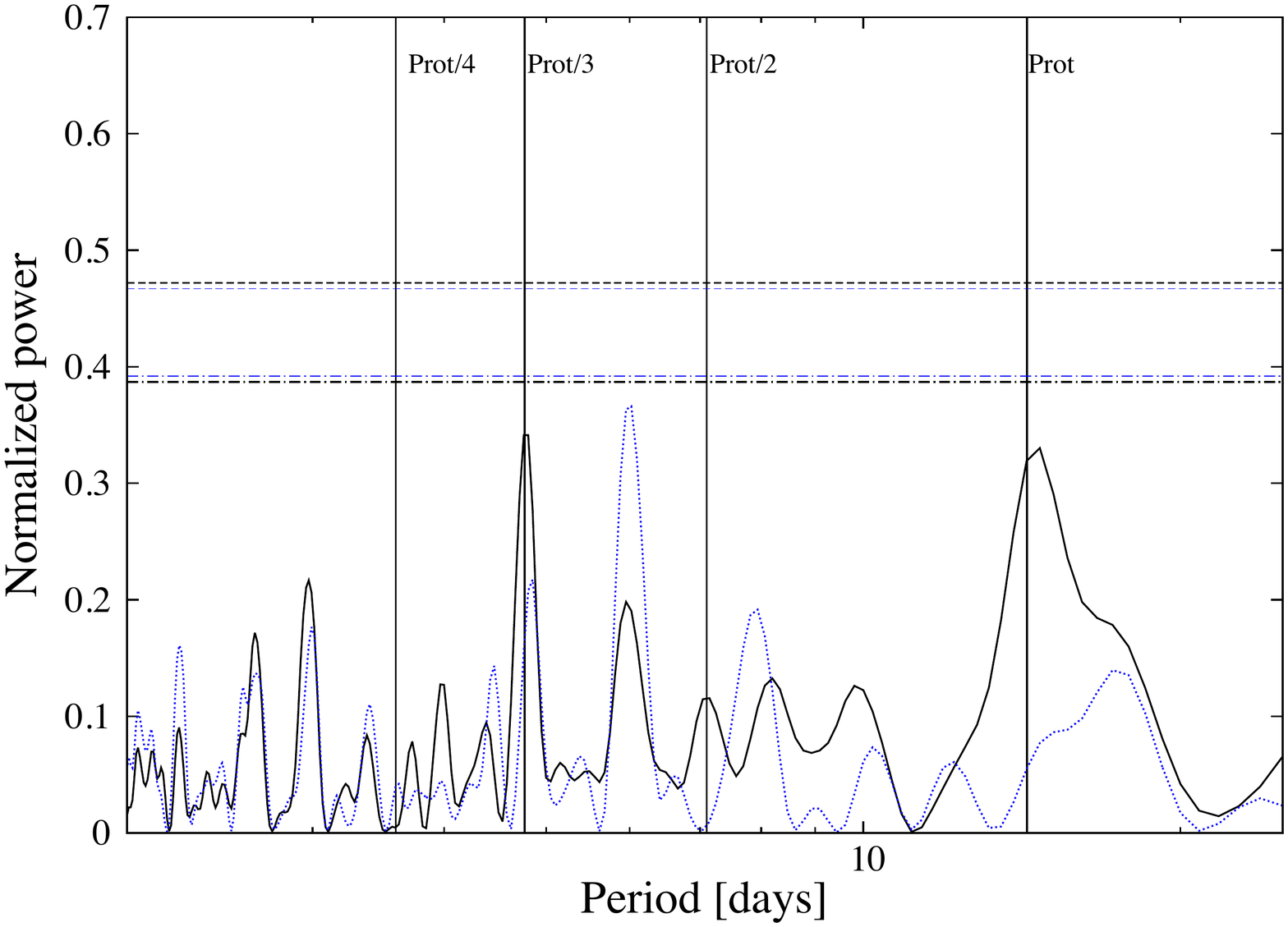}
\caption{
Same as Figure~\ref{fig:mapcq358} for GJ~410, from the whole data set.}
\label{fig:carteall_410}
\end{center}
\end{figure*}

\begin{figure*}
\begin{center}
\includegraphics[scale = 0.45]{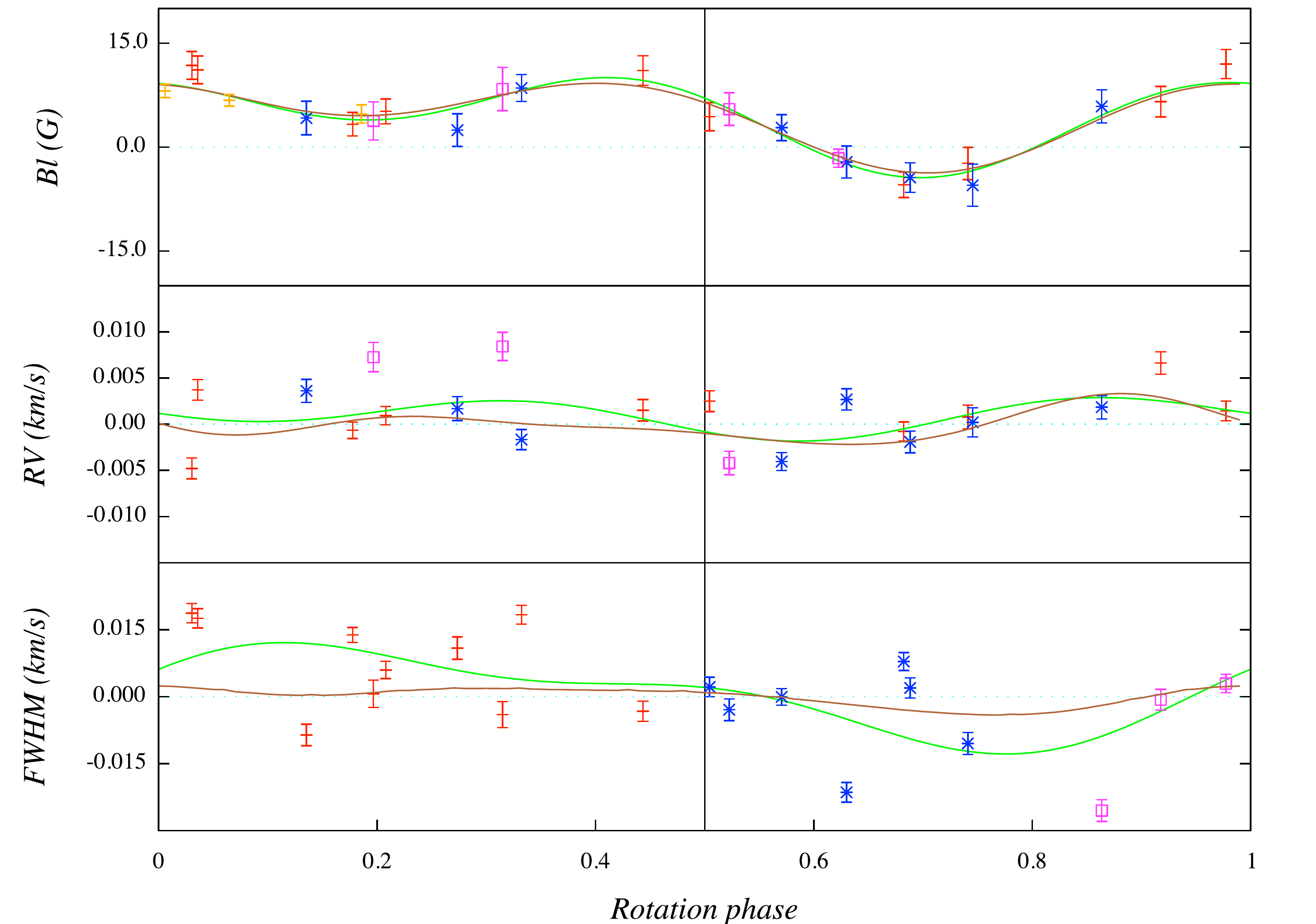}
\includegraphics[scale = 0.3]{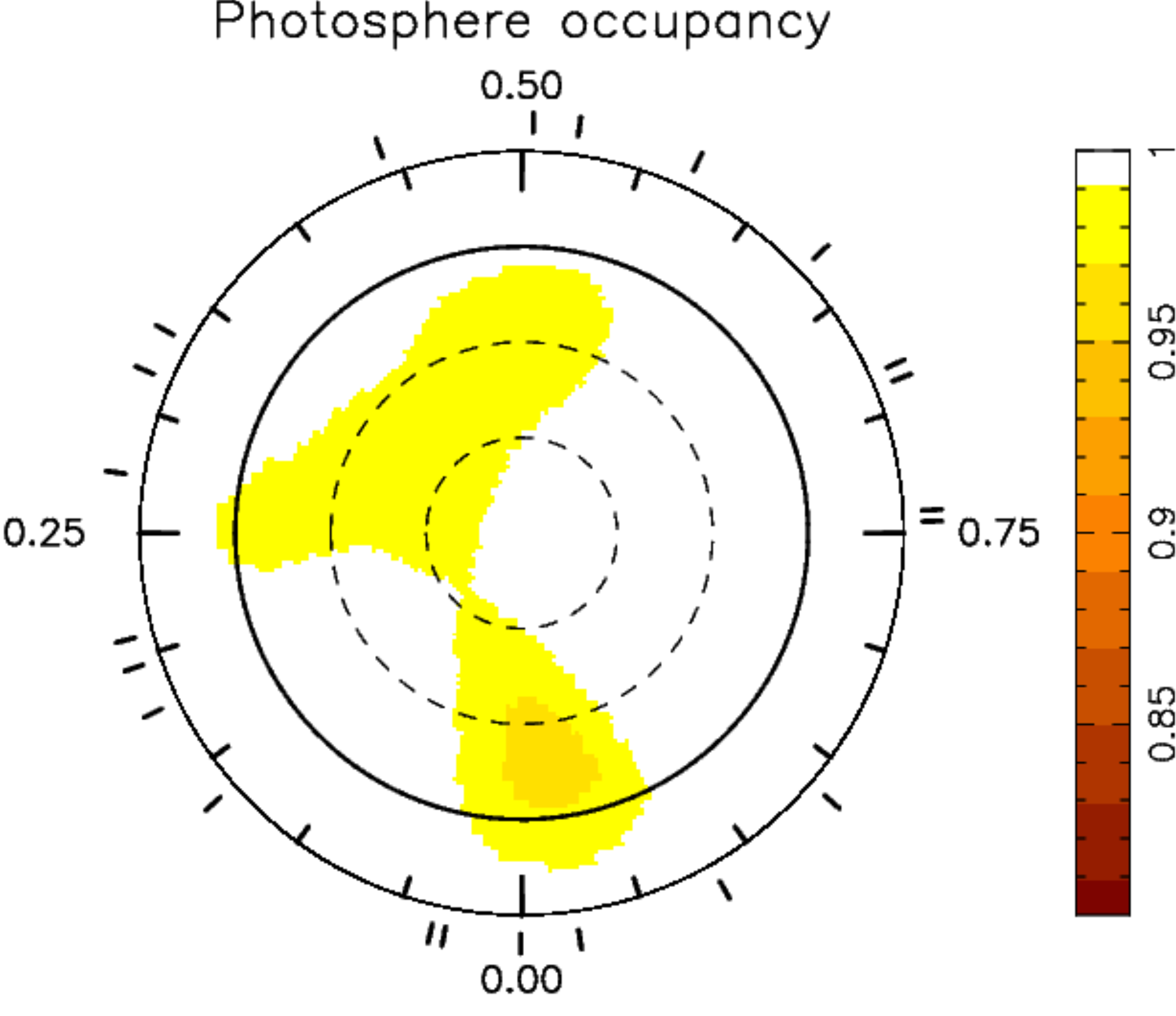}
\includegraphics[scale = 0.25]{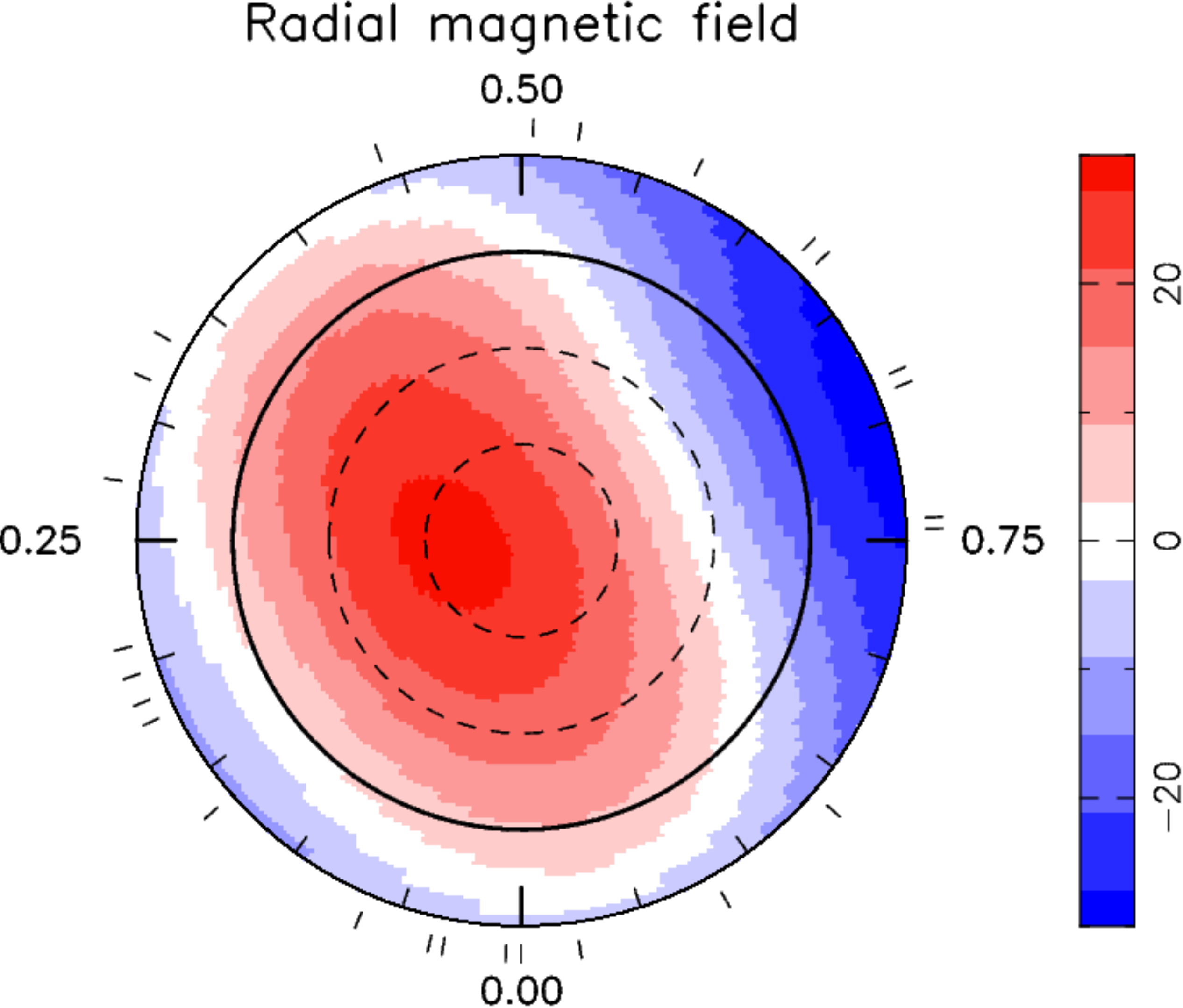}
\includegraphics[height = 6cm, width=16cm]{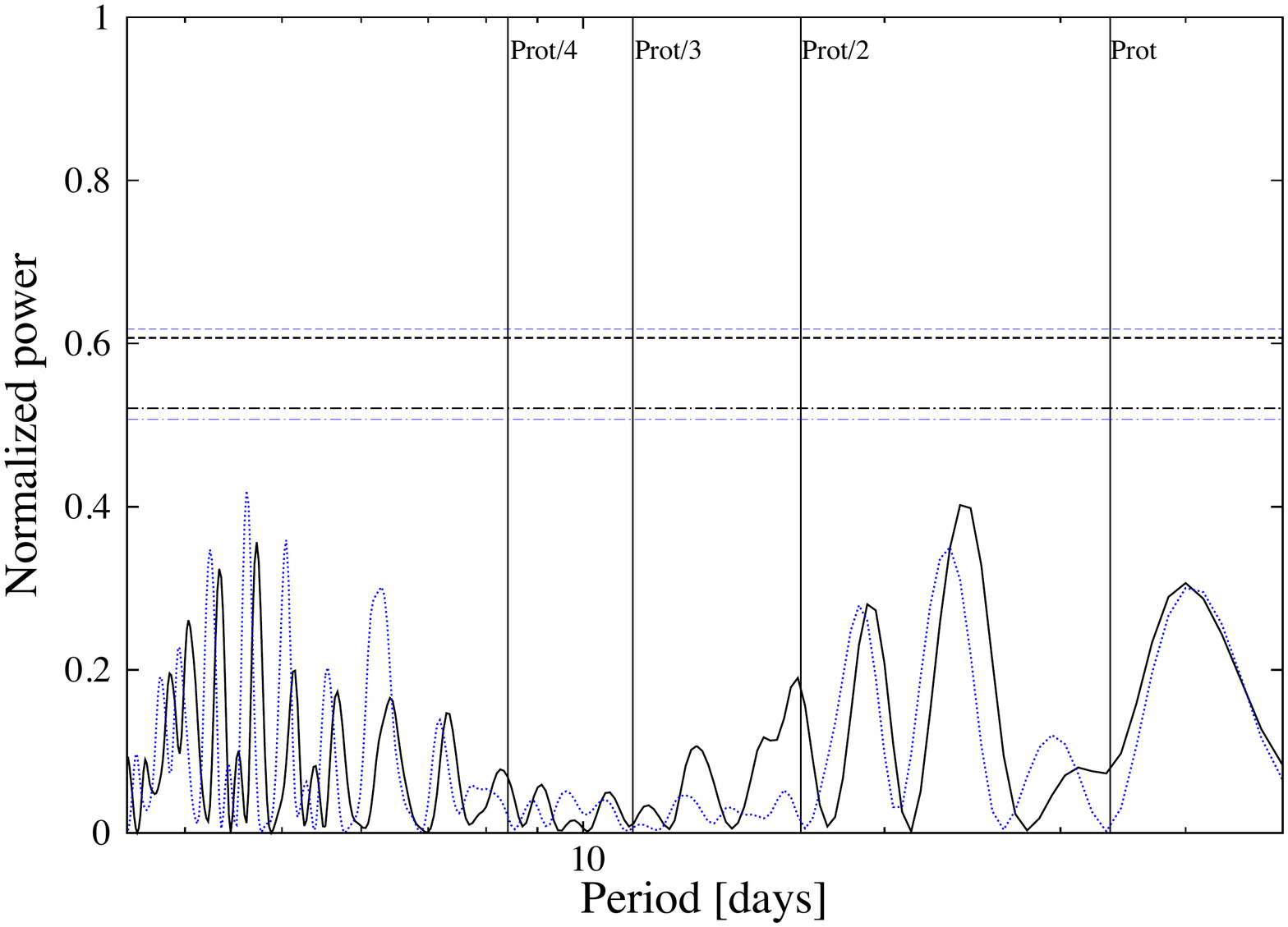}
\caption{
Same as Figure~\ref{fig:mapcq358} for GJ~205}
\label{fig:mapcq205}
\end{center}
\end{figure*}



\bsp	
\label{lastpage}
\end{document}